\documentclass{aastex631}

\hypersetup{linkcolor=red,citecolor=blue,filecolor=cyan,urlcolor=magenta}

\def\lessim{\mathrel{\hbox{\rlap{\hbox{\lower4pt\hbox{$\sim$}}}\hbox{$<$}}}}
\def\grtsim{\mathrel{\hbox{\rlap{\hbox{\lower4pt\hbox{$\sim$}}}\hbox{$>$}}}}

\usepackage{CJK}
\usepackage{amsmath}
\usepackage{amsfonts}
\usepackage{graphicx}

\shorttitle{Novae in M31}
\shortauthors{Shafter \& Hornoch}

\graphicspath{{./}{figures/}}

\begin{document}
\title{Fundamental Properties of Novae in M31}

\correspondingauthor{A. W. Shafter}
\email{ashafter@sdsu.edu}

\author[0000-0002-1276-1486]{Allen W. Shafter}
\affiliation{Department of Astronomy and Mount Laguna Observatory, San Diego State University, San Diego, CA 92182, USA}

\author[0000-0002-0835-225X]{Kamil Hornoch}
\affiliation{Astronomical Institute of the Czech Academy of Sciences, Fri\v{c}ova 298, CZ-251 65 Ond\v{r}ejov, Czech Republic}

\begin{abstract}

The peak luminosities and rates of decline
for a large sample of novae recently published by
Clark et al. have been analyzed using the
Yaron et al. nova models
to estimate fundamental properties of the M31 nova population.
The apparent white dwarf (WD) mass distribution is approximately Gaussian
with a mean $\langle M_\mathrm{WD} \rangle = 1.16\pm0.14~M_{\odot}$.
When corrected for recurrence-time bias, the mean
drops to $\langle M_\mathrm{WD} \rangle = 1.07~M_\odot$.
The average WD mass of the M31 nova sample
is found to be remarkably similar to that found by Shara et al. in their
study of 82 Galactic novae, but $\sim0.15~M_\odot$ more massive than the mean
recently determined by Schaefer in his comprehensive
study of more than 300 systems. 
As expected, the average WD mass for the
recurrent novae included in the M31 sample,
$\langle M_\mathrm{WD} \rangle = 1.33\pm0.08~M_{\odot}$,
is significantly higher than that for novae generally.
Other parameters of interest, such as the accretion rate,
velocity of the ejecta, and the predicted recurrence time, are characterized
by skewed distributions with large spreads about means of
$\langle \log \dot M (M_\odot~\mathrm{yr}^{-1}) \rangle \simeq -9.27$,
$\langle V_\mathrm{max} \rangle \simeq 1690$~km~s$^{-1}$, and
$\langle \log P_\mathrm{rec}~\mathrm{(yr)} \rangle \simeq 4.39$, respectively.
The role of hibernation in affecting the $\dot M$ and $P_\mathrm{rec}$
distributions is briefly discussed.
Finally, the nova properties were studied as a function of apparent
position (isophotal radius)
in M31, with the preponderance of evidence failing to establish
any clear dependence on stellar population.

\end{abstract}


\section{Introduction}

Nova eruptions result from a thermonuclear runaway (TNR) on the
surface of an accreting white dwarf (WD) in a close binary system.
The mass donor, commonly referred to as the secondary star
because it is the less massive component, fills it Roche lobe
and transfers material to the WD (the primary) via
and accretion disk. As the material accumulates on the
WD, the temperature and density at the base of the accreted
layer increase to the point where hydrogen burning
can commence. If the material is deposited sufficiently
slowly ($\lessim10^{-6}~M_{\odot}$~yr$^{-1}$), it
will become at least partially degenerate prior to Hydrogen
ignition, resulting in a TNR
\citep[e.g., see][and references therein]{Shara1989,Starrfield2016}.
The explosion and subsequent ejection of all or part
of the accreted material leads to an eruption that can reach
an absolute magnitude of $M_V\sim-10$, making novae among the
most luminous transient sources known.

After a nova event, accretion onto the WD eventually resumes and the
process repeats over timescales that can be as short as a year
\citep[e.g., see][]{Darnley2014,Tang2014}. \citet{Ford1978} has shown
that a significant fraction of nova progenitors must undergo multiple outbursts
in order to explain the observed nova rate in M31.
Systems with relatively
short intervals between successive eruptions, $P_\mathrm{rec}\lessim100$~yr,
and where more than one eruption has been observed, are referred to
explicitly as ``Recurrent Novae" (RNe).

The observed properties of novae, such as their peak luminosities and
subsequent rates of decline (typically parameterized by $t_2$,
the time in days to decline by 2 mag from maximum light),
the maximum velocity of the ejected material and the
the recurrence time depend critically on at least
two parameters of the progenitor binary: the mass of the WD, and
the rate of accretion onto its surface, with the latter being important
for determining the temperature of the accreted layer.
Theory and models have shown that
massive WDs accreting high rates only need to accrete a relatively small amount
of material (the ignition mass, $M_\mathrm{ign}\lessim10^{-5}~M_\odot$)
before a TNR is triggered \citep[e.g.,][]{Shara1981,Townsley2005}. The small
ignition mass leads in turn to a small ejected mass and a
rapid photometric evolution (i.e., a short $t_2$ time). In addition,
the high
accretion rate coupled with the small ignition mass leads to
a short recurrence time. Thus, such systems are characteristic of the
RNe.

Most studies of the fundamental properties of novae have involved
Galactic systems, however these studies have been often
complicated by the effects
of extinction. This is particularly true for novae in the direction
of the galactic center making the study of novae from different stellar
populations (bulge and disk) notoriously difficult. On the other hand,
the study of extragalactic novae avoids or minimizes these problems by
allowing the study of novae from different populations
in nearby, spatially-resolved spiral galaxies,
or galaxies of different Hubble types, from
late type spirals like M33 \citep{DellaValle1994,Williams2004}
and M51 \citep{Mandel2023},
to giant ellipticals in the Virgo cluster
\citet[][]{Pritchet1987,Curtin2015,Shafter2017b,Shara2016,Shara2023}.

The Andromeda Galaxy (M31), our closest neighbor galaxy at a distance of
approximately 780 kpc \citep{McConnachie2005},
offers an unparalleled
opportunity to study a large population of novae at a nearly uniform
distance, mitigating the heavy extinction and distance uncertainties 
that complicate Galactic nova studies. It has been a cornerstone
for nova studies going back to the pioneering work of Hubble and collaborators
in the early 20th century \citep{Hubble1929}.
Since Ritchey discovered the first nova in M31 on
1909 September 13 \citep{Ritchey1917}, over 1300 novae have been
cataloged to date
\citep[e.g., see][and references therein]{2019enhp.book.....S, 2020A&ARv..28....3D}.
About half these novae have been discovered serendipitously as part
of unrelated studies, or through the efforts of the increasing
number of amateur astronomers who now routinely patrol the galaxy.
The remaining half were discovered as part of a relatively
small number of targeted nova surveys that continued after Hubble's
initial survey \citep[most notably][]{Arp1956,
Rosino1964,Rosino1973,Ciardullo1987,
Shafter2001,Darnley2004,Rector2022}.
Taken together, these studies have established a nova rate
of $\sim50$~yr$^{-1}$, which is comparable to estimates for the
Milky Way \citep[e.g.][]{Shafter2017,De2021,
Kawash2022}. 

In addition to the determination of the nova rate, there have been
several attempts to study the nova populations in M31. In an early
study employing H$\alpha$ imaging (to provide better contrast against the
bright background of M31's inner bulge) to discover 40 novae,
\citet{Ciardullo1987} suggested that M31's nova population
was dominated by the galaxy's bulge component. A little over a decade later
\citet{Shafter2001} confirmed these results with a similar study of
82 novae imaged in H$\alpha$ with the 40-in reflector at Mt Laguna Observatory.

A comprehensive study of M31's nova populations was presented
by \citet{Shafter2011} who compiled broad-band photometric and spectroscopic
data for a large sample of M31 novae. With the exception of
a slight tendency for slower novae ($t_2\lessim25$~d) to be found closer to
the center of M31, the study failed to find
a significant dependence of nova properties (spectroscopic class and
$t_2$ times) on spatial position, and hence stellar population in M31.
A few years later, \citet{Shafter2015} studied the RN
population of M31 and again found no evidence that the spatial
distribution of the RN population differed from
that of novae generally.

Despite the wealth of data on the observed behavior of M31 novae,
very little is known about their fundamental properties
such as their WD masses and accretion rates.
On the other hand, despite the challenges alluded to earlier,
tremendous progress has been made in recent years
in our understanding of the fundamental properties of Galactic novae including
their distances, luminosities and WD masses
\citep[e.g.,][]{Shara2018,Schaefer2018,Selvelli2019,Schaefer2022,Schaefer2025}.

In this paper we exploit the comprehensive and uniform sample of
$R$-band light curves for the large sample of novae in M31 recently
published by \citet{Clark2024} to explore for the first time
the fundamental properties (WD masses, accretion rates,
ejecta velocities, and predicted recurrence times)
of novae in M31.
Our analysis mirrors that for Galactic novae undertaken
by \citet{Shara2018} where they estimated the masses of
a large sample of novae using the extensive grid of
nova models computed by \citet{Yaron2005}.
We conclude our study by comparing the fundamental properties
and observed behavior of M31 novae with their counterparts in the Galaxy.

\section{The M31 Light Curve Data} \label{sec:LC}
 
The $R$-band light curves from a large sample of M31 novae
collected over two decades from 2002 to 2022 have recently
been published by \citet{Clark2024}. The novae were divided into
three classes based upon their gross light-curve morphology:
``Linear" systems
that exhibited essentially monotonic declines (84 novae),
``Break" systems showning a
discontinuous change in the slope
of the decline (53 novae), and a ``Jitter" class that included
novae with relatively slow, fluctuating declines
from maximum light (27 novae).
In addition, known RNe were considered by Clark et al. as a separate
class (17 eruptions from 13 nova progenitors).
For all systems, the peak absolute magnitude in the $R$-band, $M_R$, and
the rate of decline as characterized by the $t_2$ time were
measured for a total of 177 M31 novae\footnote{One of these objects,
M31N 2015-01a, is a rare Luminous Red Nova \citep{Kurtenkov2015}
that was mistakenly included in \citet{Clark2024}. It has been replaced
in the present study by the RN M31N 2017-01e omitted in Clark et al.,
keeping the total number of novae at 177.}.

The large and homogeneous
sample of novae from the Clark et al. study is unprecedented,
and offers a unique opportunity to explore the fundamental properties
of novae in M31.
Here, we employ the nova models of
\citet{Yaron2005} to estimate a range of nova properties including
the mass and accretion rate of the WD, 
$M_\mathrm{WD}$ and
$\dot M$, as well as
the maximum ejection velocity, $V_\mathrm{max}$
and the expected recurrence time, $P_\mathrm{rec}$ for
each nova in the sample.

\section{Nova Models}

\citet{Yaron2005} have computed an extensive grid of nova
models relating observe characteristics of the eruption
to fundamental properties of the underlying progenitor binary.
In particular, their Table~3 gives properties such as the 
bolometric luminosity at the peak of the eruption,
\(L_4\) (in units of \(10^4 L_{\odot}\)), the mass loss timescale,
$t_\mathrm{ml}$, the amplitude of the outburst, $A$,
the maximum velocity
of the ejecta, $V_\mathrm{max}$, and the recurrence time between
successive eruptions, $P_\mathrm{rec}$, as functions of the WD
mass, $M_\mathrm{WD}$, accretion rate, $\dot M$,
and core WD temperature, $T_\mathrm{WD}$.

Our analysis follows a procedure
similar to that adopted by \citet{Shara2018} in their determination
of similar properties for a sample of Galactic novae.
A key difference lies in the choice of the observational input parameters.
In particular, \citet{Shara2018}
used the eruption amplitude $A~(\equiv m_\mathrm{min} - m_\mathrm{max}$)
and the mass-loss timescale $t_\mathrm{ml}$ (taken as equivalent to the $t_2$
time). However, the eruption
amplitude $A$ is highly sensitive to the evolutionary state of the donor star
(main-sequence or evolved), making its use problematic for many novae where
the nature of the secondary star is unknown or uncertain. This is a fatal
drawback in the case of novae in M31, where the nature of the secondary stars
are almost all unknown, and the great distance makes even an estimate
of the magnitude
at minimum light, and thus $A$, unavailable for the vast majority of M31 novae.
In place of the outburst amplitude,
we substitute the peak bolometric luminosity,
$L_4$, which is available from
a nova's $R$-band absolute magnitude at maximum light, $M_R$,
as an input parameter.
The high luminosity of novae at the peak of eruption makes $L_4$ insensitive
to the luminosity of the system at quiescence and thus
to the nature of the secondary star.
Then, following \citet{Shara2018}, we also
adopt the observed $t_2$ time as a proxy for $t_\mathrm{ml}$, and
assume $t_\mathrm{ml}\equiv t_2$.

To estimate the bolometric magnitude of a nova from its absolute
$R$-band magnitude requires us to specify the $V-R$ color and
the bolometric correction of a typical nova at maximum light.
We estimate the colors of our nova sample
using results from the recent study by \citet{Craig2025}
who find median colors of $B-V=0.16$~mag and $V-R=0.17$~mag
for their recommended ``Silver" sample of
Galactic novae that were observed within 1~mag of peak brightness.
To estimate the bolometric correction, we draw on the study by
\citet{Pecaut2013}. From their Table~5 for the intrinsic colors, effective
temperatures and bolometric corrections for 09 -- M9 dwarfs, we
find that $B-V\sim0.16$ corresponds to an effective temperature
of $T_\mathrm{eff,peak}\sim8000$~K, and a bolometric correction,
$\mathrm{BC}=-0.02$.
Thus, our expression for the bolometric luminosity of a nova with
absolute magnitude, $M_R$, at maximum light
(in units of $10^4~L_{\odot}$) is given by:

\begin{equation}
L_4 = 10^{(-4-0.4[M_R + (V-R) + \mathrm{BC} - M_\mathrm{bol,\odot}])},
\end{equation}
where
$M_\mathrm{bol,\odot}=4.74$ \citep{Torres2010}, and we have adopted
$V-R=0.17$ (median of the Galactic nova sample) and
$\mathrm{BC}=-0.02$ (characteristic of main-sequence stars with $B-V=0.17$).
The uncertainty in the luminosity of a given nova, $\sigma_{L_4}$,
has been determined by propagating the uncertainties in $M_R$ and $V-R$.
Specifically: $\sigma_{L_4} = 0.4~\mathrm{ln}(10)~L_4~(\sigma_{M_R}^2 +
\sigma_{(V-R)}^2)^{0.5}$.

\subsection{Choice of $T_\mathrm{WD}$}

Another slight difference between our analysis and that of \citet{Shara2018}
concerns the adopted value of $T_\mathrm{WD}$, which must be specified
for the \citet{Yaron2005} models. 
Specifically,
the model grid is divided into three representative WD
temperature regimes:
\( T_\mathrm{WD} = 10,\ 30,\ \mathrm{and}\ 50 \times
10^6 \, \mathrm{K} \) thought to be appropriate for accreting
WDs in nova binaries.

The choice of the appropriate temperature is not immediately obvious.
Steady-burning \( T_\mathrm{WD} \) values (\( \sim 5-8 \times 10^7 \,
\mathrm{K} \))
are too high given that the accretion rates
\( \dot{M} \) (\( \sim 10^{-11} \) to 
\( 10^{-8} \, M_\odot \, \mathrm{yr}^{-1} \)) are typically
well below the critical rate for
steady hydrogen burning (e.g., \( \dot{M}_\mathrm{crit} \sim 1-2 \times 10^{-7} \,
M_\odot \, \mathrm{yr}^{-1} \) for a 1.0 \( M_\odot \) WD; \citealt{Nomoto2007}).
In this regime, \(^3\)He-driven limit cycles dominate, triggering premature
thermonuclear runaways that prevent significant core heating
\citep{Townsley2004}.

In their Galactic nova study,
\citet{Shara2018} arbitrarily adopted \( T_\mathrm{WD} = 30 \times
10^6 \, \text{K} \), while noting that nova outburst models are relatively
insensitive to the choice of \( T_{\text{WD}} \). However, we selected
the lower
\( T_{\text{WD}} = 10 \times 10^6 \, \text{K} \) 
value for our analysis, motivated by \citet{Townsley2004}.
In particular, their Figure~8 (upper panel) shows that for accretion rates
\(\log \dot{M} (M_\odot~\mathrm{yr}^{-1}) \lessim -8\) typical of nova systems, WD
temperatures are expected to be \(\lessim 10 \times 10^6 \, \text{K}\).
Since our analysis focuses on \(\log \dot{M}\) from \(-12.0\) to
\(-7.0\), the \( T_{\text{WD}} = 10^7 \, \text{K} \) models best
match the physical conditions of nova outbursts in our study. The
insensitivity of the models to $T_\mathrm{WD}$ noted by \citet{Shara2018},
coupled with our own experimentation, suggests our results are
robust despite our choice of the lower WD temperature.

\subsection{Interpolation of the Model Grid}

The model grid given in Table~3 of \citet{Yaron2005} is rather coarse,
consistently spanning six values of $\log \dot{M}$ (from
$-12$ to $-7$) for just four representative values of $M_\mathrm{WD}$
(0.65, 1.00, 1.25, 1.40 $M_\odot$).
To enable high-resolution mapping between the observable quantities
($L_4$, $t_2$, $V_\mathrm{max}$, $P_\mathrm{rec}$) and the
relevant physical parameters (log ${\dot M}$ and $M_\mathrm{WD}$),
we have interpolated the $T_\mathrm{WD}=10\times10^6$~K
model grid (reproduced in Table~\ref{tab1}),
using second-order polynomial surface fits similar to the approach
taken by \citet{Shara2018}.
Specifically, for each model nova parameter $Z$
(i.e., $L_4$, $t_\mathrm{ml}$, $V_\mathrm{max}$, $P_\mathrm{rec}$),
our interpolation model fits the following expression:
\begin{equation}
\log Z(X,Y) = c_1 + c_2 X + c_3 Y + c_4 X^2 + c_5 XY + c_6 Y^2,
\end{equation}
where $X \equiv \log \dot{M}$ and $Y \equiv M_\mathrm{WD}$. The coefficients
$c_1, \dots, c_6$ are determined via least-squares minimization using
the normal equations, solved with Gaussian elimination.
The fit was performed in $\log Z$ to ensure positivity, and
is appropriate for quantities spanning many
orders of magnitude. The resulting interpolation was evaluated on a dense grid:
500 points in $\log \dot{M}$ ($\Delta X = 0.01$) and 750 points in
$M_\mathrm{WD}$ ($\Delta Y = 0.001\,M_\odot$), yielding a total of $N = 375{,}000$
interpolated points per model parameter. Where necessary
(e.g., for $\log \dot M=-12; M_\mathrm{WD}=1.40$) to fill out the
complete grid, the fit was extrapolated.
The uncertainty in the model fit
has been assessed by computing the typical residual error in
$\log Z$ from the squared differences between observed and
predicted values at the original grid points.
This error is then propagated through the fit coefficients to
estimate the standard error in $\log Z$ at any interpolated point.

\begin{figure}
\plottwo{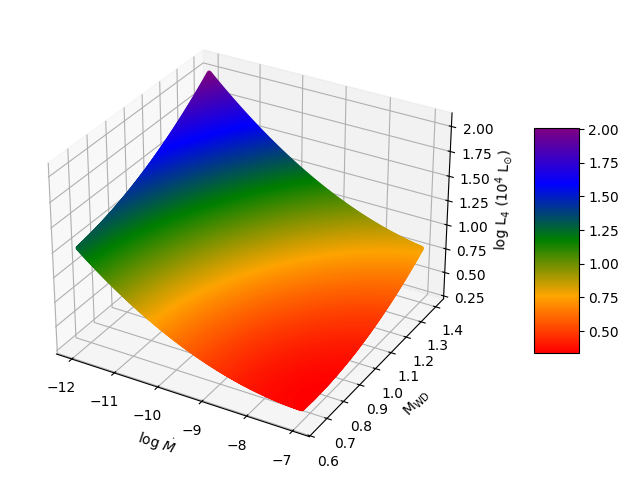}{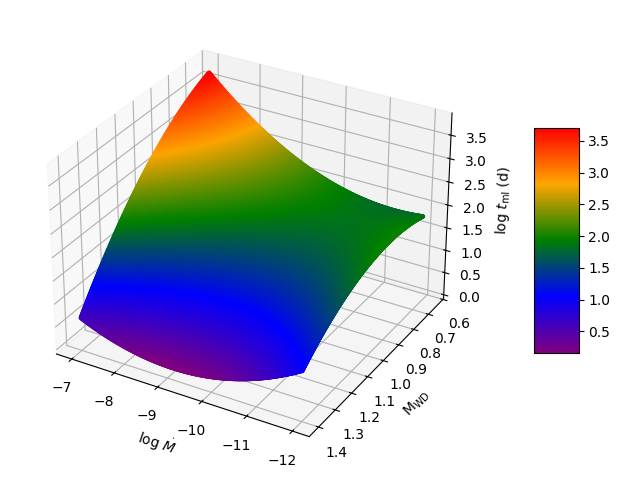}
\caption{The interpolated model grids for our key input
parameters, the bolometric luminosity
at the peak of the eruption, $L_4$, (left panel), and
the mass-loss timescale,
$t_\mathrm{ml}$ (Right Panel) as functions of
$\log \dot M$ ($M_{\odot}$~yr$^{-1}$) and
$M_\mathrm{WD}$ ($M_{\odot}$). The interpolation was performed
by fitting second-order polynomials
to the nova models of \citet{Yaron2005} as described in section 3.2.
}
\label{fig1}
\end{figure}

\begin{figure}
\plottwo{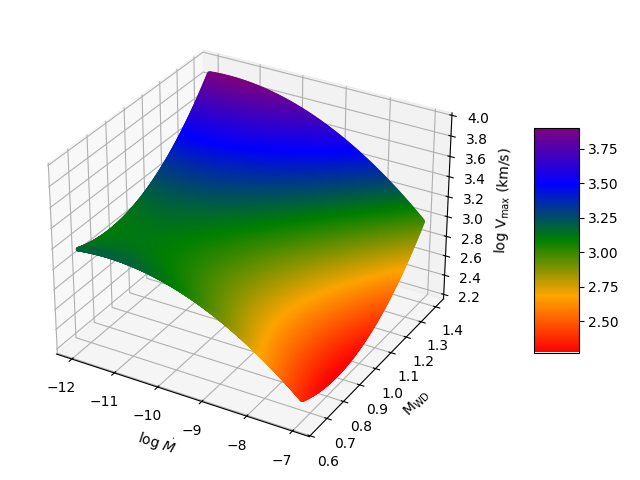}{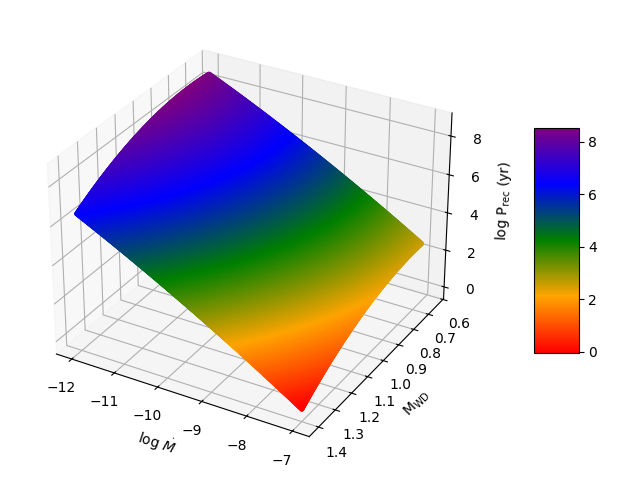}
\caption{The same as in Fig. 1 except showing the model interpolations for
the auxiliary parameters, the maximum expansion velocity
of the nova ejecta, $V_\mathrm{max}$ (left panel), and the predicted
recurrence time, $\log P_\mathrm{rec}$ (right panel).
}
\label{fig2}
\end{figure}

The interpolated grids are shown in Figures~\ref{fig1} and \ref{fig2}.

\bigskip
\section{Grid inversion and determination of nova parameters}

Log accretion rates (X) and WD masses (Y) have been estimated
for each M31 nova by
inverting the interpolated $L_4(X,Y)$ and $t_\mathrm{ml}(X,Y)$ grids for
the observed values of $L_4$ (through Eqn.~1)
and the observed $t_2$ time, with
observational uncertainties $\sigma_{L_4}$ and $\sigma_{t_2}$ and
model interpolation uncertainties $\delta_{L_4,i}$ and
$\delta_{t_\mathrm{ml},i}$ at a given grid point $i$. The effective
uncertainty in $L_{4,i}$ and $t_{\mathrm{ml},i}$ at this grid point is then
\begin{equation}
\Delta_{L_{4,i}} = \sqrt{(\sigma_{L_4})^2 + \delta_{L_{4,i}}^2},
\quad
\Delta_{t_{{\mathrm ml},i}} = \sqrt{(\sigma_{t_2})^2 + \delta_{t_{{\mathrm ml},i}}^2}.
\end{equation}

The identification of the best estimates of $\log \dot{M}$ and $M_\mathrm{WD}$
begins by searching the grid for all points in an effective error box
that simultaneously satisfies the following constraints:

\begin{equation}
|L_{4,i} - L_4| < \Delta_{L_{4,i}} \quad \text{and} \quad
|t_{{\mathrm{ml},i}} - t_2| < \Delta t_{{\mathrm{ml},i}}.
\end{equation}

In rare cases where no points satisfied these criteria,
the effective uncertainties ($\Delta L_{4,i}$ and $\Delta t_{{\mathrm ml},i}$) were
increased by incrementing the observational uncertainties
($\sigma_{L_4}$ and $\sigma_{t_2}$) by their original values
until a solution was found.

Once a set of potential solutions (an error box) for each nova
were identified, the the best estimates of
$\log \dot{M}$ and $M_{\text{WD}}$ were found by
minimizing the weighted Euclidean norm,

\begin{equation}
\chi_i = \sqrt{ \left( \frac{L_{4,i} - L_4}
{\Delta_{L_{4,i}}} \right)^2 + \left(
\frac{t_{\mathrm{ml},i} - t_2}
{\Delta_{t_{{\mathrm ml},i}}} \right)^2 },
\end{equation}
for all points within the error box.
The grid point \(i_{\text{min}}\) corresponding to the smallest $\chi_i$ then
provides our optimal solutions for
the mass accretion rate and the WD mass,
$\log \dot{M}(i_\mathrm{min}$) and $M_\mathrm{WD}(i_\mathrm{min}$).

Uncertainties in the adopted values of $\log \dot{M}$ and $M_\mathrm{WD}$ have
been estimated as the root-mean-square (RMS) deviations from the optimum value
of all points satisfying our uncertainty constraints.
This non-parametric RMS method measures the true spread of all
acceptable solutions inside effective the error box, while accounting for
how tightly or loosely the points cluster.
The method works well even when the allowed region is long, narrow, or
irregularly shaped. It gives more weight to solutions near the best
fit, and it includes both observational errors and model
interpolation uncertainties without assuming any specific error
distribution. We have done our best to estimate uncertainties in our
derived parameters given the available observational error
estimates and model grid fit errors, but these estimates may not capture
the full extent of the parameter uncertainties due to any unaccounted-for
systematic errors.

\subsection{Derived Quantities, $V_\mathrm{max}$ and $P_\mathrm{rec}$}

For a given M31 nova in our sample,
we are now in a position to predict the values of other potentially
observable quantities such as
the maximum velocity of its ejecta, as well as the likely interval
between successive eruptions. Specifically,
armed with the optimum solutions for the model
parameters $M_\mathrm{WD}$, log $\dot M$, and their
associated uncertainties, we have employed the
interpolated model grids shown in Figure~\ref{fig2}
to estimate the optimum values (and their uncertainties)
for both $V_\mathrm{max}$ and $P_\mathrm{rec}$.

\section{Observed Distributions of Nova Parameters}

Our best estimates of
$M_\mathrm{WD}$, $\log \dot{M}$, $V_\mathrm{max}$ and $P_\mathrm{rec}$ 
for the M31 novae presented in \citet{Clark2024} are given
in Tables~\ref{tab2} - \ref{tab5},
where for consistency we have
maintained the distinction between the various light curve morphologies:
Linear, Break, and Jitter, and considered the 14 known RNe with available light curve data separately.
Available data for all 22 known and suspected M31 RNe are summarized in Appendix~A.
Finally, for completeness, we also have included in the tables 
values for the model input light curve parameters $L_4$ and $t_2$, and
retained the original quality assessment (Gold, Silver, Bronze)
from \citet{Clark2024}.

\begin{figure}
\plottwo{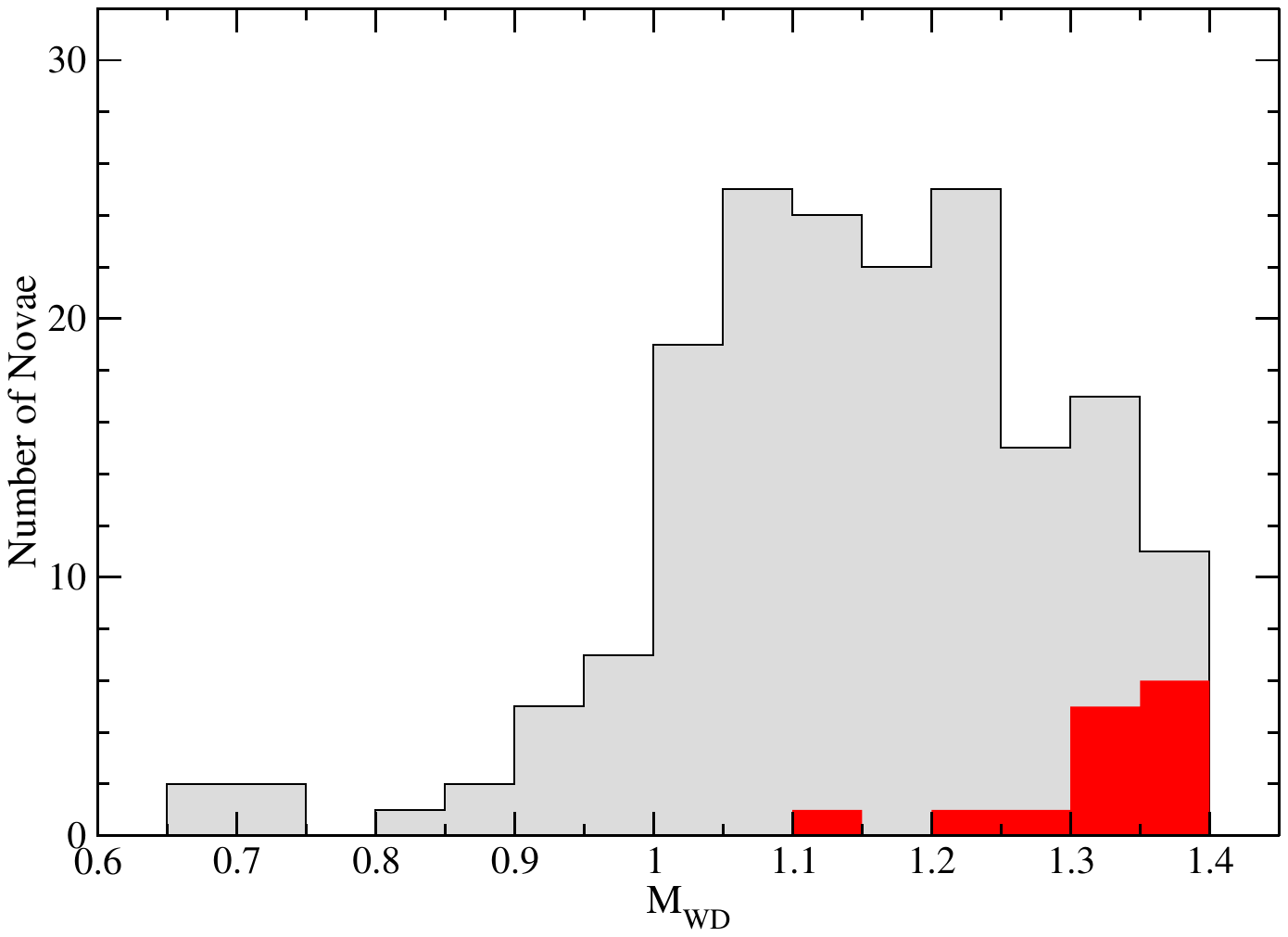}{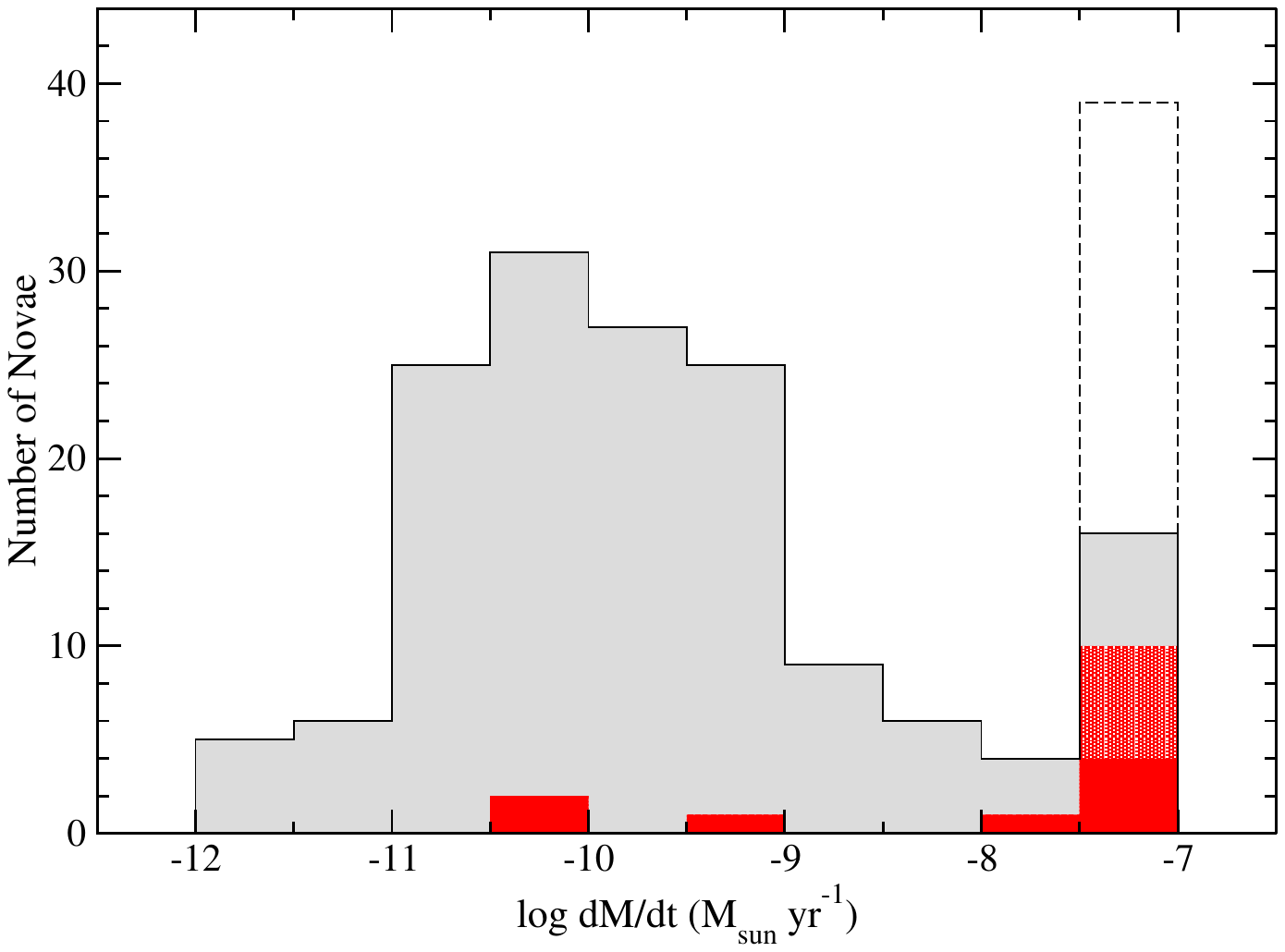}
\plottwo{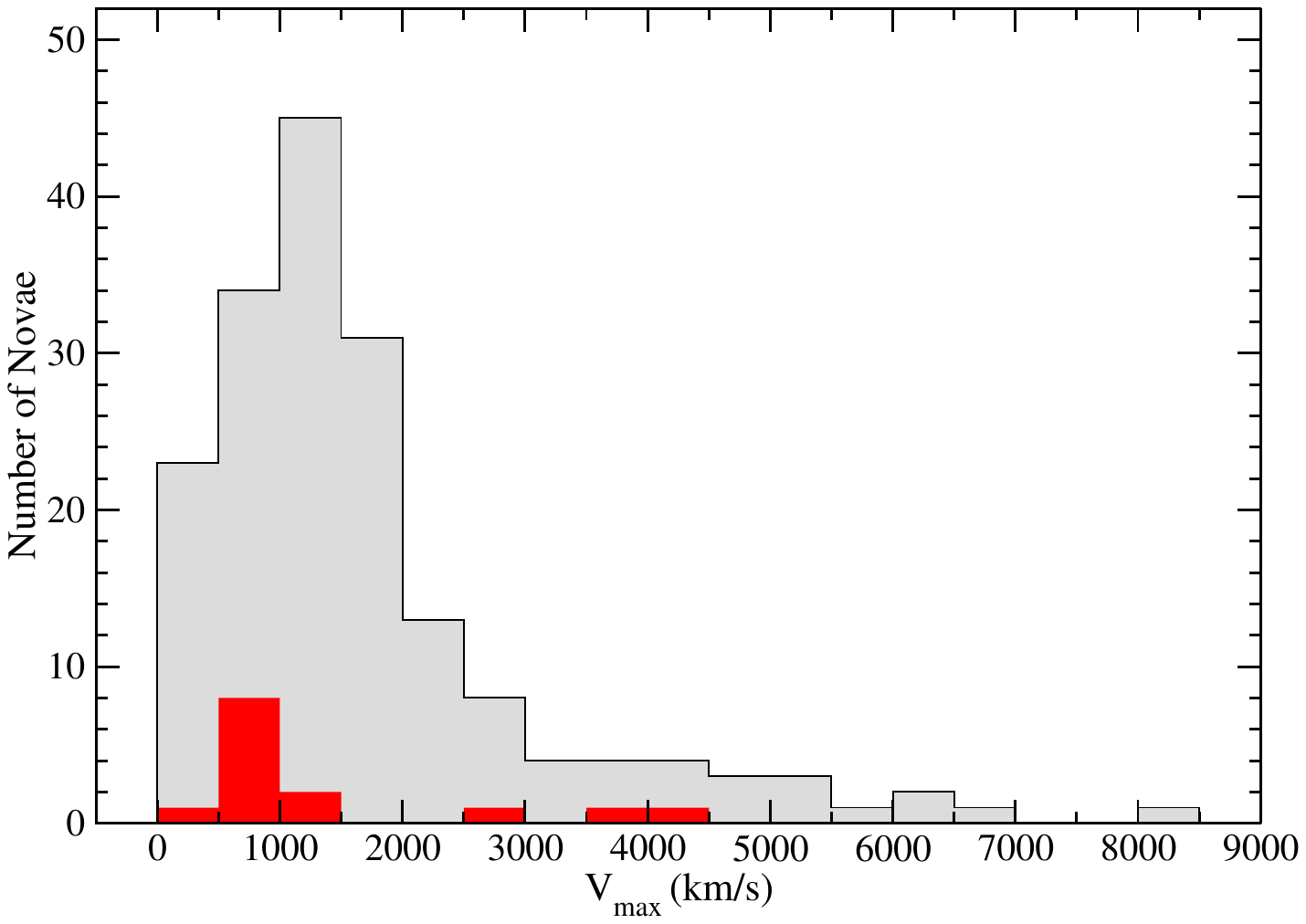}{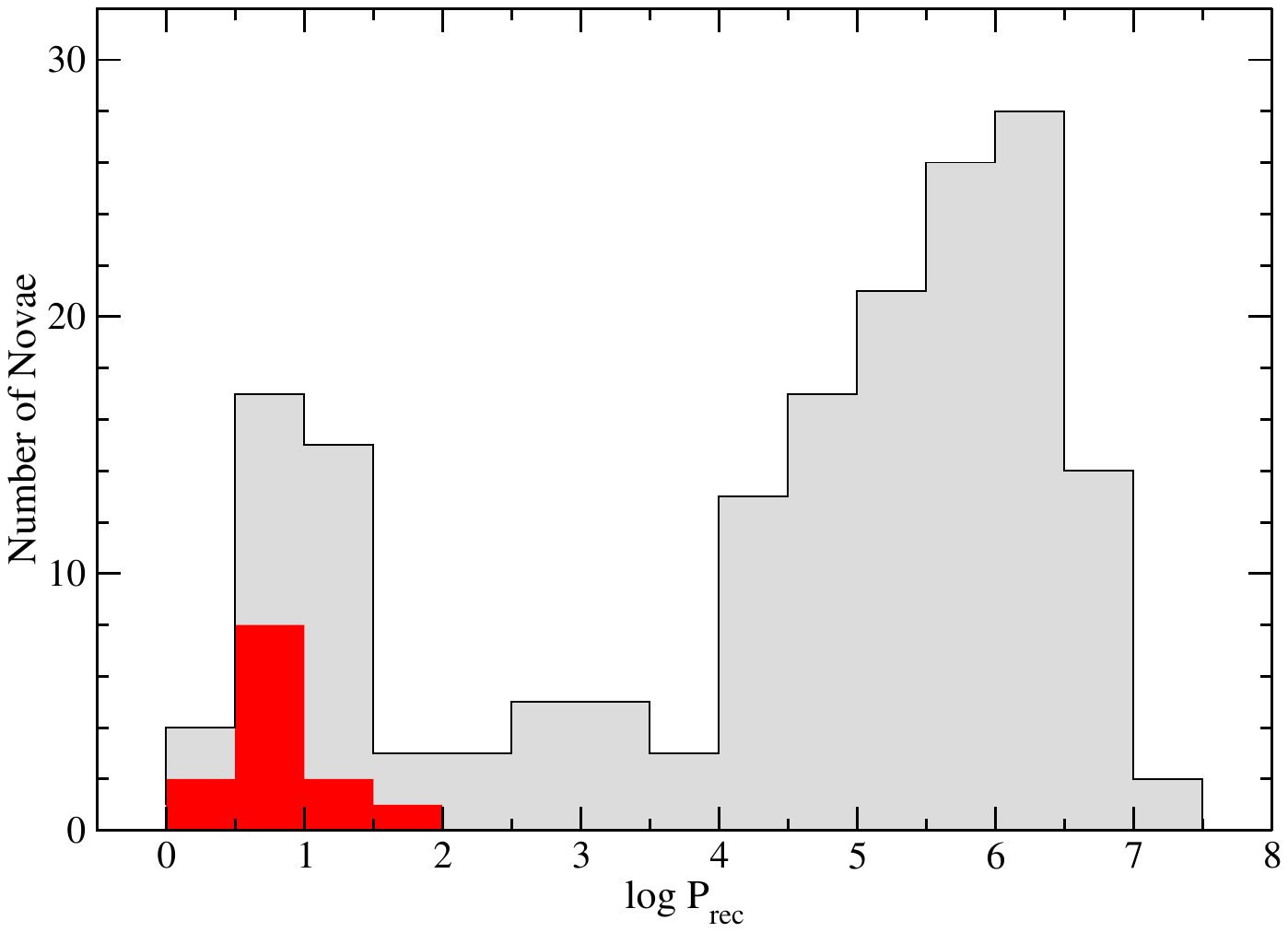}
\caption{Distributions of the observed nova properties
based on the model predictions from Tables~\ref{tab2} - \ref{tab5},
Top Left: The observed WD mass, $M_\mathrm{WD}$, distribution
(known RNe shown in red).
Top Right: The mass accretion rate,
$\log \dot M$, distribution. The dashed region in the brightest bin
shows the contribution of all novae (RNe in light red), including those
with uncertain estimates of $\log \dot M = -7.0$, while the
grey and dark red regions omit these novae.
Bottom Left: The maximum ejecta velocity, $V_\mathrm{max}$, distribution
(known RNe shown in red).
Bottom Right: The
recurrence time, log $P_\mathrm{rec}$, distribution.
For known RNe, observed values
of the recurrence times are shown.
}
\label{fig3}
\end{figure}

Figure~\ref{fig3} shows the observed distributions of WD mass, $M_\mathrm{WD}$,
mass accretion rate, $\log {\dot M}$, maximum ejection velocity,
$V_\mathrm{max}$, and recurrence times, $\log P_\mathrm{rec}$,
for our sample of 177 M31 novae.
The mean values and the RMS errors for each observed distribution,
light curve morphology and quality rating
are summarized in Table~\ref{tab6}.
Generally, the distribution means are consistent
across the various light curve morphologies and quality designations.
Thus, for simplicity, we will only consider the properties
of the aggregate sample of novae in the remaining discussion.

Each distribution is discussed in more detail below.

\subsection{The WD mass distribution}

The WD mass distribution is characterized by a mean
of $1.15$~M$_{\odot}$ with an RMS deviation about the mean,
RMSD$(M_\mathrm{WD})=0.14$.
Overall, the distribution appears to be very similar to
that found by \citet{Shara2018} in their
study of 82 Galactic novae, where they found
the observed distribution of WD masses could be approximated by
a Gaussian with a mean of $1.13~M_\odot$ and a FWHM = $0.29~M_\odot$
($\sigma=0.12$).

Not surprisingly, the known RNe in our M31 sample
(shown in red) are strongly concentrated toward
the right tail of the observed WD mass distribution, as is the case
for Galactic novae. Models have consistently shown that
massive WDs and relatively high accretion rates are necessary
to produce recurrence times of order 100~yr, or less
\citep[e.g.,][]{Wolf2013,Kato2014}.

\begin{figure}
\plotone{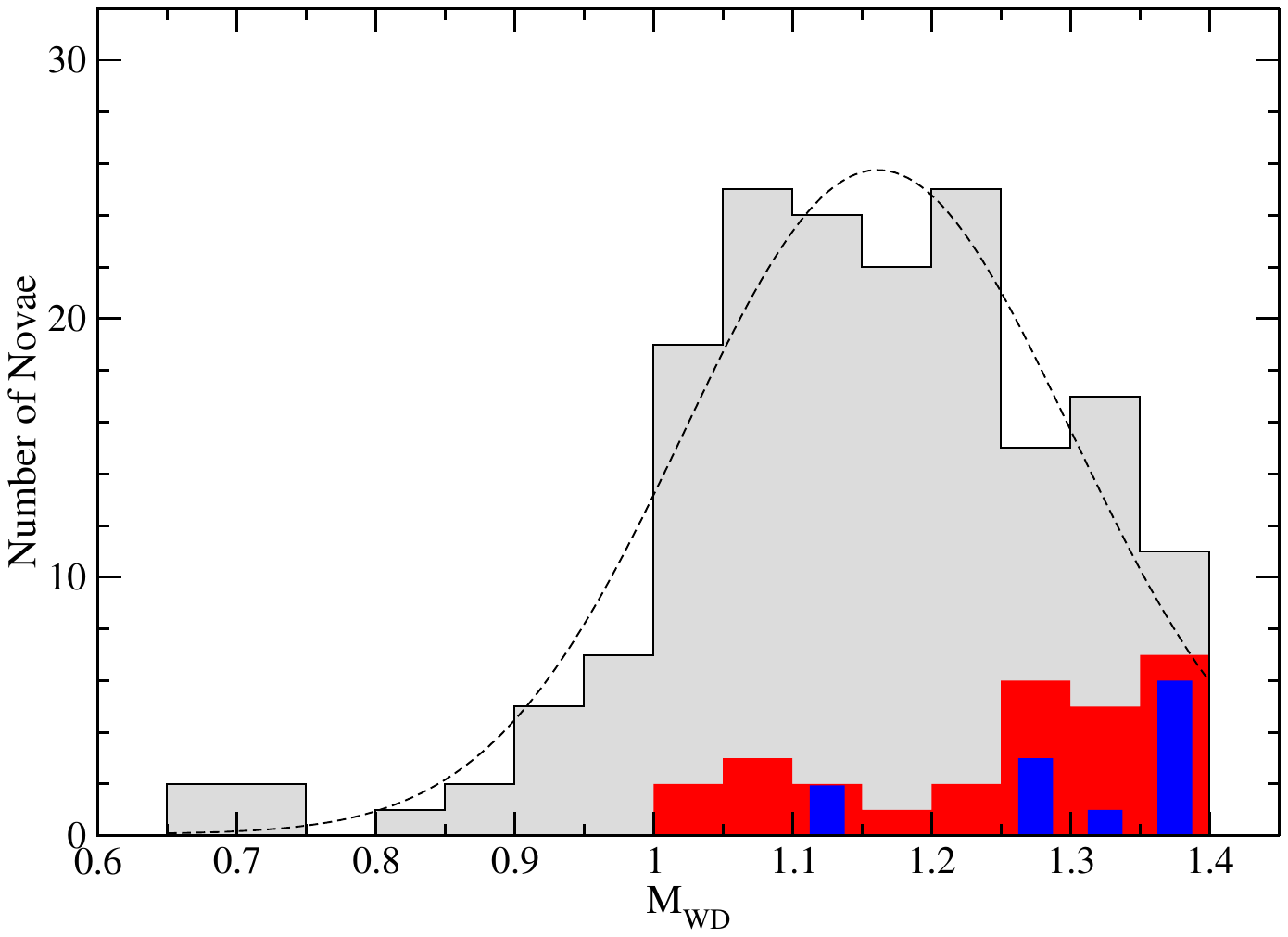}
\caption{The observed distribution of the WD mass for He/N and \ion{Fe}{2}b
M31 novae compared with the overall WD mass distribution. The red distribution
show all known and suspected He/N and \ion{Fe}{2}b novae,
while the blue distribution
shows firmly established He/N novae. The average WD mass,
$\langle M_\mathrm{WD} \rangle = 1.15\pm0.14$~M$_{\odot}$,
$1.26\pm0.12$~M$_{\odot}$, and $1.32\pm0.09$~M$_{\odot}$ for all novae,
known and suspected He/N + \ion{Fe}{2}b novae, and firmly established He/N
novae, respectively.}
\label{fig4}
\end{figure}

The spectra of RN shortly after eruption usually exhibit broad
Hydrogen Balmer, Helium (\ion{He}{1}, \ion{He}{2}), and Nitrogen (\ion{N}{2},
\ion{N}{3}) emission lines characteristic of the
He/N spectroscopic class in the scheme of \citet{Williams1992}.
To explore the dependence of spectroscopic class on WD
mass, we have reproduced the 
observed WD mass distribution in
Figure~\ref{fig4}, showing where the He/N and broad-lined \ion{Fe}{2} (\ion{Fe}{2}b) novae fall within the overall distribution. 
As expected, the He/N and \ion{Fe}{2}b novae, of which the RNe are a subset,
are concentrated at higher mass.
If we approximate the overall distribution
by a Gaussian (dashed line, truncated at $1.4~M_{\odot}$) we find a
mean of $\langle M_\mathrm{WD} \rangle = 1.16$ and
standard deviation, $\sigma=0.14$, which are essentially identical to
the sample mean and RMS deviation determined from the individual
WD mass estimates.

\subsection{The observed $\log \dot M$ distribution}

The distribution for $\log \dot M$ is surprisingly bimodal with the main
concentration of novae clustered around an estimated accretion rate
of $\sim10^{-10}~M_{\odot}$~yr$^{-1}$. However, there is also a spike of
novae with accretion rates
near $10^{-7}~M_{\odot}$~yr$^{-1}$, where the majority of
the RNe are found. The excess of novae at the high
end of the accretion rate distribution possibly results
from truncation of the model grid, which does not extend to
accretion rates above $\log \dot M = -7$.
As a result, the best-fit solutions for any novae
accreting at rates higher than this will accumulate at this end of the
distribution. Novae falling at the edge of the grid
with accretion rate estimates of
exactly $\log \dot M = -7$ (indicating that the true values may
lie above this accretion rate) are marked as uncertain in
Tables~\ref{tab2} -- \ref{tab5} and
are represented by the unshaded portion of the histogram. Not
coincidentally, a significant fraction of these novae turned out
to be ones that failed to converge using the initial error estimates
during the grid inversion process. In these cases,
convergence was achieved only after incrementing the initial values
of the error estimates for $L_4$ and $t_2$. It is therefore reasonable
to assume that the derived $\log \dot M$ values for these
systems are particularly uncertain.
It is also possible that many of these systems are
unrecognized RN systems with evolved secondary stars producing
high accretion rates.

\subsection{The predicted $V_\mathrm{max}$ distribution}

The distribution for the predicted maximum ejecta velocity
is characterized by a strong peak between 1000 and 1500~km~s$^{-1}$.
This is typical of the FWHM of the Balmer emission lines
seen in the vast majority of novae, most of which are members of
the \ion{Fe}{2} spectroscopic class \citep{Williams1992}. Surprisingly,
the distribution of the maximum expansion velocities for
known RNe generally follows that of the overall nova distribution,
which is dominated by relatively narrow-lined \ion{Fe}{2} novae.
However, as mentioned earlier, most recurrent novae are members
of the He/N class with emission lines typically characterized
by FWHM $\grtsim 2500$~km~s$^{-1}$. The root of this
discrepancy can be traced
back to the model grid  for $V_\mathrm{max}$ (see Fig~2, left panel).
Expansion velocities in excess of 2000~km~s$^{-1}$ can only
be produced for high mass WDs accreting at rates
$\dot M \lessim 10^{-8}~M_\odot$~yr$^{-1}$,
whereas most RNe have accretion rates $\sim10^{-7}~M_\odot$~yr$^{-1}$.

\begin{figure}
\centering
\includegraphics[angle=0,scale=0.40]{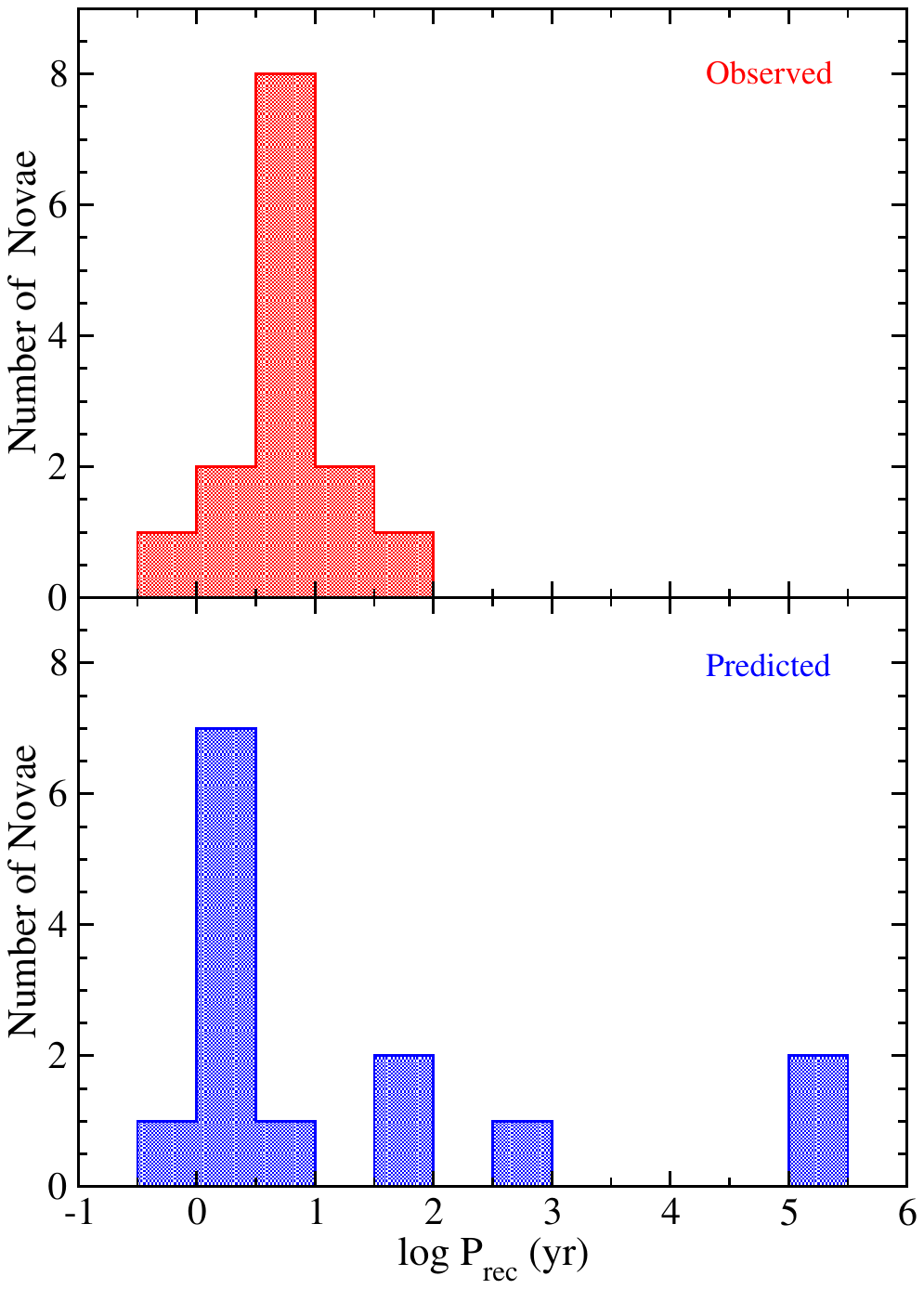}
\caption{The observed distribution of the RN recurrence times (top panel)
compared with the predicted RN recurrence time distribution based
on the model fits (bottom panel).
The model fails to accurately predict the recurrence times for
three RNe (M31N 1984-07a, 2006-11c, 2013-10c).
}
\label{fig5}
\end{figure}

\subsection{The predicted recurrence time distribution}

As with the observed $\log \dot M$ distribution,
the distribution for the predicted recurrence times is also
bimodal, with one peak near a recurrence time of $\sim10$~yr,
and the other at considerably longer intervals
of order several $\times 10^5$~yr. All of the novae in the former
group with $P_\mathrm{rec}<100$~yr are either known or unrecognized
RNe. It is tempting to speculate that the bimodal nature of the
distribution results from differences in the evolutionary
state of the mass donor, and that the peak at recurrence times
$\sim10$~yr is formed
by systems harboring evolved secondary stars.

Given that the true recurrence times (or possibly upper limits in some cases)
are known for the RN sample offers an opportunity
to check the predictive power of the models themselves.
Figure~\ref{fig5} shows the distributions of the observed recurrence times 
compared with the model predictions. For the majority of known RN, the models
correctly assign recurrence times, $P_\mathrm{rec} \lessim 100$~yr.
However, for four of the 17 observed RNe eruptions (M31N 2012-09a, 2013-10c,
2015-02b, and 2022-11b)
the predicted recurrence times are considerably longer. The two
eruptions with the longest predicted recurrence times
(2012-09a and 2022-11b) are believed to be subsequent eruptions of
M31N 1984-07a. This nova appears very close to the nucleus of M31
making it difficult to firmly establish that the novae arise from
the same progenitor. It is therefore possible, although unlikely, that
the object may not be a RN, but instead the chance positional
near-coincidence of three unrelated novae
\citep[see the discussion in][]{Shafter2015} for further details.
The other two RN eruptions occur further from the nucleus
of M31, and the RN designations are thus much more secure.
In all cases, the \citet{Yaron2005} models require 
accretion rates $\lessim 10^{-9}~M_\odot$~yr$^{-1}$ to achieve
the relatively high luminosity of the observed eruptions ($M_R < -8$).
In such systems, the recurrence times are predicted to be
$\grtsim 500$~yrs even for the most massive WDs.

\section{Intrinsic Distributions for $M_\mathrm{WD}$ and $\log \dot M$}

As emphasized by \citet{Shara2018} in their study of the Galactic
nova population, the observed distributions of $M_\mathrm{WD}$ and
$\log {\dot M}$ are strongly biased toward systems with short
recurrence times. Given our estimates of $P_\mathrm{rec}$ for
the individual novae, it is straightforward to estimate
the intrinsic (unbiased) mean
for the $M_\mathrm{WD}$ and $\log {\dot M}$
distributions by simply by weighting the observed values of the WD
mass, $M_{\mathrm{WD},i}$, and log accretion rate, $[\log \dot M]_i$,
by $P_\mathrm{rec}$. Specifically, for recurrence time weights
$w_{\mathrm{rec},i} = P_{\mathrm{rec},i}/\sum_{i=1}^{N} P_{\mathrm{rec},i}$, we have:
$\langle M_\mathrm{WD} \rangle _\mathrm{int} = \sum_{i=1}^{N} w_{\mathrm{rec},i}~M_{\mathrm{WD},i}$ and
$\langle \log \dot M \rangle _\mathrm{int} = \sum_{i=1}^{N} w_{\mathrm{rec},i}~[\log \dot M]_i$.
When applied to the full sample of $N=177$ novae in
Tables~\ref{tab2} -- \ref{tab5} we find
$\langle M_\mathrm{WD} \rangle _\mathrm{int} = 1.07~M_{\odot}$,
and
$\langle \dot M \rangle _\mathrm{int} = 2\times10^{-11}~M_{\odot}$~yr$^{-1}$.

To determine the intrinsic distributions for $M_\mathrm{WD}$ and $\log \dot M$,
we follow the procedure outlined in section~5 of \citet{Shara2018}.
Briefly, if the distribution of WD masses are divided into $n_i$ bins,
$x_i$, and the distribution of $\log \dot M$ values are divided into
$n_j$ bins, $y_j$, where
$P_\mathrm{x,obs}(x_i)$ is the observed fraction novae in WD mass bin $x_i$
and $P_\mathrm{y,obs}(y_j)$ is the observed
fraction novae in $\log \dot M$ bin $y_j$,
then the intrinsic fraction of novae in each WD mass and $\log \dot M$ bin is
given by:

\begin{equation}
\begin{aligned}
P_\mathrm{x,int}(x_i) = P_\mathrm{x,obs}(x_i) \Bigg[ \sum_{j=1}^{n_j} {P_\mathrm{y,obs}(y_j) \over P_\mathrm{rec}(x_i,y_j)} \Bigg]^{-1} \\\\
P_\mathrm{y,int}(y_j) = P_\mathrm{y,obs}(y_j) \Bigg[ \sum_{i=1}^{n_i} {P_\mathrm{x,obs}(x_i) \over P_\mathrm{rec}(x_i,y_j)} \Bigg]^{-1},
\end{aligned}
\end{equation}
where the weighting function
$P_\mathrm{rec}(x_i,y_j)$ is obtained from the model
interpolation for the recurrence time shown in the right panel
of Figure~\ref{fig2}.

\begin{figure}
\plottwo{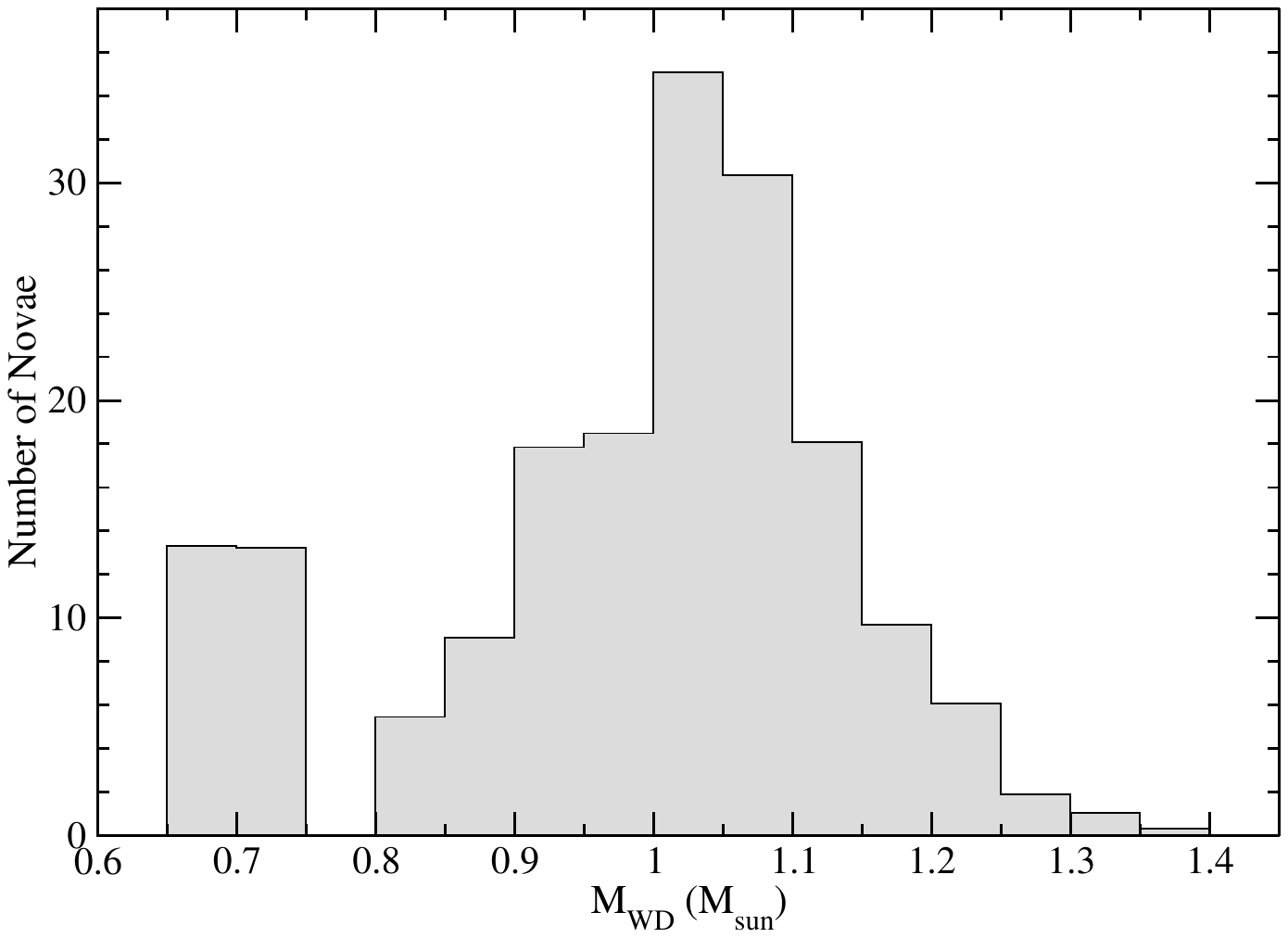}{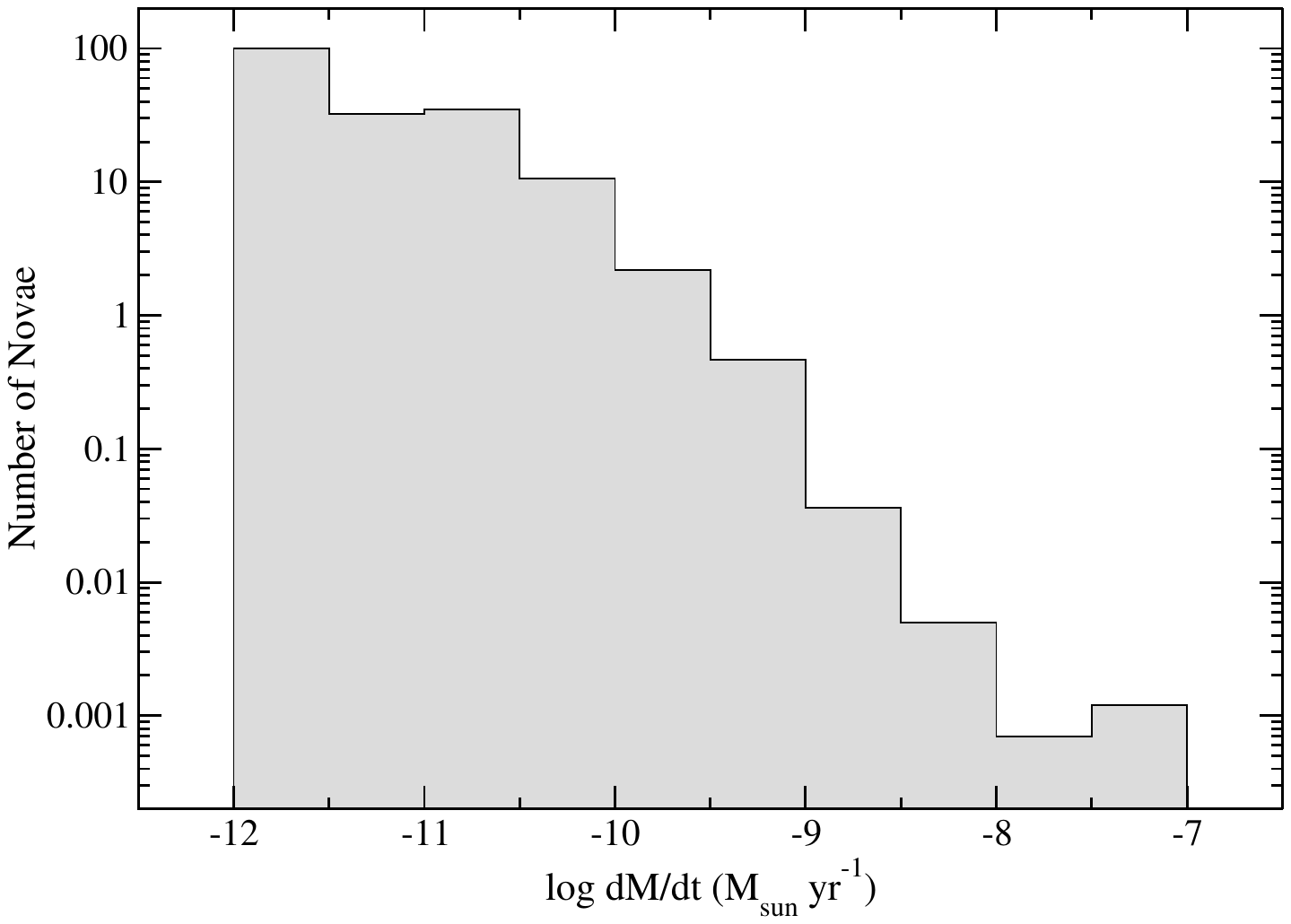}
\caption{The intrinsic distributions of the WD mass (left panel)
and log $\dot M$ (right panel). The mean of the WD mass
distribution is $\langle M_\mathrm{WD} \rangle_\mathrm{dist} = 1.01~M_{\odot}$,
while that for the $\log \dot M$ distribution
(shown as log number of novae for clarity)
is given by
$\langle \log \dot M \rangle_\mathrm{dist} = -11.1$,
corresponding to
$\langle \dot M \rangle_\mathrm{dist} = 8\times10^{-12}~M_{\odot}$~yr$^{-1}$.
The underlying nova population is dominated by systems with
accretion rates $\lessim 10^{-10}~M_{\odot}$~yr$^{-1}$.}
\label{fig6}
\end{figure}

The resulting intrinsic distributions, normalized to
$\sum_{x_i=1}^{n_i} P_\mathrm{x,int}(x_i) = \sum_{j=1}^{n_j} P_\mathrm{y,int}(y_j) = N$,
are shown in Figure~\ref{fig6}.
The means of the distributions are given by
$\langle M_\mathrm{WD} \rangle_\mathrm{dist} = 1.01~M_{\odot}$ and
$\langle \dot M \rangle_\mathrm{dist} = 8\times10^{-12}~M_{\odot}$~yr$^{-1}$.
The mean of the WD mass distribution is close to that found
in the direct recurrence-time-weighted average of the
individual M31 novae, as well as to the true $M_\mathrm{WD}$ mean
computed by \citet{Shara2018} for the Galactic novae. On the other hand,
the intrinsic accretion rate mean for M31 novae
is about an order of magnitude lower than the corresponding
Galactic determination. The source of the discrepancy is unclear, but
may possibly be related to the choice of nova amplitude, rather than
peak luminosity, as a model input. The amplitude is quite sensitive
to both the mass accretion rate (which determines the quiescent luminosity)
and the evolutionary state of the secondary star. For systems with
luminous secondaries, the eruption amplitudes will be reduced,
resulting in an increase in the inferred accretion rates.
Regardless of whether our M31 $\dot M$ distribution can be brought in line
with the Shara et al.'s Galactic distribution, the mass accretion
rates remain surprisingly low and the recurrence times surprisingly long.

\subsection{Evidence for Hibernation}

Our analysis reveals an M31 nova population characterized by a high
average WD mass ($\langle M_\mathrm{WD} \rangle \approx 1.15~M_{\odot}$),
a low average accretion rate
($\langle \dot M \rangle \approx 10^{-10} M_\odot~\mathrm{yr}^{-1}$),
and long average recurrence times
($\langle P_\mathrm{rec} \rangle \approx 10^6~\mathrm{yrs}$).
This result is internally consistent with the physics underlying
TNR models \citep[e.g.,][]{Townsley2005,Wolf2013},
which show that this low average accretion rate
forces the system to accumulate a large ignition mass
($M_\mathrm{ign} \approx 10^{-4} M_\odot$) prior to the next eruption.
This large ignition mass
coupled with low accretion rates results in the long predicted recurrence times,
$P_\mathrm{rec}~(\equiv M_\mathrm{ign}/{\dot M})$.

The low derived accretion rates ($\dot M \lessim 10^{-10}~M_\odot$~yr$^{-1}$)
directly conflicts with secular evolution models for cataclysmic variables
\citep[e.g.,][]{Rappaport1983,Knigge2011,Zorotovic2020},
which predict much higher cycle-averaged nova accretion
rates ($\sim 10^{-9}$ to $10^{-8} M_\odot~\mathrm{yr}^{-1}$)
driven by standard angular momentum loss mechanisms.
The irradiation-driven hibernation hypothesis originally proposed
by M. Shara four decades ago \citep{Shara1986,Hillman2020}
provides the physical mechanism to bridge this gap.
Following each eruption, strong irradiation of the secondary star
by the hot white dwarf can force the secondary star out of
thermal equilibrium and temporarily drive high mass
transfer ($\sim10^{-8}$~--~$10^{-7}\,M_\odot\,{\rm yr}^{-1}$)
for of order $\sim10^2$~--~$10^3$\,yr.  As the white dwarf cools,
the irradiation fades and the secondary contracts toward its
equilibrium radius, causing the
accretion rate to drop to $\lesssim10^{-11}\,M_\odot\,{\rm yr}^{-1}$
for the remainder of a prolonged ($\sim10^5$~--~$10^6$\,yr)
inter-outburst period. Thus,
the low average accretion rates we derive from the \citet{Yaron2005}
models reflects the system's dominant physical state: a
prolonged ($\sim 10^6~\mathrm{yr}$) low-accretion
hibernation tail ($\lessim 10^{-10} M_\odot~\mathrm{yr}^{-1}$)
following the brief post-nova high state.

\section{The effect of stellar population}

One of the unresolved questions in nova studies concerns whether
observed properties
(and thus the fundamental properties of the progenitor binary)
vary systematically with changes in the underlying stellar population.
Two observational characteristics that are expected to be particularly
sensitive to stellar population are:
(1) the average speed class, as measured
by the $t_2$ time, and
(2) the dominant spectroscopic class of the nova population
(e.g., \ion{Fe}{2} or He/N), as introduced by \citet{Williams1992}.
Studies of Galactic novae have suggested that so-called ``disk novae"
(novae close to the Galactic plane) are faster and
brighter than are ``bulge novae" (novae observed
further from the plane)
\citep[e.g., see][and references therein]{DellaValle2020,Cohen2025}.
Convincing arguments have also been put forward that the traditional
spectroscopic class of a nova \citep[see][]{Williams1992} varies
with stellar population in the Galaxy \citep{DellaValle1998}.

The justification for the variation of
speed class comes directly from models of TNRs on WDs which
demonstrate that for a given accretion rate massive WDs have
lower ignition masses \citep[e.g.,][]{Townsley2005}. Under the
assumption that the ejected masses are proportional to the
ignition mass, novae with massive WDs will produce lower
mass ejecta that rapidly become optically thin,
leading to a faster photometric evolution.

The physics underlying the formation of novae with differing
spectroscopic class is less well understood. Indeed, the very
notion that a nova belongs to a given spectroscopic class
has been called into question \citep{Aydi2024}. These authors
present compelling evidence showing that
the spectral evolution many Galactic novae
display both \ion{Fe}{2} and He/N lines at various times, and thus
appear to transition between spectroscopic classes
on the declining branch of their eruptions. This is no doubt
the case, but it remains true that the line widths, in particular,
shortly after maximum light appear to be markedly different
in the classic He/N novae compared with the relatively narrow-lines
and P-Cygni profiles that characterize their \ion{Fe}{2} counterparts.
That said, clearly the concept of spectroscopic classes of novae
needs to be refined going forward.

\subsection{The unique role of M31}

As a nearby and spatially-resolved galaxy, M31 has played a key role
in the study of nova populations.
In their comprehensive study
of the spectroscopic and photometric properties of novae in M31,
\citet{Shafter2011} looked at the spatial distribution of a large sample
of 91 M31 novae with spectroscopic data sufficient for classification
and found no significant variation of
spectroscopic class (\ion{Fe}{2} vs He/N) with distance from the
center of the galaxy as measured by the $R$-band isophotal radius. In light
of the Galactic studies, this is perhaps surprising given that
novae far from the center of M31 arise overwhelmingly from the galaxy's disk
population. \citet{Shafter2011}
did however find a weak dependence of a nova's rate of decline
from maximum light -- the $t_2$ time -- with distance from the center of M31,
in the sense that the ``faster"
novae ($t_2\leq25$~d) were slightly more spatially
extended compared with their slower 
counterparts with $t_2$ values longer than 25~d.

\begin{figure}
%plotone{f7.pdf}
\centering
\includegraphics[angle=0,scale=0.36]{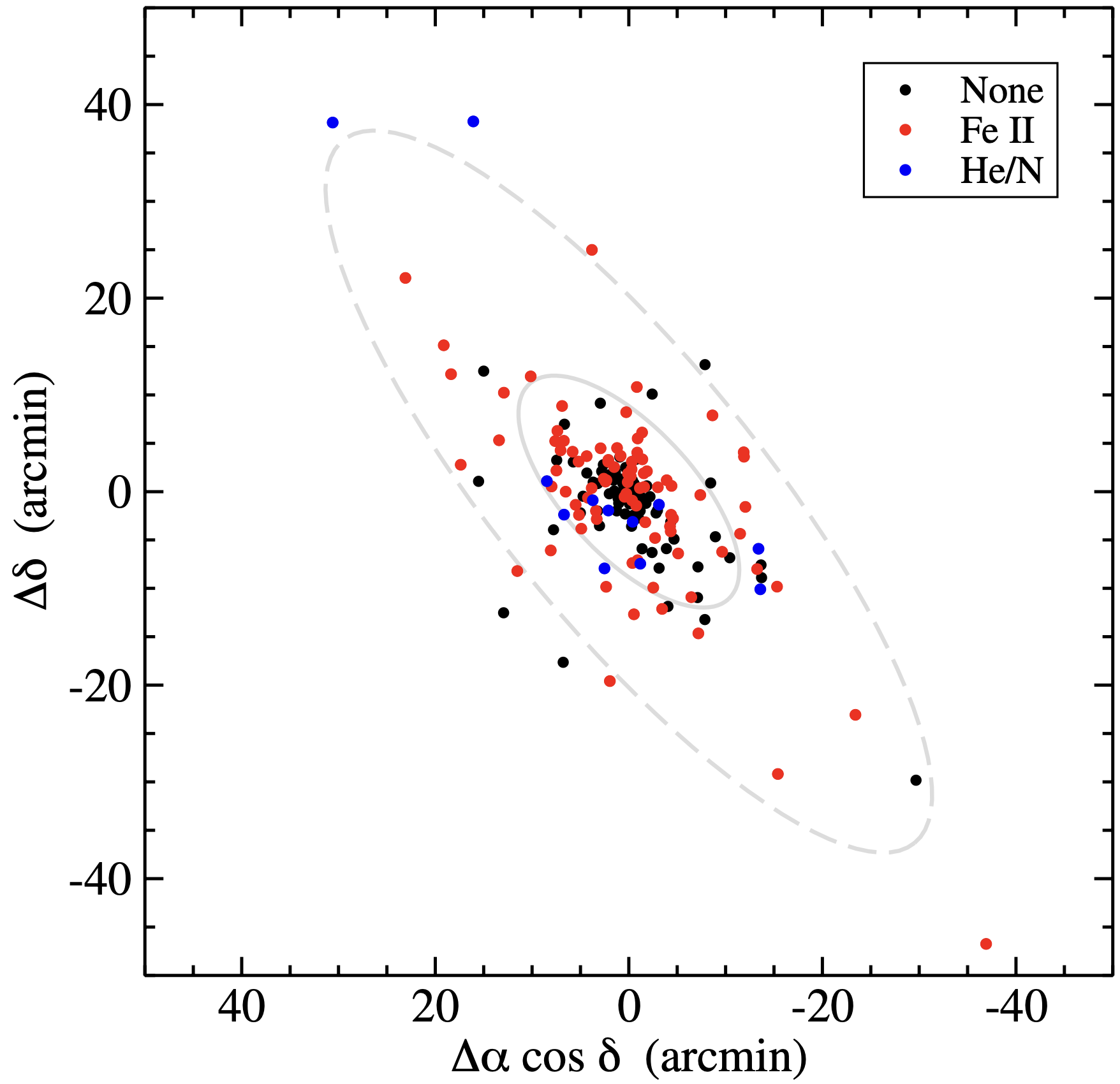}
\caption{The spatial distribution of the 177 novae in our M31 sample
(black filled circles). Known \ion{Fe}{2} novae are shown in red, while
firmly-established He/N novae are shown in blue. The
solid and dashed grey contours show the $a=15'$ and $a=50'$ $R$-band
isophotes, respectively. The majority of novae in our sample
are part of M31's bulge population.
}
\label{fig7}
\end{figure}

\begin{figure}
\plottwo{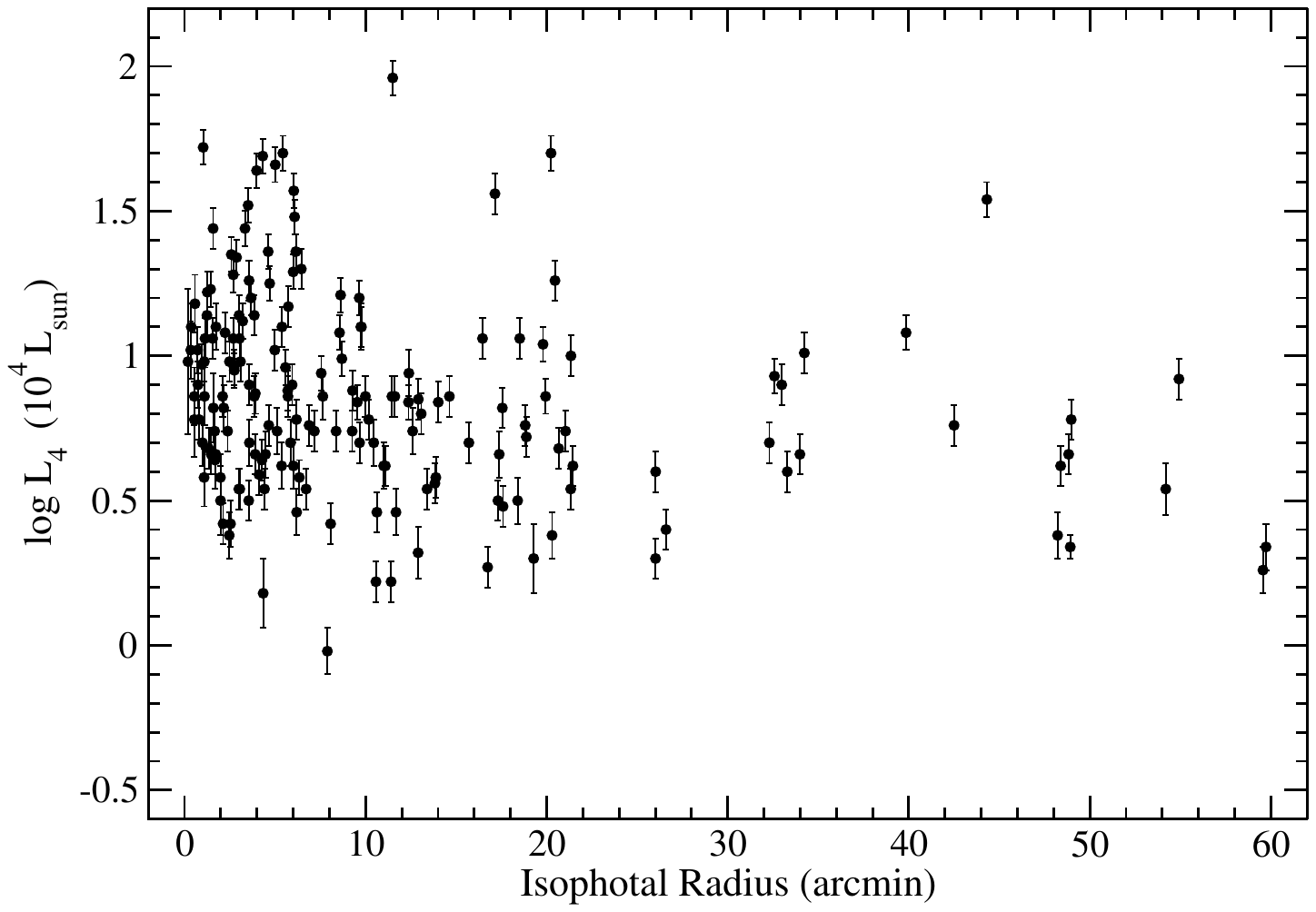}{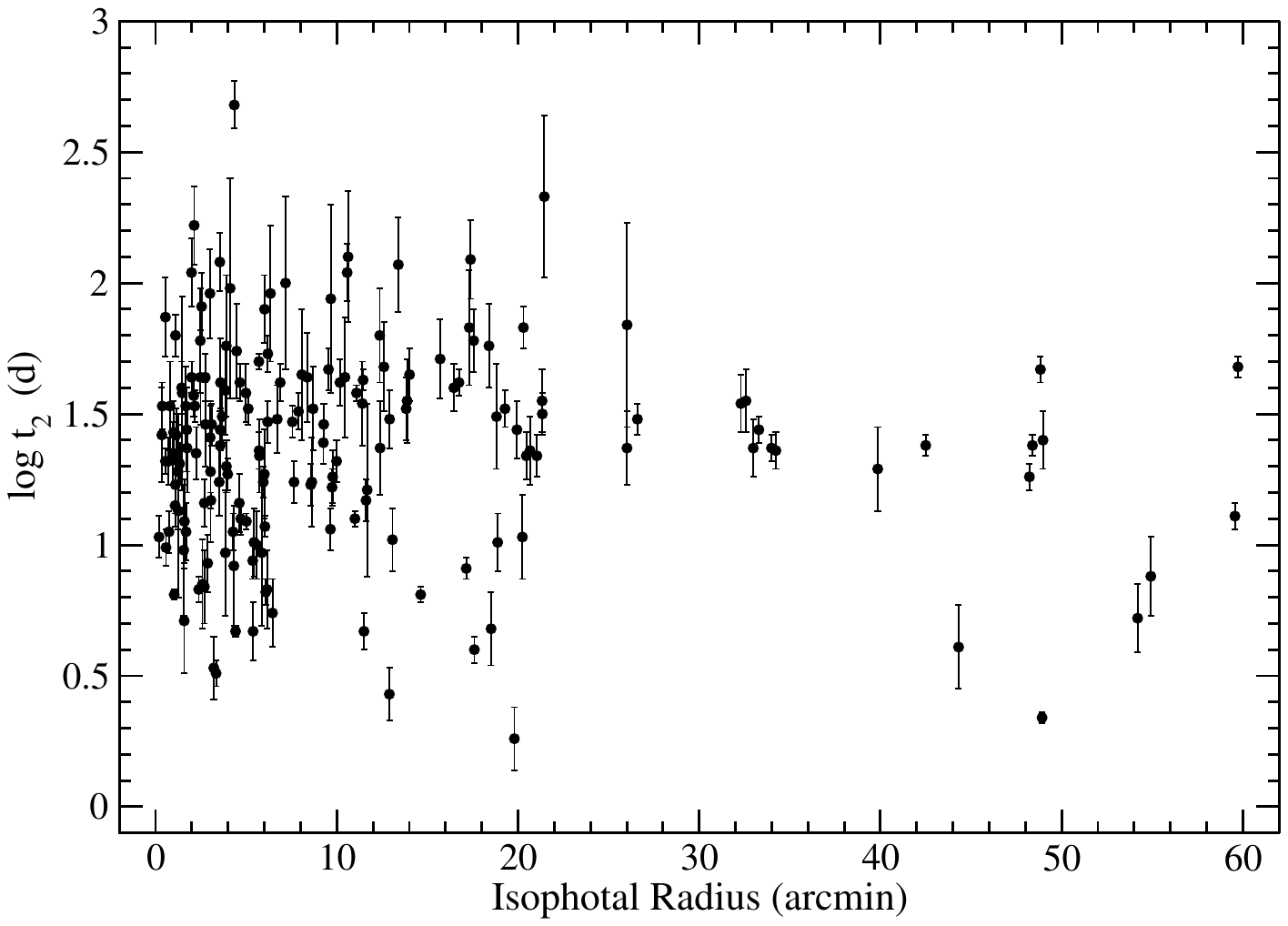}
\plottwo{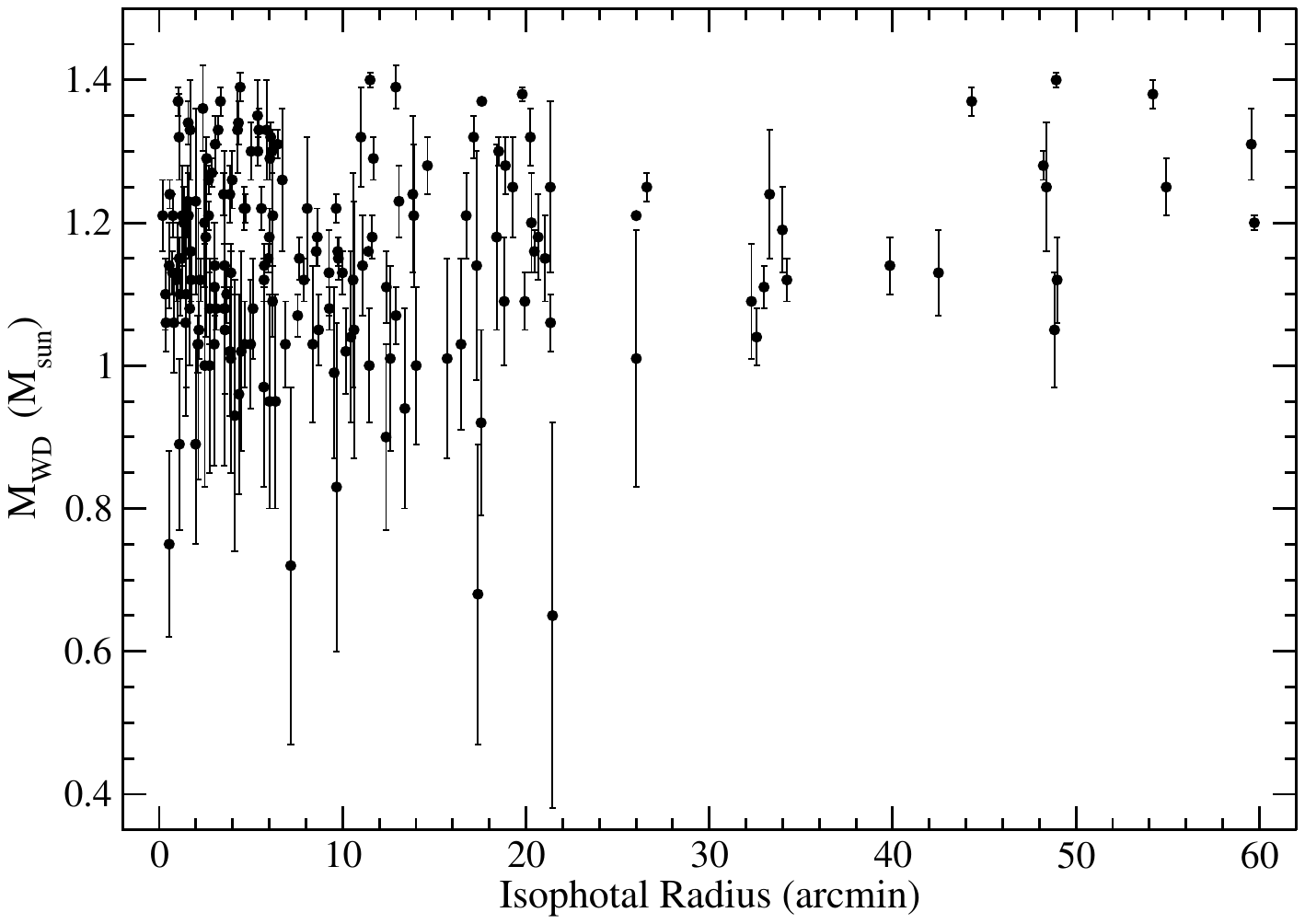}{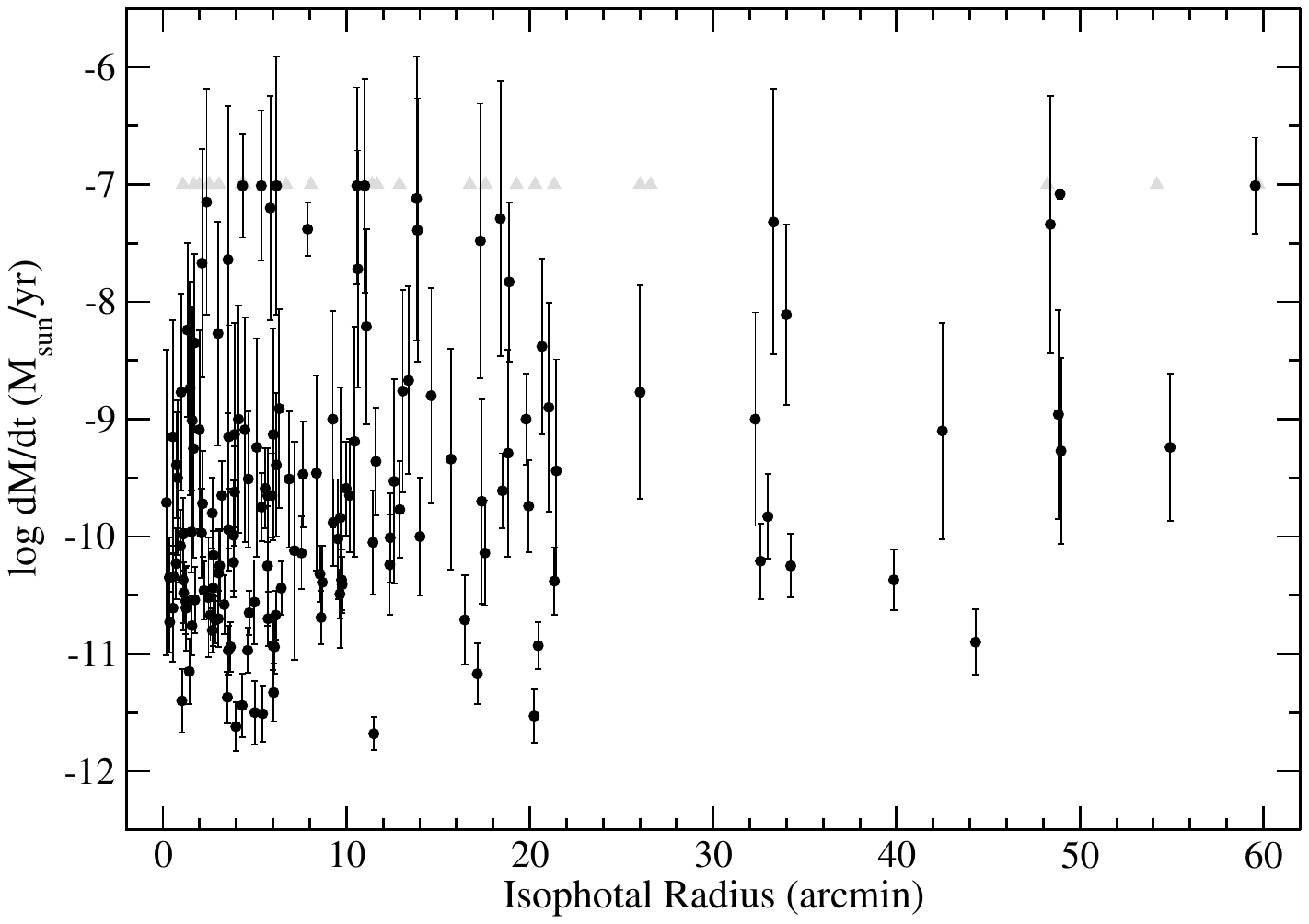}
\plottwo{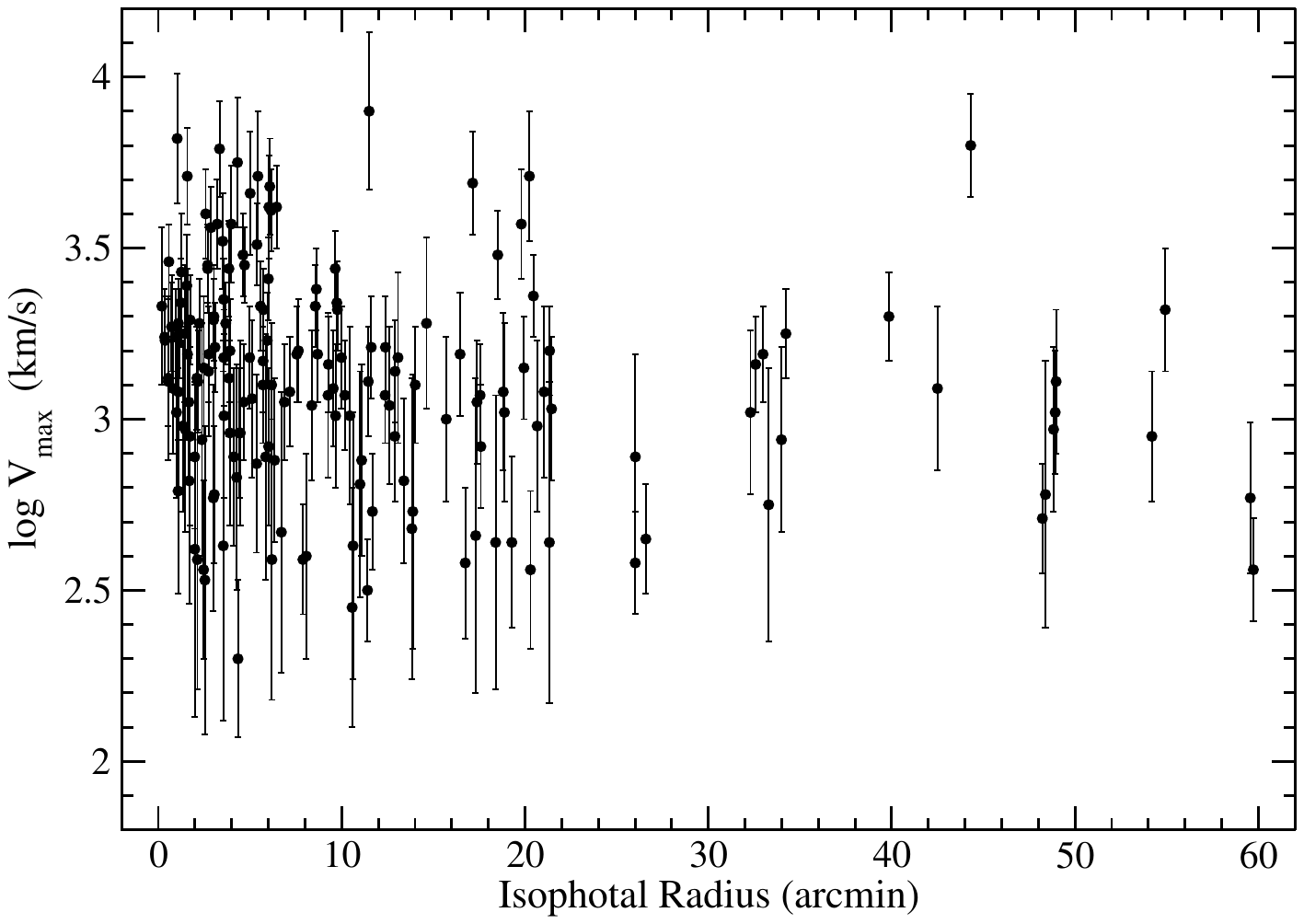}{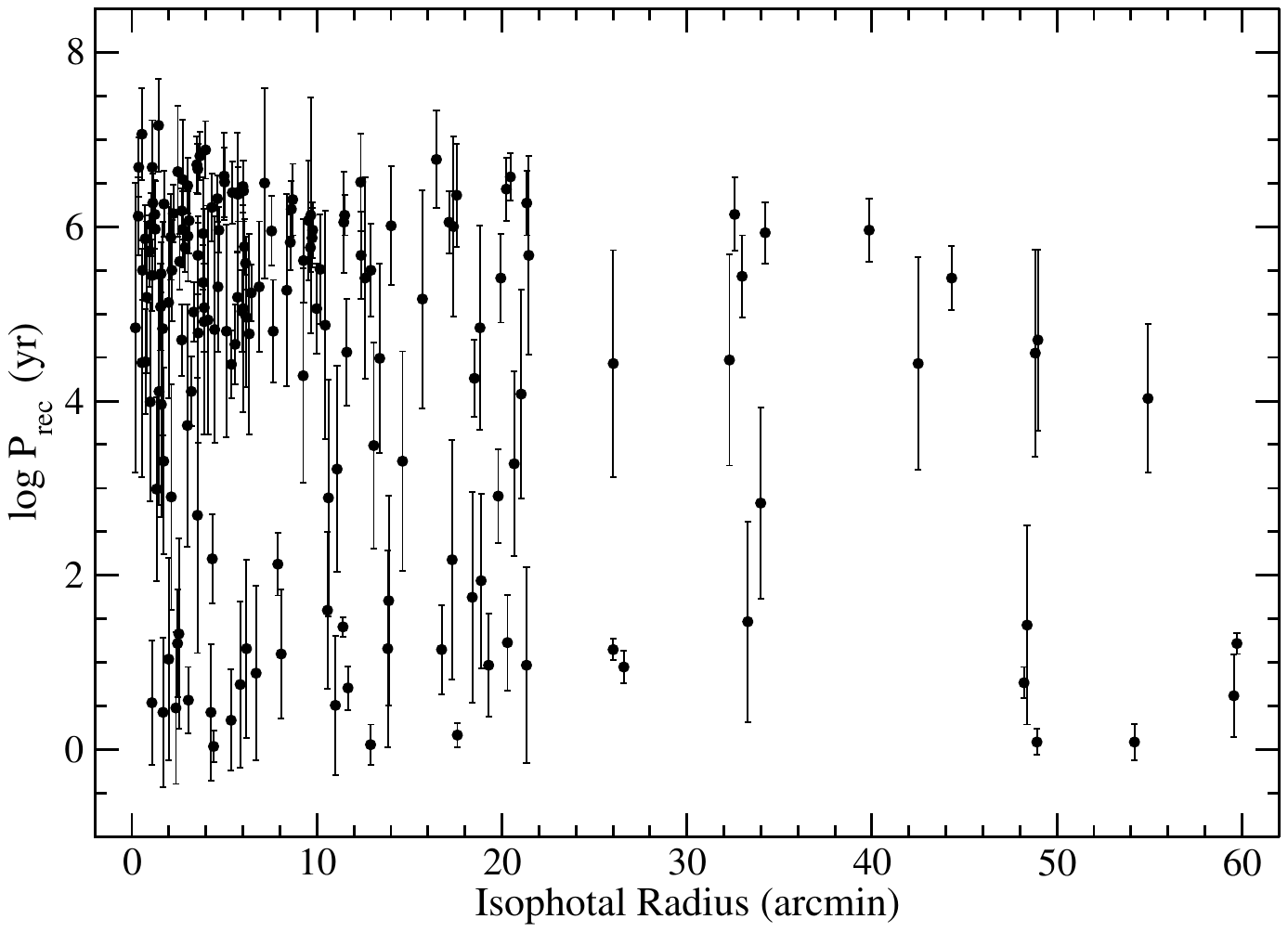}
\caption{Dependence of nova properties on the isophotal radius, $a$, within M31.
Top left: The peak luminosity, $L_4$;
Top right: The $t_2$ time;
Middle left: The observed WD mass, $M_\mathrm{WD}$;
Middle right: the accretion rate, $\log \dot M$. The grey upward
pointing triangles represent novae with $\log \dot M = -7$,
which we consider lower limits on the true accretion rates;
Bottom left: The maximum ejecta velocity, $V_\mathrm{max}$;
Bottom right: The predicted recurrence time, log $P_\mathrm{rec}$.
None of the obsered properties show an obvious dependence on
isophotal radius.
}
\label{fig8}
\end{figure}

We are now in a position to re-visit the photometric analysis using our
updated sample of 177 M31 nova light curves,
of which 143 are new (post 2009) and
were not included in the analysis presented in \citet{Shafter2011}. 
Figure~\ref{fig7} shows the spatial distribution of the novae
in our M31 sample. The spectral types are indicated (red points: \ion{Fe}{2}
and blue points: He/N) when known. The
inner isophote has a semimajor axis, $a=15'$, which at the distance
of M31 corresponds to a radius of $R\simeq3.4$~kpc -- slightly larger
than the radius of M31's bulge component. Novae outside this radius
are assumed to arise from M31's disk population. The observed spatial
distribution does not accurately reflect the true spatial distribution of
M31 novae because our observed sample is biased by more frequent
observations of the inner bulge regions of M31. Thus,
the relative number of novae with $a<15'$ and novae with $a>15'$ does not
reflect the true ratio of bulge to disk novae. Nevertheless,
despite the limited number
of novae that are unambiguously from M31's disk,
it is still possible to explore whether there are any obvious
differences between the speed classes and spectral types of the
bulge and disk novae in our sample.

Figure~\ref{fig8} shows the full slate of nova parameters
plotted as a function of the isophotal radius, $a$. 
In addition to the derived nova parameters ($M_\mathrm{WD}$, $\log {\dot M}$,
$V_\mathrm{max}$, and $P_\mathrm{rec}$), we have also included the principal
input parameters, $L_4$ and $t_2$.
None of the parameters show an obvious trend with distance
from the nucleus of M31. However, we can test this impression
by comparing the means for each parameter in the
bulge and disk regions of M31 separately.
Although the high inclination of M31
to our line of sight makes it difficult to
unambiguously separate the bulge and disk components, if we assume
that M31's bulge and disk components dominate at isophotal radii,
$a\leq15'$ and $a>15'$, respectively, we find no evidence
that the $t_2$ time or $M_\mathrm{WD}$ vary with $a$. However,
the mean
values for the remaining parameters appear to differ at the $>2\sigma$
level (see Table~\ref{tab7}).  In particular, there is a hint 
that the bulge novae may on average have marginally higher peak
luminosities and accretion rates and longer recurrence times
compared with novae with $a>15'$ that we associate with the disk of M31.
These results must be viewed with extreme caution, however, as
all or part of this difference could result from our somewhat arbitrary
choice of $a=15'$ to divide the bulge and disk populations, or from
other systematic effects that we have been unable to take into
account, such as the spatial variation of extinction within M31.

\begin{figure}
\plottwo{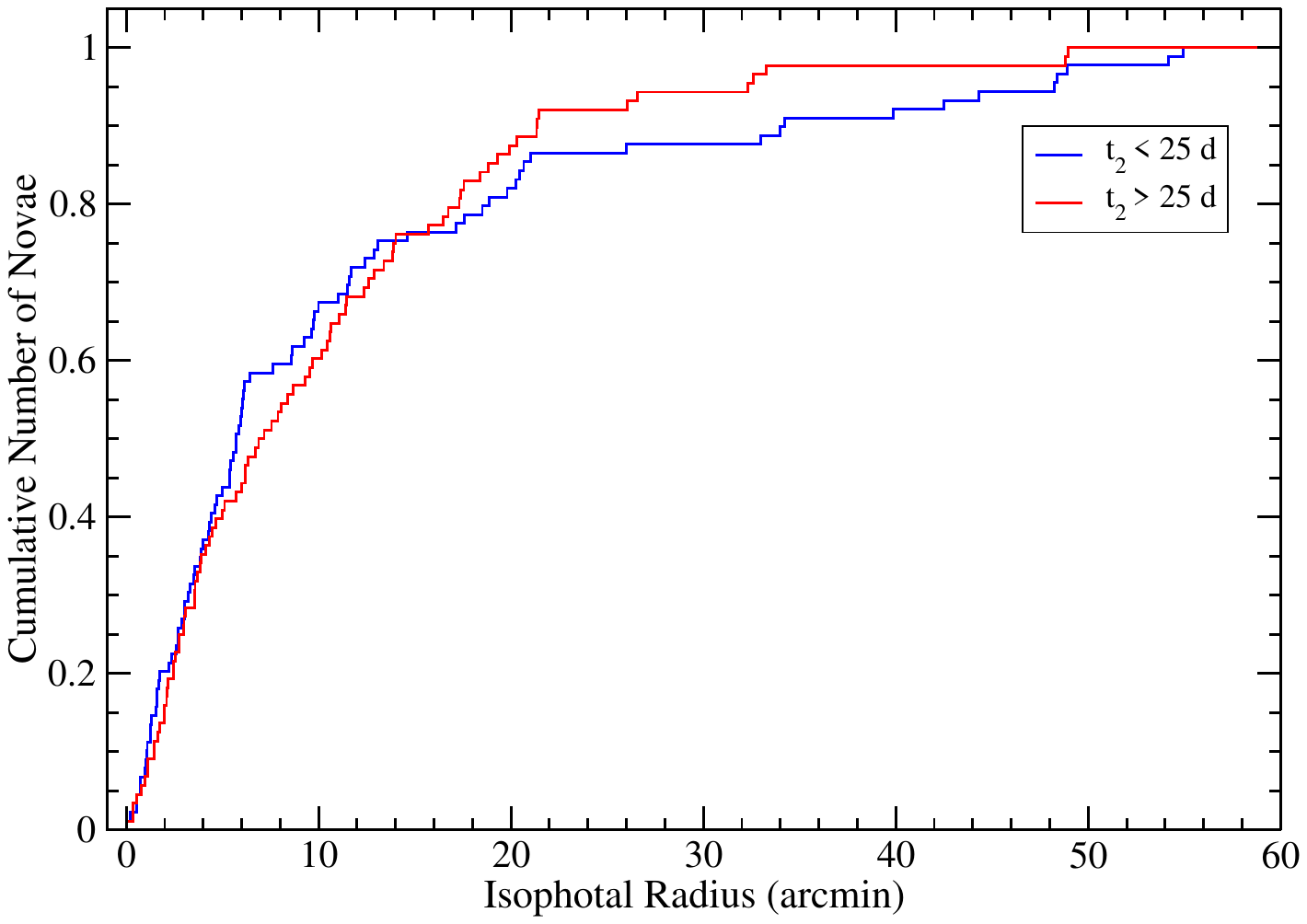}{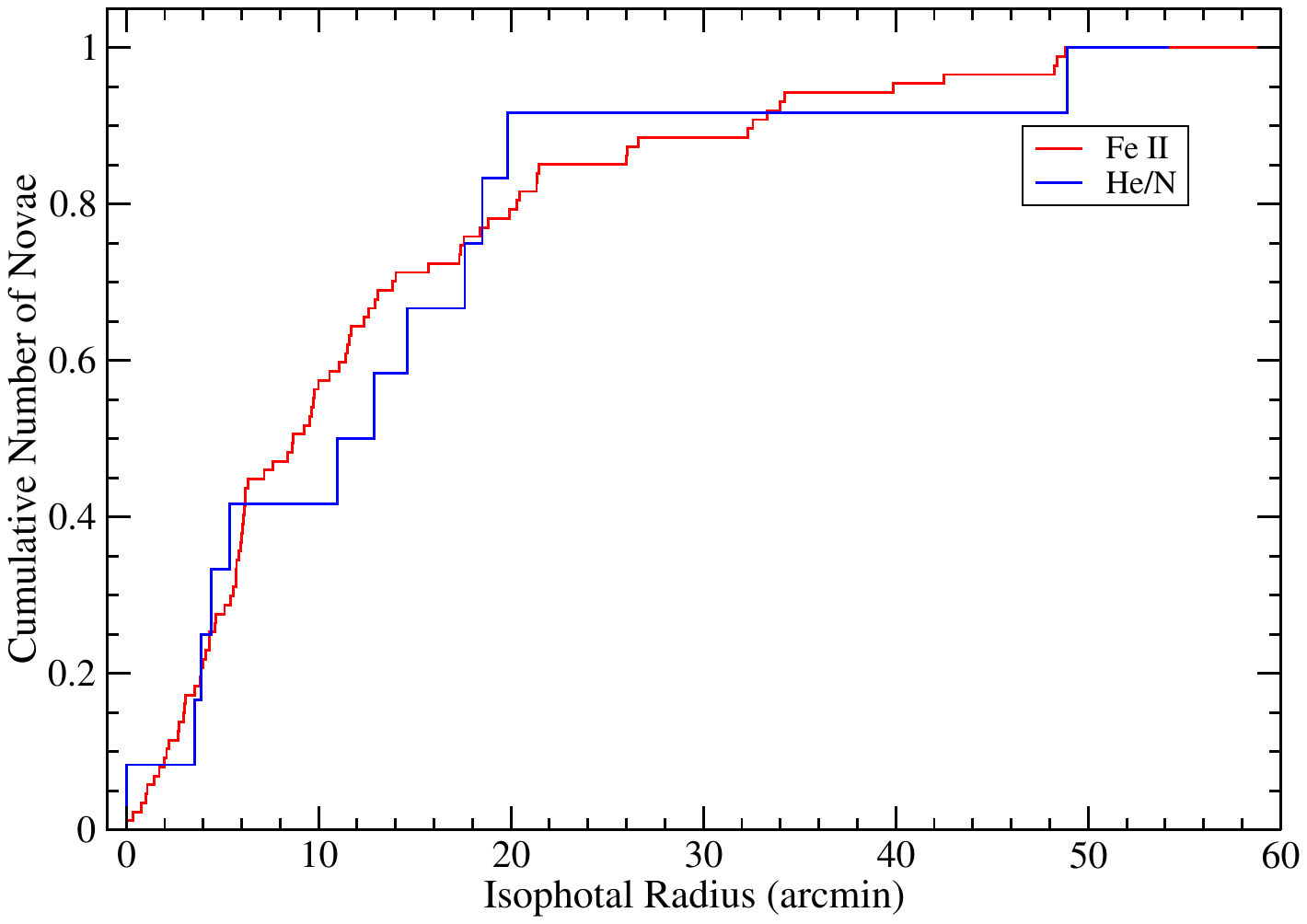}
\caption{Left panel:
The cumulative distributions of the isophotal radius, $a$,
for novae with $t_2\leq25$~d (blue line) compared with novae
with $t_2>25$~d (red line). A Kolmogorov-Smirnov
test yields K-S statistic of K-S $p=0.54$ demonstrating that there
is no significant difference between the two distributions.
Right panel:
The cumulative distributions of the isophotal radius
for novae belonging to the He/N spectroscopic class
(blue line) compared with \ion{Fe}{2} novae
(red line). A K-S
test yields $p=0.86$ demonstrating that there
is no significant difference between the two distributions.
}
\label{fig9}
\end{figure}

A better approach for analyzing spatial variation of nova
properties, such as speed and spectroscopic class,
is to compare their cumulative distributions as
functions of isophotal radius.
Figure~\ref{fig9} shows cumulative distributions for both the speed class
(``fast" vs. ``slow") and the spectroscopic class (He/N vs. \ion{Fe}{2}) for the
total of 116 novae in our sample for which it is known.
Specifically, in the left panel we show
the cumulative distributions of novae with $t_2\leq25$~d, which
following \citet{Warner2003} we refer to as ``fast novae",
compared with novae characterized by $t_2>25$~d (here referred to
as ``slow novae").
The blue line shows the distribution for 93
objects in our ``fast" sample
while the red line shows the distribution for the remaining 87 ``slow" novae.
A Kolmogorov-Smirnov (K-S) test yields a K-S statistic $p=0.55$,
indicating that there is no significant difference between the two nova
samples.

The right panel of Figure~\ref{fig9} shows the cumulative distributions
for novae with known spectroscopic class. As with the speed class, a
K-S test confirms that there is no significant difference between
the spatial distributions of the 87 \ion{Fe}{2} and 29 He/N novae in our sample.

Below we probe the stellar population of novae by
comparing the aggregate properties of novae in M31 with their
Galactic counterparts.

\subsection{Comparison with the Galaxy}

The Andromeda Galaxy (M31) is classified as SA(s)b \citep{Devaucouleurs1991},
with a
prominent classical bulge and tightly wound arms. In contrast, the Milky Way
is classified as SAB(rs)bc \citep{Bland-Hawthorn2016},
featuring a bar-driven pseudo-bulge and fragmented spiral arms.
These structural differences, coupled with divergent star formation rate (SFR)
histories can affect the
present rates of novae and their observed properties.

As pointed out nearly three decades ago by \citet{Yungelson1997},
stellar populations with more recent star formation are expected
to produce a higher frequency of nova eruptions.
This is primarily a consequence of the fact that
the average mass of the WDs in {\it active\/} nova systems is
expected to decline with time since the formation of the progenitor binary.
The progenitor binaries containing
massive WDs that formed long ago in older stellar populations
have by now evolved to states with very low mass transfer rates and
no longer contribute significantly to the active nova population
(i.e., the recurrence times become extremely long).
On the other hand,
populations with more recent star formation are expected to be dominated
by active novae containing more massive WDs, leading to
eruptions that are on average
brighter, faster, with shorter recurrence times
compared with older stellar populations.

M31's current SFR is $\sim$0.25\,$M_\odot$\,yr$^{-1}$ \citep{Ford2013},
largely confined to a 10\,kpc ring that may have been triggered
$\sim$100 -- 200\,Myr ago from an interaction with its
dwarf elliptical companion, M32 \citep{Williams2015}.
The Milky Way has a significantly higher SFR of
$\sim$1.7\,$M_\odot$\,yr$^{-1}$, which is
distributed across the disk \citep{Licquia2015}. M31's star formation
peaked 8 -- 10\,Gyr ago, yielding an
older mean stellar age, while the Milky Way's more 
extended SF history has produced younger populations.
In view of these differences, one can expect the nova rate per unit mass
in stars to be higher in the Galaxy compared with M31, and to contain,
on average, brighter and faster novae.

Current estimates of the nova rate in M31 range from
$R_\mathrm{M31}=40$~yr$^{-1}$
\citep{Rector2022} to $R_\mathrm{M31}=65$~yr$^{-1}$ \citep{Darnley2006},
while the most recent Galactic nova rate estimates have mostly fallen in the
range of $R_\mathrm{MW}\sim 40-50$~yr$^{-1}$ \citep{Shafter2017,De2021,
Kawash2022}. To first order, given the uncertainties,
the best estimates of the nova rates in M31 and the Galaxy appear to
be comparable at a rate of $\sim50\pm15$~yr.
On the other hand, estimates of the mass (in stars) of the two galaxies
are significantly
different, with M31 containing $\sim1.5 - 2$ times the stellar mass
of the Galaxy \citep{Sick2015,Licquia2015}.
Thus, M31 apparently has a lower nova rate per unit mass in stars, in
agreement with the expectations for a relatively quiescent
galaxy with relatively little recent star formation.

\begin{figure}
\plottwo{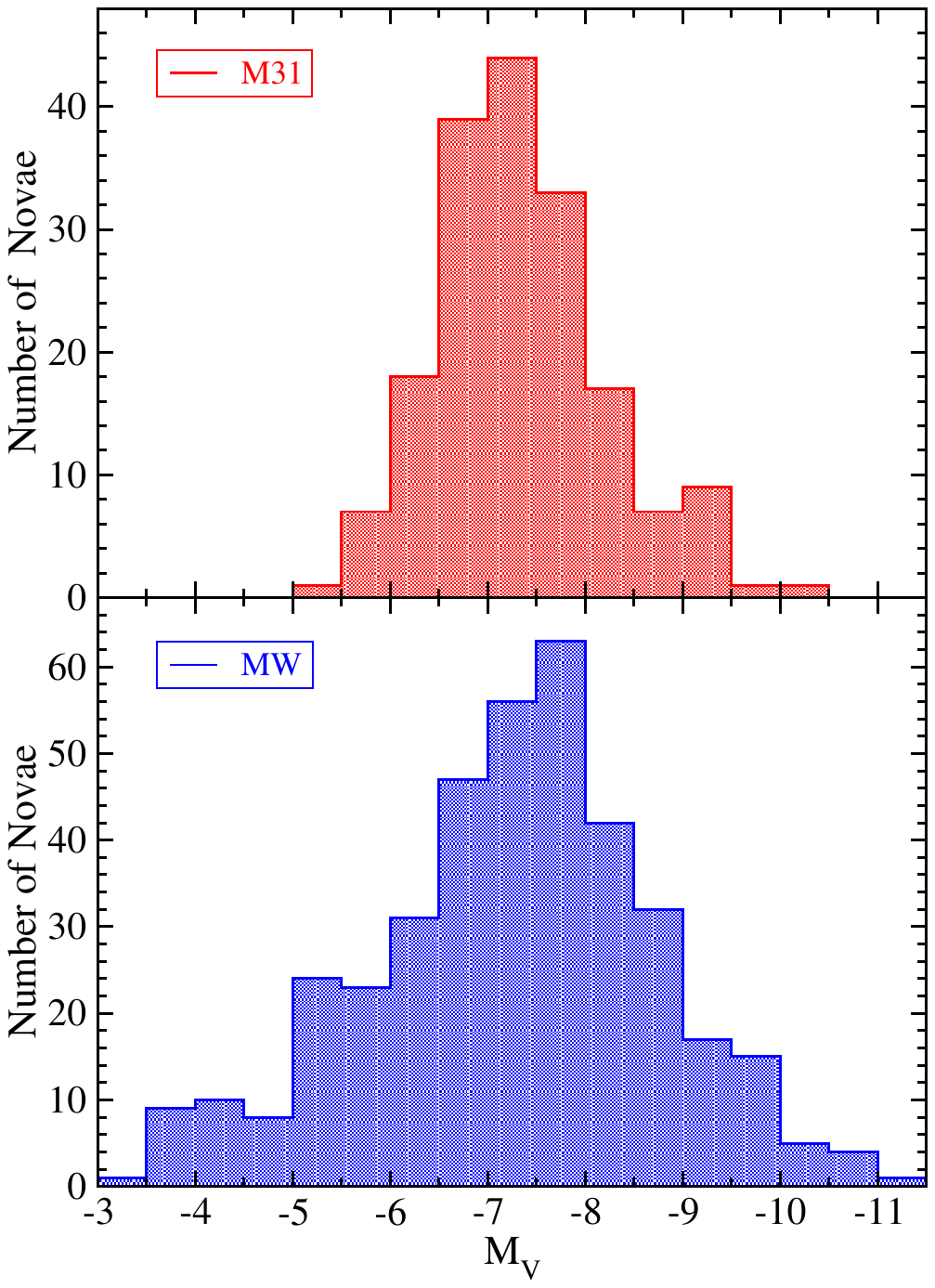}{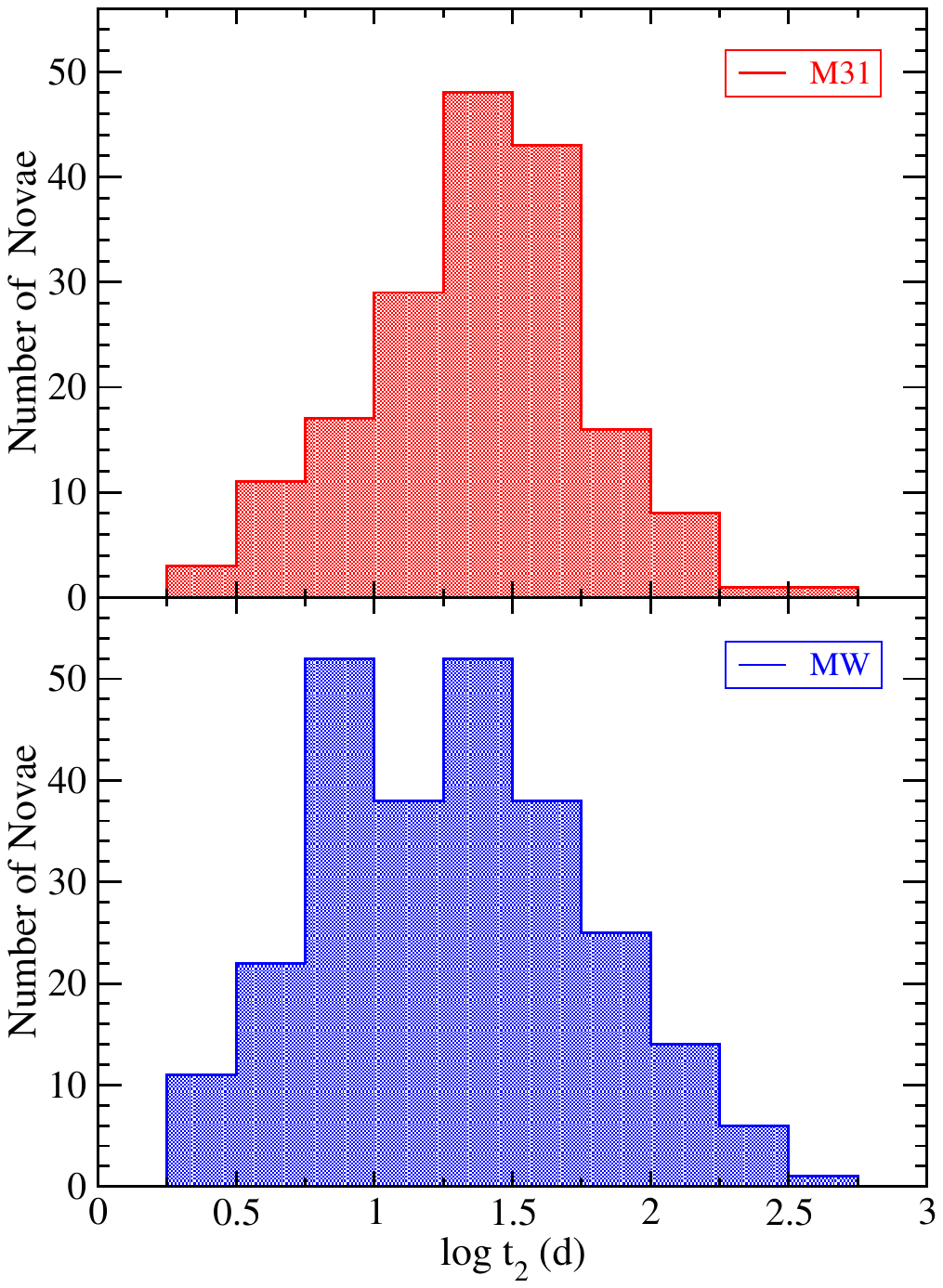}
\caption{Left panel: The absolute magnitude at maximum light, $M_V$,
for our M31 nova
sample (red histogram) compared with the Galactic nova sample
(blue histogram) of \citet{Schaefer2025}. Despite the
low luminosity tail seen in the
Galactic distribution, the absolute magnitude distributions of
M31 and Galactic novae are remarkably similar.
Right panel: The distribution of $t_2$ times for our M31 nova
sample ($M_R$, red histogram) compared with the Galactic nova sample
($M_V$, blue histogram) from \citet{Schaefer2025}. There is
no significant difference in the means for the two distributions.
}
\label{fig10}
\end{figure}

\begin{figure}
\plotone{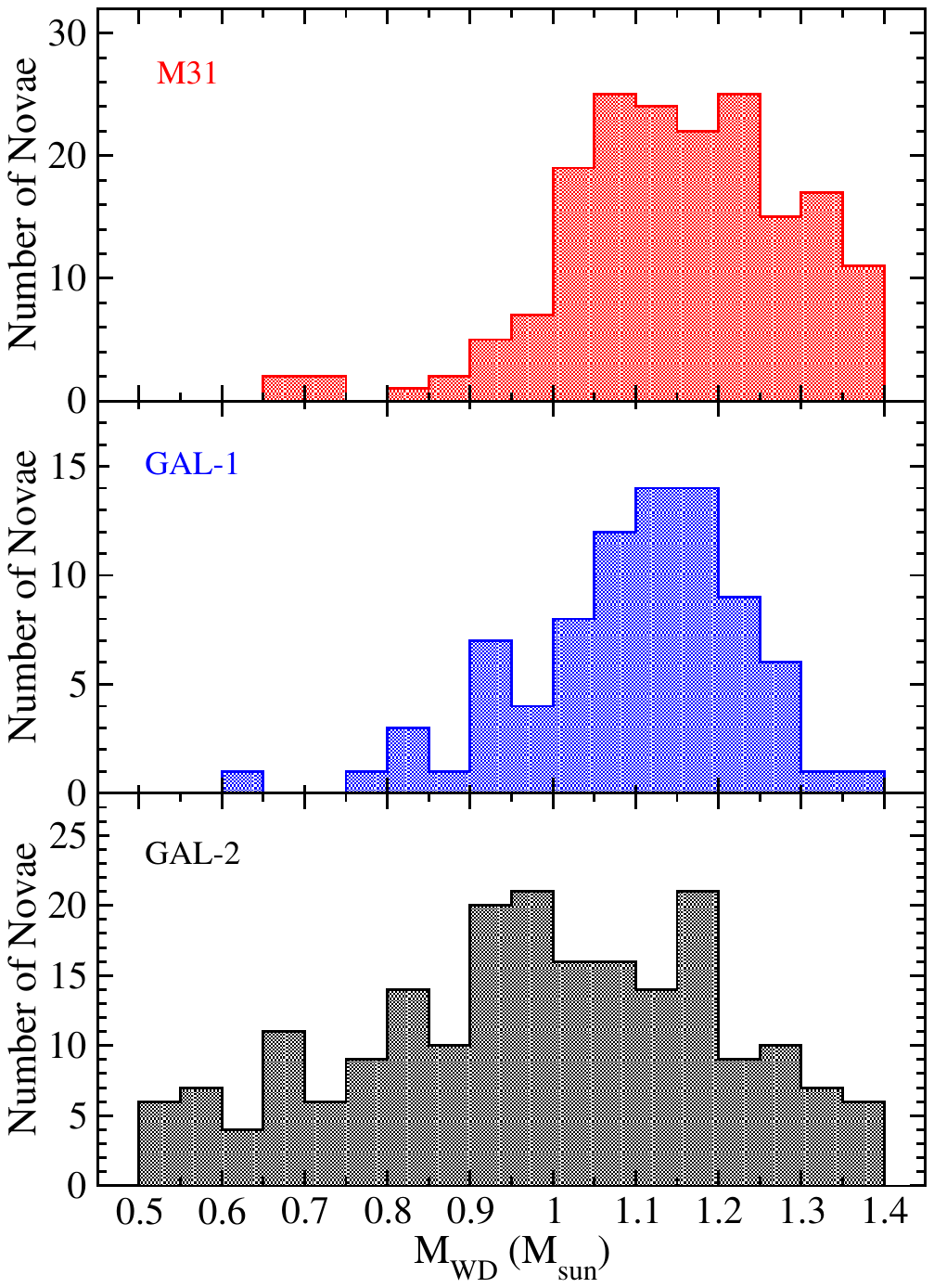}
\caption{The distribution of apparent WD masses (uncorrected for
recurrence time)
from our M31 sample (top panel) compared with the Galactic nova
samples of \citet{Shara2018} and \citet{Schaefer2025}.
The M31 and Galactic sample from \citet{Shara2018} are similar
with mean WD masses of 1.16 and $1.13~M_\odot$, respectively;
however, both of these distributions differ from that of \citet{Schaefer2025}
who finds a significantly lower
mean WD mass for Galactic novae of $\sim1~M_\odot$.
}
\label{fig11}
\end{figure}

In addition to considering the overall nova rates,
we can also compare the observed and model parameters
of the nova populations in M31 and the Milky Way.
Figure~\ref{fig10} shows a comparison of the $V$-band
absolute magnitudes\footnote{We have transformed our values
of $M_R$ to $M_V$ using the expected $V-R$ color of novae
near peak brightness given in \citet{Craig2025}.}
from our M31 nova sample compared with
that from the recent Galactic study by
\citet{Schaefer2025}. With the exception of the faint tail
(which, if present, would not be detected in our M31 survey), the two
distributions appear similar. Indeed, if we only consider
novae brighter than $M_V=-5$ we find $\langle M_V\rangle=-7.49\pm0.0653$
and $\langle M_V\rangle=-7.38\pm0.0659$,
for the Galactic and M31 samples, respectively,
which differ only at the $\sim1\sigma$
level\footnote{Errors reported are standard errors of the mean}.
However, if we restrict the comparison to novae
with $M_V<6.0$, where we expect our M31 sample to be
mostly complete, we find $\langle M_V\rangle=-7.78\pm0.0588$ for the
Galactic sample, which
is significantly brighter (at $>3\sigma$) than the mean of
our M31 nova sample, $\langle M_V\rangle=-7.46\pm0.0632$,
in agreement with expectations based on the differences in stellar population.

We can also compare the rates of decline ($t_2$ times) between the same
M31 and Galactic nova samples,
as shown in the right panel of Figure~\ref{fig10}.
The mean of the M31 $\log t_2$ distribution is
$\langle \log t_2\rangle=1.37\pm0.030$, while that for the Galactic sample is
$\langle \log t_2\rangle=1.28\pm0.030$. Although the mean of the
M31 $\log t_2$ distribution is slightly greater than that
for the Galactic distribution, as expected if the M31 novae
were on average ``slower" than their Galactic counterparts, the difference
is barely significant ($\sim2.1\sigma$).

Now that the distribution of WD masses for a large sample of M31 novae
is available, it is possible for the first time to
directly compare the observed WD mass distributions for M31 and
the Galaxy. Figure~\ref{fig11} shows the apparent M31 WD mass distribution
along with the Galactic distributions from the work of \citet{Shara2018}
and \citet{Schaefer2025}. As previously noted, the M31 WD mass distribution
is remarkably similar to that determined by \citet{Shara2018} for the Galaxy.
Although the apparent distributions (uncorrected for recurrence time bias)
are shown here, the intrinsic distributions for M31 and the Galaxy
also have very similar means and dispersions. The Galactic
WD mass distribution based on the data from
\citet{Schaefer2025} is broader than that observed in M31, with a
more prominent left tail towards lower WD masses. As noted by
Schaefer in his discussion of the Shara masses, the source of the
discrepancy -- whether it lies with the Galactic nova data, or with the
nova models, or both -- is unclear.

\section{Conclusions}

We have analyzed the
$R$-band light curves for a large and homogeneous sample of
177 novae in M31 using the \citet{Yaron2005} models
to estimate fundamental properties of the
progenitor binaries.
To test for possible dependencies on stellar population,
we explored whether any of the resulting nova properties,
including the peak luminosity, rate of decline,
WD mass, accretion rate, maximum ejection velocity, and
recurrence time, varied systematically with deprojected
distance (isophotal radius) from the center of M31.
Finally, the properties of our M31 nova sample were compared with
those for the Galactic
nova population.
Our principal conclusions can be summarized as follows:

{\bf (1)} The observed distribution of WD masses in M31 novae
is approximately Gaussian (with the right tail truncated at $1.4~M_{\odot}$).
The best fit Gaussian is characterized by
$\langle M_\mathrm{WD} \rangle = 1.16\pm0.14~M_{\odot}$,
which is virtually identical to the mean and RMS deviation
(1.15, 0.14) derived directly from the individual masses.
Both the mean and dispersion of the observed
WD mass distribution are remarkably
similar to those found by \citet{Shara2018} for Galactic novae,
with the average observed WD mass of M31 novae being just
$\sim0.03~M_{\odot}$ larger than the Galactic average.
However, the mean WD mass for our M31 nova sample is
$\sim0.15~M_\odot$ higher than that recently found by \citet{Schaefer2025}
in his comprehensive study of more than 300 Galactic novae.

{\bf (2)} As expected, the estimated WD masses for known RNe
cluster toward the high mass end of the distribution. The mean
observed WD mass for RNe was found to be
$\langle M_\mathrm{WD}(\mathrm{RN}) \rangle = 1.33\pm0.08~M_{\odot}$, while
the mean intrinsic WD mass was found to be
$\langle M_\mathrm{WD,int}(\mathrm{RN}) \rangle = 1.27\pm0.10~M_{\odot}$.

{\bf (3)} The observed accretion rate distribution is highly skewed
towards lower accretion rates and is characterized
by $\langle \log \dot M \rangle = -9.3\pm1.4$, corresponding
to an accretion rate of $5\times10^{-10}~M_\odot$~yr$^{-1}$.
An excess of systems with estimated
accretion rates near $10^{-7}~M_\odot$~yr$^{-1}$ is seen, and
is possibly an artefact caused by the truncation of the model
grid for $\log \dot M > -7$. It is also possible that
the spike near $\log \dot M = -7$ is real, reflecting a subpopulation of
novae with evolved secondary stars producing higher accretion rates.

{\bf (4)} As noted by \citet{Shara2018} in their study of
Galactic novae, the observed distributions for $M_\mathrm{WD}$ and
$\log \dot M$ will be biased toward systems with short recurrence times.
The intrinsic (unbiased) WD mass and $\log \dot M$ distributions,
which were estimated
by weighting the observed distributions by the recurrence times
of the novae, are characterized by means of
$\langle M_\mathrm{WD} \rangle = 1.07~M_\odot$ and
$\langle \log \dot M \rangle = -10.7$.
The mean of the intrinsic WD mass distribution is $\sim0.1~M_\odot$
lower than the observed mean, which is similar to the
$0.07~M_\odot$ difference
between the observed and intrinsic means
found by \citet{Shara2018} in their study of Galactic novae. 
The estimated mean intrinsic accretion rate for M31 novae appears to be
remarkably low, of order $10^{-11}~M_\odot$~yr$^{-1}$, which is
about an order of magnitude lower than that estimated
by \citet{Shara2018} for Galactic novae. The source of the
discrepancy is unclear, but if the eruption amplitude is used as
a model input, inferred accretion rates can be overestimated
in low-outburst-amplitude systems with evolved (luminous) secondary stars.

{\bf (5)} The low accretion rates and long recurrence times
inferred for both our M31 sample and for the Galactic novae
studied by \citet{Shara2018} are likely the result of novae 
spending most of their inter-eruption time
in hibernation \citep[e.g., see][]{Shara1986}.

{\bf (6)} The isophotal radius of each M31
nova was computed and used to explore the potential variation of
nova properties with distance from the center of M31. No
systematic variation with isophotal radius
was found for WD masses or $t_2$ times for the novae in our sample.
However, there was weak evidence suggesting that the peak luminosity,
accretion rate, recurrence time, and possibly the maximum ejection
velocity might vary with distance from the center of M31. 
We do not consider this evidence to be compelling given uncertainties
in distinguishing bulge and disk populations from $a$ alone, and
because of the unknown effects of extinction across M31.
This skepticism is reinforced by considering the
cumulative distributions of
``fast" ($t_2\leq25$~d) and ``slow" ($t_2>25$~d)
novae, and of \ion{Fe}{2} and He/N novae in cases where the spectroscopic type
is known, which show no significant variation with isophotal radius.
Overall, we find no compelling evidence that the properties of novae
are sensitive to stellar population (bulge vs. disk) in M31.

{\bf (7)} The M31 and Galactic nova populations appear to
be qualitatively similar in terms of the global nova rate, but differ
slightly in terms of average light curve properties
($M_V$ and $t_2$ time distributions).
In particular, the Galactic novae
are on average brighter ($>3\sigma$ significance) and faster
($1.9\sigma$ significance) than their M31 counterparts.
This matches expectations for novae arising in a galaxy
with a slightly later Hubble type and more recent star formation.

The principal focus of our study has been to compare
the fundamental properties of M31 novae --
particularly the WD mass and accretion rate distributions --
with their Galactic counterparts.
The mean mass accretion rate for M31 novae is surprisingly low,
of order $10^{-10}~M_\odot$~yr$^{-1}$ and $10^{-11}~M_\odot$~yr$^{-1}$
for the apparent and intrinsic distributions, respectively. Both of these
numbers are approximately an order of magnitude lower than those
found for Galactic novae. On the other hand,
the M31 WD mass distribution is
remarkably similar to that found by \citet{Shara2018} for
a sample of 82 Galactic novae, but it is about 10 to 15\% higher
than that recently found by \citet{Schaefer2025} in his comprehensive
study of Galactic novae. In the latter case, the key difference being
the dearth of M31 novae with $M_\mathrm{WD}<1~M_\odot$, or 11\%, compared
with $\sim50$\% in the \citet{Schaefer2025} sample.
A more reliable comparison of the M31 and Galactic nova populations
will require more accurate observations (more complete light curves)
coupled with more detailed nova models.

\bigskip
\noindent
%\begin{acknowledgments}
We thank the anonymous referee for constructive suggestions
that helped improve our paper.
K.H. was supported by the RVO:67985815.
%\end{acknowledgments}

\newpage

\begin{deluxetable}{cccrrc}
\tablenum{1}
\tablecolumns{6}
\tabletypesize{\scriptsize}
\tablecaption{Nova Model Grid\tablenotemark{a}\label{tab1}}
\tablehead{\colhead{$\log {\dot M}$} & \colhead{$M_\mathrm{WD}$} & \colhead{$L_4$} & \colhead{$t_\mathrm{ml}$} & \colhead{$V_\mathrm{max}$} & \colhead{$P_\mathrm{rec}$} \\ \colhead{(M$_{\odot}$~yr$^{-1}$)} & \colhead{(M$_{\odot}$)} & \colhead{($10^4$L$_{\odot}$)} & \colhead{(days)} & \colhead{(km~s$^{-1}$)} & \colhead{(yr)}}
\startdata
$  -7 $&$ 0.65 $&$   1.48 $&$ \dots $&$\dots $&  2.45E+02  \cr
$  -7 $&$ 1.00 $&$   3.45 $&$   210 $&$  265 $&  8.96E+01  \cr
$  -7 $&$ 1.25 $&$   4.84 $&$  65.1 $&$  414 $&  1.92E+01  \cr
$  -7 $&$ 1.40 $&$   6.03 $&$  4.13 $&$ 1410 $&  7.71E$-$01\cr
$  -8 $&$ 0.65 $&$   1.52 $&$  1170 $&$  156 $&  1.01E+04  \cr
$  -8 $&$ 1.00 $&$   3.26 $&$   127 $&$  351 $&  2.06E+03  \cr
$  -8 $&$ 1.25 $&$   6.38 $&$  20.2 $&$ 1110 $&  3.67E+02  \cr
$  -8 $&$ 1.40 $&$   9.27 $&$  1.57 $&$ 3060 $&  1.64E+01  \cr
$  -9 $&$ 0.65 $&$   4.76 $&$   264 $&$ 2590 $&  1.61E+05  \cr
$  -9 $&$ 1.00 $&$   3.88 $&$  93.5 $&$  525 $&  4.66E+04  \cr
$  -9 $&$ 1.25 $&$   6.67 $&$  9.56 $&$ 1480 $&  9.27E+03  \cr
$  -9 $&$ 1.40 $&$   9.23 $&$  1.39 $&$ 2850 $&  4.12E+02  \cr
$ -10 $&$ 0.65 $&$   13.7 $&$   117 $&$ 4210 $&  2.55E+06  \cr
$ -10 $&$ 1.00 $&$   11.3 $&$  33.6 $&$ 1920 $&  8.40E+05  \cr
$ -10 $&$ 1.25 $&$   7.14 $&$  24.3 $&$ 2230 $&  1.91E+05  \cr
$ -10 $&$ 1.40 $&$   10.1 $&$ 0.678 $&$ 5270 $&  5.90E+03  \cr
$ -11 $&$ 0.65 $&$   5.98 $&$  27.6 $&$ 1300 $&  2.58E+07  \cr
$ -11 $&$ 1.00 $&$   6.08 $&$  35.4 $&$ 1250 $&  8.72E+06  \cr
$ -11 $&$ 1.25 $&$   69.4 $&$  9.31 $&$ 4240 $&  2.97E+06  \cr
$ -11 $&$ 1.40 $&$   37.5 $&$  2.95 $&$ 4460 $&  2.59E+05  \cr
$ -12 $&$ 0.65 $&$   20.7 $&$   110 $&$  682 $&  3.94E+08  \cr
$ -12 $&$ 1.00 $&$   9.80 $&$  55.3 $&$ 1230 $&  9.28E+07  \cr
$ -12 $&$ 1.25 $&$    184 $&$  58.6 $&$ 9680 $&  3.22E+07  \cr
$ -12 $&$ 1.40 $&$ \dots  $&$ \dots $&$ \dots$&$ \dots$    \cr
\enddata
\tablenotetext{a}{From \citet{Yaron2005}, Table~3, $T_\mathrm{WD}=10\times10^6$~K.}
\end{deluxetable}

\startlongtable
\begin{deluxetable*}{llrrccccl}
\tablenum{2}
\tablecolumns{9}
\tabletypesize{\scriptsize}
\tablecaption{Linear Lightcurve Nova Parameters\label{tab2}}
\tablehead{\colhead{Nova} & \colhead{} & \colhead{$L_4$} & \colhead{$t_2$} & \colhead{log $\dot M$} & \colhead{$M_\mathrm{WD}$} & \colhead{$V_\mathrm{max}$} & \colhead{log $P_\mathrm{rec}$} & \colhead{} \\ \colhead{(M31N)} & \colhead{Quality} & \colhead{($10^4$L$_{\odot}$)} & \colhead{(days)} & \colhead{(M$_{\odot}$~yr$^{-1}$)}  & \colhead{(M$_{\odot}$)} & \colhead{(km~s$^{-1}$)} & \colhead{(yr)} & \colhead{Type}}
\startdata
2002-08a&Silver&$  6.92\pm 1.10$&$   44.4\pm  10.3$&$-10.00\pm 0.50$&$ 1.00\pm 0.11$&$  1256\pm 495$&$  6.01\pm0.68$&FeII: \cr
2004-08b&Silver&$  5.50\pm 1.04$&$   47.4\pm  18.0$&$ -9.53\pm 0.87$&$ 1.01\pm 0.13$&$  1091\pm 567$&$  5.41\pm1.15$&FeII  \cr
2004-09a&Silver&$  4.57\pm 0.72$&$   27.4\pm  10.1$&$ -8.35\pm 0.76$&$ 1.16\pm 0.07$&$   895\pm 527$&$  3.31\pm1.07$&FeII  \cr
2004-11a&Silver&$ 11.48\pm 2.18$&$   19.0\pm  12.0$&$-10.31\pm 0.37$&$ 1.14\pm 0.06$&$  1995\pm 705$&$  5.89\pm0.52$&FeII  \cr
2004-11b&Silver&$ 10.47\pm 1.66$&$   37.7\pm   9.7$&$-10.56\pm 0.36$&$ 1.03\pm 0.09$&$  1531\pm 541$&$  6.58\pm0.50$&FeIIb \cr
2005-01a&Silver&$ 43.65\pm 5.97$&$   18.7\pm   2.4$&$-11.62\pm 0.21$&$ 1.26\pm 0.04$&$  3673\pm1461$&$  6.88\pm0.33$&FeII  \cr
2005-07a&Bronze&$  5.01\pm 0.95$&$    9.4\pm   6.1$&$ -7.20\pm 0.96$&$ 1.33\pm 0.07$&$   785\pm 641$&$  0.75\pm0.95$&FeII  \cr
2006-06a&Gold  &$  4.17\pm 0.66$&$   37.6\pm   2.8$&$ -8.21\pm 0.83$&$ 1.14\pm 0.07$&$   760\pm 490$&$  3.22\pm1.18$&FeII  \cr
2006-09c&Gold  &$  7.24\pm 1.15$&$   14.9\pm   2.7$&$ -9.36\pm 0.46$&$ 1.18\pm 0.03$&$  1633\pm 576$&$  4.56\pm0.61$&FeII  \cr
2006-12a&Gold  &$  5.50\pm 1.04$&$   33.1\pm   4.6$&$ -9.24\pm 0.93$&$ 1.08\pm 0.07$&$  1140\pm 607$&$  4.80\pm1.22$&FeII  \cr
2007-02b&Gold  &$  9.91\pm 1.57$&$   31.7\pm   5.7$&$-10.38\pm 0.29$&$ 1.06\pm 0.03$&$  1589\pm 485$&$  6.27\pm0.37$&FeII: \cr
2007-07e&Silver&$  3.16\pm 0.84$&$   43.5\pm   5.8$&$ -7.00:       $&$ 1.23\pm 0.13$&$   414\pm 464$&$  1.04\pm1.16$&FeII  \cr
2007-11b&Bronze&$  2.19\pm 0.41$&$   47.4\pm   4.4$&$ -7.00:       $&$ 1.20\pm 0.01$&$   363\pm 127$&$  1.22\pm0.12$&He/Nn \cr
2007-12b&Bronze&$  7.24\pm 1.15$&$    6.5\pm   0.5$&$ -8.80\pm 0.92$&$ 1.28\pm 0.04$&$  1897\pm1107$&$  3.31\pm1.26$&He/N  \cr
2008-05c&Silver&$  6.03\pm 0.95$&$   29.4\pm   5.1$&$ -9.39\pm 0.61$&$ 1.09\pm 0.05$&$  1247\pm 513$&$  4.96\pm0.80$&FeII  \cr
2008-06b&Silver&$ 18.20\pm 2.88$&$   23.8\pm   6.9$&$-10.97\pm 0.21$&$ 1.14\pm 0.03$&$  2239\pm 625$&$  6.66\pm0.29$&He/N  \cr
2008-07b&Silver&$  2.00\pm 0.32$&$   23.5\pm   7.7$&$ -7.00:       $&$ 1.21\pm 0.00$&$   382\pm 134$&$  1.15\pm0.12$&FeII  \cr
2008-08a&Silver&$  9.55\pm 1.51$&$   17.1\pm   4.4$&$ -9.98\pm 0.31$&$ 1.15\pm 0.03$&$  1888\pm 573$&$  5.44\pm0.41$&FeII  \cr
2008-10b&Bronze&$  2.88\pm 0.55$&$   54.1\pm   9.3$&$ -7.01\pm 1.10$&$ 1.21\pm 0.12$&$   386\pm 366$&$  1.16\pm1.02$&FeII  \cr
2008-11a&Bronze&$ 11.48\pm 1.82$&$    4.8\pm   1.5$&$ -9.61\pm 0.32$&$ 1.30\pm 0.02$&$  3048\pm 910$&$  4.26\pm0.44$&He/N  \cr
2009-08d&Bronze&$  6.03\pm 0.95$&$   33.9\pm  13.3$&$ -9.50\pm 0.66$&$ 1.06\pm 0.07$&$  1219\pm 528$&$  5.19\pm0.87$&FeII  \cr
2009-08e&Silver&$  3.16\pm 0.50$&$  118.9\pm  30.3$&$ -7.64\pm 1.31$&$ 1.08\pm 0.22$&$   422\pm 498$&$  2.69\pm1.58$&FeII  \cr
2009-09a&Bronze&$  4.57\pm 0.87$&$  124.2\pm  43.1$&$ -9.70\pm 0.87$&$ 0.68\pm 0.21$&$  1130\pm 466$&$  6.00\pm1.03$&FeII  \cr
2009-10c&Gold  &$ 12.59\pm 2.38$&$   33.6\pm   6.7$&$-10.73\pm 0.26$&$ 1.06\pm 0.04$&$  1698\pm 497$&$  6.68\pm0.34$&FeII  \cr
2009-11a&Gold  &$  4.17\pm 0.66$&$   23.9\pm   2.1$&$ -7.34\pm 1.10$&$ 1.25\pm 0.09$&$   605\pm 547$&$  1.43\pm1.14$&FeII  \cr
2009-11c&Gold  &$  7.05\pm 1.12$&$   30.1\pm   7.7$&$ -9.77\pm 0.41$&$ 1.07\pm 0.04$&$  1377\pm 477$&$  5.50\pm0.53$&FeII  \cr
2009-11d&Gold  &$  5.70\pm 0.90$&$   23.9\pm   2.2$&$ -9.10\pm 0.92$&$ 1.13\pm 0.06$&$  1233\pm 676$&$  4.43\pm1.22$&FeII  \cr
2009-11e&Bronze&$  7.24\pm 1.15$&$   38.8\pm  15.1$&$ -9.99\pm 0.47$&$ 1.02\pm 0.09$&$  1318\pm 514$&$  5.92\pm0.64$&FeII  \cr
2010-05a&Gold  &$  7.24\pm 1.15$&$   37.5\pm   6.8$&$ -9.97\pm 0.38$&$ 1.03\pm 0.04$&$  1327\pm 443$&$  5.88\pm0.49$&FeII  \cr
2010-09b&Gold  &$  2.40\pm 0.45$&$   18.0\pm   2.2$&$ -7.00:       $&$ 1.28\pm 0.02$&$   512\pm 188$&$  0.77\pm0.18$&FeII  \cr
2010-10c&Gold  &$  2.51\pm 0.40$&$   30.1\pm   3.9$&$ -7.00:       $&$ 1.25\pm 0.02$&$   443\pm 162$&$  0.95\pm0.19$&FeII  \cr
2011-01a&Gold  &$ 49.20\pm 6.73$&$    8.4\pm   4.4$&$-11.44\pm 0.27$&$ 1.34\pm 0.03$&$  5636\pm2424$&$  6.22\pm0.39$&FeII  \cr
2011-02a&Bronze&$  6.03\pm 0.95$&$   24.9\pm   6.5$&$ -9.27\pm 0.79$&$ 1.12\pm 0.06$&$  1279\pm 622$&$  4.70\pm1.04$&\dots \cr
2011-02b&Bronze&$  3.80\pm 0.85$&$   14.0\pm   2.6$&$ -7.00:       $&$ 1.32\pm 0.06$&$   614\pm 418$&$  0.54\pm0.71$&\dots \cr
2011-02c&Bronze&$  7.94\pm 1.50$&$   11.3\pm   2.1$&$ -9.39\pm 0.45$&$ 1.21\pm 0.03$&$  1866\pm 633$&$  4.45\pm0.60$&\dots \cr
2011-02d&Gold  &$  7.94\pm 1.26$&$   23.3\pm   5.8$&$ -9.83\pm 0.36$&$ 1.11\pm 0.03$&$  1563\pm 515$&$  5.43\pm0.47$&\dots \cr
2011-06b&Gold  &$  1.85\pm 0.29$&$   42.1\pm   4.7$&$ -7.00:       $&$ 1.21\pm 0.06$&$   383\pm 197$&$  1.15\pm0.51$&\dots \cr
2011-06d&Gold  &$  7.94\pm 1.26$&$   17.2\pm   2.2$&$ -9.65\pm 0.36$&$ 1.15\pm 0.02$&$  1687\pm 539$&$  5.03\pm0.47$&FeII  \cr
2011-07a&Silver&$ 12.02\pm 1.81$&$   22.6\pm   5.4$&$-10.46\pm 0.25$&$ 1.12\pm 0.03$&$  1923\pm 556$&$  6.15\pm0.33$&FeII  \cr
2011-07b&Gold  &$  4.57\pm 0.72$&$   23.4\pm   2.6$&$ -8.11\pm 0.77$&$ 1.19\pm 0.06$&$   861\pm 532$&$  2.83\pm1.10$&FeII  \cr
2011-08b&Gold  &$  8.55\pm 1.17$&$   35.2\pm   9.4$&$-10.21\pm 0.32$&$ 1.04\pm 0.04$&$  1449\pm 466$&$  6.14\pm0.42$&FeII  \cr
2011-12b&Bronze&$  2.40\pm 0.45$&$   67.5\pm  12.5$&$ -7.00:       $&$ 1.20\pm 0.07$&$   361\pm 193$&$  1.23\pm0.55$&FeII  \cr
2012-01a&Silver&$  5.01\pm 0.79$&$   34.8\pm   8.7$&$ -9.00\pm 0.91$&$ 1.09\pm 0.08$&$  1038\pm 584$&$  4.47\pm1.21$&FeII  \cr
2012-02c&Gold  &$  7.59\pm 1.20$&$   28.6\pm   5.3$&$ -9.88\pm 0.37$&$ 1.08\pm 0.03$&$  1452\pm 482$&$  5.61\pm0.48$&\dots \cr
2012-03b&Bronze&$  5.01\pm 0.95$&$   43.5\pm  22.7$&$ -9.19\pm 0.98$&$ 1.04\pm 0.12$&$  1021\pm 610$&$  4.87\pm1.31$&\dots \cr
2012-03c&Bronze&$ 11.48\pm 1.82$&$   39.7\pm   8.1$&$-10.71\pm 0.38$&$ 1.03\pm 0.12$&$  1552\pm 626$&$  6.77\pm0.56$&\dots \cr
2012-05c&Gold  &$  5.50\pm 0.87$&$   21.7\pm   4.0$&$ -8.90\pm 0.89$&$ 1.15\pm 0.06$&$  1194\pm 677$&$  4.08\pm1.20$&\dots \cr
2012-05d&Gold  &$  2.00\pm 0.53$&$   33.0\pm   5.6$&$ -7.00:       $&$ 1.25\pm 0.07$&$   440\pm 252$&$  0.97\pm0.59$&\dots \cr
2012-06a&Silver&$  9.82\pm 1.41$&$   33.2\pm  12.1$&$-10.39\pm 0.31$&$ 1.05\pm 0.05$&$  1560\pm 499$&$  6.31\pm0.41$&FeII  \cr
2012-06f&Bronze&$  1.66\pm 0.26$&$   34.4\pm  12.7$&$ -7.00:       $&$ 1.16\pm 0.00$&$   318\pm 113$&$  1.41\pm0.11$&FeII  \cr
2012-06g&Gold  &$  7.45\pm 1.12$&$   20.0\pm   4.6$&$ -9.62\pm 0.39$&$ 1.13\pm 0.03$&$  1567\pm 526$&$  5.07\pm0.51$&He/N  \cr
2012-07a&Bronze&$  3.16\pm 0.60$&$   58.0\pm  21.9$&$ -7.29\pm 1.17$&$ 1.18\pm 0.13$&$   434\pm 432$&$  1.75\pm1.21$&FeII  \cr
2012-07b&Silver&$  8.71\pm 1.25$&$   29.5\pm   4.0$&$-10.14\pm 0.31$&$ 1.07\pm 0.03$&$  1538\pm 483$&$  5.95\pm0.40$&\dots \cr
2013-02b&Bronze&$  3.47\pm 0.55$&$   14.9\pm   1.1$&$ -7.00:       $&$ 1.31\pm 0.04$&$   596\pm 268$&$  0.57\pm0.38$&\dots \cr
2013-03b&Bronze&$  9.55\pm 1.51$&$   43.6\pm  17.6$&$-10.52\pm 0.51$&$ 1.00\pm 0.17$&$  1419\pm 685$&$  6.63\pm0.75$&\dots \cr
2013-06a&Bronze&$  6.61\pm 1.75$&$   12.3\pm   4.6$&$ -9.01\pm 0.96$&$ 1.21\pm 0.06$&$  1563\pm 908$&$  3.96\pm1.29$&\dots \cr
2013-08a&Gold  &$  4.17\pm 0.79$&$    8.7\pm   1.4$&$ -7.01\pm 0.64$&$ 1.35\pm 0.05$&$   733\pm 435$&$  0.34\pm0.58$&He/N: \cr
2013-10g&Bronze&$  6.85\pm 0.99$&$   47.1\pm   8.2$&$-10.02\pm 0.51$&$ 0.99\pm 0.12$&$  1233\pm 491$&$  6.07\pm0.69$&FeII  \cr
2014-01c&Silver&$  6.61\pm 1.25$&$   34.2\pm   4.6$&$ -9.72\pm 0.45$&$ 1.05\pm 0.04$&$  1291\pm 456$&$  5.50\pm0.58$&\dots \cr
2014-06a&Gold  &$ 10.47\pm 1.98$&$   21.1\pm   4.4$&$-10.23\pm 0.30$&$ 1.13\pm 0.03$&$  1849\pm 551$&$  5.86\pm0.39$&\dots \cr
2014-07a&Gold  &$  4.57\pm 0.72$&$   47.3\pm   5.8$&$ -8.96\pm 0.89$&$ 1.05\pm 0.08$&$   925\pm 516$&$  4.55\pm1.19$&FeII  \cr
2014-09a&Gold  &$ 12.59\pm 2.38$&$   16.7\pm   2.6$&$-10.37\pm 0.26$&$ 1.16\pm 0.02$&$  2178\pm 599$&$  5.87\pm0.34$&FeII  \cr
2015-05b&Silver&$  7.24\pm 1.15$&$   21.0\pm   3.7$&$ -9.59\pm 0.40$&$ 1.13\pm 0.03$&$  1517\pm 513$&$  5.06\pm0.52$&FeII  \cr
2015-06a&Bronze&$ 52.00\pm 7.11$&$    6.5\pm   0.3$&$-11.40\pm 0.27$&$ 1.37\pm 0.02$&$  6546\pm2817$&$  6.02\pm0.36$&FeII  \cr
2015-07d&Silver&$ 30.20\pm 4.34$&$    6.6\pm   0.7$&$-10.94\pm 0.23$&$ 1.32\pm 0.02$&$  4819\pm1506$&$  5.77\pm0.32$&FeII  \cr
2015-08a&Silver&$  5.70\pm 0.90$&$   41.5\pm   7.1$&$ -9.51\pm 0.58$&$ 1.03\pm 0.06$&$  1135\pm 450$&$  5.31\pm0.75$&FeII  \cr
2015-09a&Bronze&$ 45.29\pm 6.19$&$   12.2\pm   0.8$&$-11.50\pm 0.27$&$ 1.30\pm 0.04$&$  4560\pm1891$&$  6.51\pm0.40$&\dots \cr
2015-10a&Silver&$ 35.97\pm 5.43$&$    8.1\pm   0.8$&$-11.17\pm 0.26$&$ 1.32\pm 0.03$&$  4875\pm1711$&$  6.05\pm0.36$&FeIIb \cr
2015-11b&Bronze&$  0.95\pm 0.18$&$   32.2\pm   5.5$&$ -7.38\pm 0.23$&$ 1.12\pm 0.03$&$   390\pm 142$&$  2.13\pm0.36$&\dots \cr
2016-03b&Bronze&$  5.70\pm 0.90$&$   41.5\pm   7.1$&$ -9.51\pm 0.58$&$ 1.03\pm 0.06$&$  1135\pm 450$&$  5.31\pm0.75$&\dots \cr
2016-10b&Silver&$  9.46\pm 1.43$&$   28.6\pm   5.5$&$-10.25\pm 0.30$&$ 1.08\pm 0.03$&$  1607\pm 494$&$  6.07\pm0.38$&FeII  \cr
2016-10c&Silver&$ 10.47\pm 2.33$&$   26.3\pm  10.9$&$-10.35\pm 0.34$&$ 1.10\pm 0.05$&$  1722\pm 551$&$  6.12\pm0.45$&\dots \cr
2016-12b&Bronze&$  5.01\pm 0.95$&$   41.4\pm  16.0$&$ -9.15\pm 0.95$&$ 1.05\pm 0.09$&$  1026\pm 577$&$  4.78\pm1.26$&\dots \cr
2016-12d&Silver&$ 15.85\pm 2.28$&$   30.8\pm   7.0$&$-10.94\pm 0.21$&$ 1.10\pm 0.04$&$  1905\pm 544$&$  6.81\pm0.28$&FeIIb \cr
2017-01a&Bronze&$  9.55\pm 5.43$&$   10.6\pm   2.0$&$ -9.71\pm 1.30$&$ 1.21\pm 0.05$&$  2153\pm1147$&$  4.84\pm1.66$&\dots \cr
2017-01c&Silver&$ 19.59\pm 2.82$&$   18.7\pm   7.3$&$-10.93\pm 0.21$&$ 1.18\pm 0.04$&$  2541\pm 723$&$  6.46\pm0.30$&FeII  \cr
2017-02a&Bronze&$  5.50\pm 0.87$&$    6.7\pm   0.7$&$ -7.15\pm 0.96$&$ 1.36\pm 0.06$&$   879\pm 685$&$  0.48\pm0.87$&\dots \cr
2017-03a&Silver&$  7.24\pm 1.37$&$   17.5\pm   3.3$&$ -9.47\pm 0.45$&$ 1.15\pm 0.03$&$  1581\pm 553$&$  4.80\pm0.59$&FeII  \cr
2017-03b&Silver&$  6.03\pm 1.86$&$   21.1\pm   2.3$&$ -9.15\pm 0.99$&$ 1.14\pm 0.06$&$  1318\pm 740$&$  4.44\pm1.31$&\dots \cr
2017-04b&Gold  &$ 13.80\pm 2.18$&$   25.9\pm   7.0$&$-10.70\pm 0.24$&$ 1.11\pm 0.03$&$  1936\pm 555$&$  6.47\pm0.32$&\dots \cr
2017-06a&Gold  &$ 11.48\pm 1.82$&$   26.6\pm   6.1$&$-10.48\pm 0.26$&$ 1.10\pm 0.03$&$  1782\pm 525$&$  6.27\pm0.34$&\dots \cr
2017-08a&Gold  &$  4.79\pm 0.76$&$   20.6\pm   3.7$&$ -8.24\pm 0.74$&$ 1.20\pm 0.05$&$   951\pm 552$&$  2.99\pm1.05$&\dots \cr
2017-10b&Silver&$  5.50\pm 1.04$&$   33.6\pm  11.6$&$ -9.25\pm 0.93$&$ 1.08\pm 0.08$&$  1135\pm 614$&$  4.83\pm1.22$&\dots \cr
2019-03a&Gold  &$ 49.66\pm 6.79$&$   10.2\pm   3.0$&$-11.51\pm 0.24$&$ 1.33\pm 0.03$&$  5188\pm2223$&$  6.39\pm0.36$&FeII  \cr
\enddata
\end{deluxetable*}

\startlongtable
\begin{deluxetable*}{llrrccccl}
\tablenum{3}
\tablecolumns{9}
\tabletypesize{\scriptsize}
\tablecaption{Break Lightcurve Nova Parameters\label{tab3}}
\tablehead{\colhead{Nova} & \colhead{} & \colhead{$L_4$} & \colhead{$t_2$} & \colhead{log $\dot M$} & \colhead{$M_\mathrm{WD}$} & \colhead{$V_\mathrm{max}$} & \colhead{log $P_\mathrm{rec}$} & \colhead{} \\ \colhead{(M31N)} & \colhead{Quality} & \colhead{($10^4$L$_{\odot}$)} & \colhead{(days)} & \colhead{(M$_{\odot}$~yr$^{-1}$)}  & \colhead{(M$_{\odot}$)} & \colhead{(km~s$^{-1}$)} & \colhead{(yr)} & \colhead{Type}}
\startdata
2006-11a&Gold  &$ 18.20\pm 2.88$&$   21.8\pm   4.5$&$-10.93\pm 0.20$&$ 1.16\pm 0.03$&$  2317\pm 628$&$  6.57\pm0.27$&FeII  \cr
2009-10b&Silver&$ 22.70\pm 3.10$&$    6.8\pm   2.3$&$-10.67\pm 0.21$&$ 1.30\pm 0.03$&$  4102\pm1166$&$  5.58\pm0.30$&FeII  \cr
2010-01a&Gold  &$ 19.23\pm 2.63$&$   14.3\pm   3.1$&$-10.80\pm 0.19$&$ 1.21\pm 0.02$&$  2812\pm 735$&$  6.18\pm0.26$&FeII  \cr
2010-01b&Gold  &$  8.71\pm 1.65$&$   23.7\pm   9.7$&$-10.01\pm 0.38$&$ 1.11\pm 0.05$&$  1633\pm 548$&$  5.67\pm0.50$&\dots \cr
2010-01c&Bronze&$  8.32\pm 1.26$&$    7.6\pm   2.6$&$ -9.24\pm 0.63$&$ 1.25\pm 0.04$&$  2109\pm 893$&$  4.03\pm0.85$&\dots \cr
2010-02a&Gold  &$ 27.80\pm 4.19$&$    5.1\pm   2.3$&$-10.76\pm 0.25$&$ 1.34\pm 0.03$&$  5129\pm1706$&$  5.46\pm0.36$&\dots \cr
2010-03a&Silver&$  7.24\pm 1.61$&$   63.7\pm  11.6$&$-10.37\pm 0.43$&$ 0.89\pm 0.12$&$  1205\pm 377$&$  6.68\pm0.54$&\dots \cr
2010-06a&Gold  &$  7.59\pm 1.20$&$   49.8\pm   3.9$&$-10.25\pm 0.51$&$ 0.97\pm 0.14$&$  1271\pm 505$&$  6.39\pm0.69$&FeII  \cr
2010-06c&Gold  &$  4.79\pm 0.76$&$   23.0\pm   6.8$&$ -8.38\pm 0.75$&$ 1.18\pm 0.06$&$   966\pm 553$&$  3.28\pm1.06$&\dots \cr
2010-07a&Gold  &$ 16.14\pm 2.32$&$   17.3\pm   6.6$&$-10.69\pm 0.23$&$ 1.18\pm 0.03$&$  2415\pm 677$&$  6.20\pm0.32$&FeII  \cr
2010-10a&Gold  &$  6.37\pm 1.01$&$   10.4\pm   2.9$&$ -8.76\pm 0.86$&$ 1.23\pm 0.05$&$  1524\pm 869$&$  3.49\pm1.18$&FeII  \cr
2010-10d&Gold  &$  7.24\pm 1.15$&$   22.9\pm   3.9$&$ -9.65\pm 0.40$&$ 1.12\pm 0.03$&$  1489\pm 502$&$  5.19\pm0.52$&FeII  \cr
2010-12c&Gold  &$ 11.48\pm 1.82$&$    6.9\pm   2.2$&$ -9.80\pm 0.30$&$ 1.26\pm 0.02$&$  2754\pm 793$&$  4.70\pm0.41$&\dots \cr
2011-01b&Gold  &$ 13.06\pm 1.79$&$    3.4\pm   0.9$&$ -9.65\pm 0.29$&$ 1.33\pm 0.02$&$  3698\pm1088$&$  4.11\pm0.40$&\dots \cr
2011-06a&Gold  &$  3.47\pm 0.55$&$   35.8\pm   9.5$&$ -7.00:       $&$ 1.25\pm 0.12$&$   440\pm 473$&$  0.97\pm1.12$&FeII  \cr
2011-08a&Gold  &$  3.98\pm 0.63$&$   27.7\pm   3.0$&$ -7.32\pm 1.13$&$ 1.24\pm 0.09$&$   568\pm 526$&$  1.47\pm1.15$&FeII  \cr
2011-09a&Gold  &$ 11.91\pm 1.63$&$   19.5\pm   7.4$&$-10.37\pm 0.26$&$ 1.14\pm 0.04$&$  2014\pm 599$&$  5.96\pm0.36$&FeII  \cr
2011-09b&Gold  &$  9.12\pm 1.31$&$    9.9\pm   2.9$&$ -9.59\pm 0.34$&$ 1.22\pm 0.03$&$  2128\pm 658$&$  4.65\pm0.46$&FeII  \cr
2011-10a&Bronze&$  9.04\pm 1.24$&$   43.9\pm   9.4$&$-10.44\pm 0.49$&$ 1.00\pm 0.15$&$  1393\pm 598$&$  6.54\pm0.69$&FeII  \cr
2011-10d&Gold  &$  8.87\pm 1.27$&$   29.0\pm  10.4$&$-10.16\pm 0.33$&$ 1.08\pm 0.05$&$  1556\pm 508$&$  5.97\pm0.44$&FeIIb \cr
2011-11c&Gold  &$ 10.28\pm 1.55$&$   22.7\pm   3.6$&$-10.25\pm 0.27$&$ 1.12\pm 0.03$&$  1795\pm 531$&$  5.93\pm0.35$&FeII  \cr
2011-11e&Silver&$ 17.06\pm 2.45$&$   39.5\pm   9.5$&$-11.15\pm 0.28$&$ 1.06\pm 0.13$&$  1770\pm 830$&$  7.16\pm0.53$&FeII  \cr
2011-12a&Gold  &$  5.70\pm 0.86$&$   30.8\pm  14.3$&$ -9.29\pm 0.88$&$ 1.09\pm 0.09$&$  1189\pm 634$&$  4.84\pm1.17$&FeII  \cr
2012-02b&Gold  &$ 34.67\pm 4.51$&$    4.1\pm   1.5$&$-10.90\pm 0.28$&$ 1.37\pm 0.02$&$  6353\pm2262$&$  5.41\pm0.37$&\dots \cr
2012-05a&Gold  &$ 32.81\pm 4.72$&$   17.4\pm   5.2$&$-11.37\pm 0.22$&$ 1.24\pm 0.03$&$  3311\pm1082$&$  6.71\pm0.32$&\dots \cr
2012-06e&Gold  &$  2.88\pm 0.55$&$   16.1\pm  12.1$&$ -7.00:       $&$ 1.29\pm 0.03$&$   535\pm 211$&$  0.71\pm0.25$&FeII  \cr
2012-07c&Gold  &$  7.18\pm 1.03$&$   27.4\pm   7.2$&$ -9.74\pm 0.39$&$ 1.09\pm 0.04$&$  1419\pm 487$&$  5.41\pm0.51$&FeII  \cr
2012-09b&Gold  &$ 27.54\pm 3.77$&$    3.2\pm   0.4$&$-10.58\pm 0.25$&$ 1.37\pm 0.02$&$  6109\pm1979$&$  5.02\pm0.34$&FeIIb \cr
2013-04a&Gold  &$ 13.80\pm 2.18$&$    9.4\pm   5.1$&$-10.22\pm 0.30$&$ 1.24\pm 0.04$&$  2786\pm 875$&$  5.36\pm0.43$&\dots \cr
2013-06b&Silver&$ 14.72\pm 2.22$&$   21.9\pm   7.0$&$-10.70\pm 0.23$&$ 1.14\pm 0.03$&$  2113\pm 598$&$  6.37\pm0.31$&FeII  \cr
2013-08b&Silver&$ 12.59\pm 2.38$&$   23.4\pm   9.0$&$-10.54\pm 0.28$&$ 1.12\pm 0.04$&$  1932\pm 580$&$  6.26\pm0.38$&\dots \cr
2013-09a&Silver&$  3.47\pm 0.55$&$   29.9\pm   8.8$&$ -7.00:       $&$ 1.26\pm 0.10$&$   467\pm 443$&$  0.88\pm1.00$&\dots \cr
2013-09c&Gold  &$ 12.13\pm 1.74$&$   16.8\pm   3.0$&$-10.32\pm 0.24$&$ 1.16\pm 0.02$&$  2138\pm 594$&$  5.82\pm0.32$&\dots \cr
2013-09d&Bronze&$  6.85\pm 0.99$&$   62.7\pm  26.0$&$-10.24\pm 0.43$&$ 0.90\pm 0.13$&$  1180\pm 386$&$  6.51\pm0.56$&FeII  \cr
2013-10a&Gold  &$ 12.59\pm 1.90$&$   18.3\pm   4.3$&$-10.41\pm 0.24$&$ 1.15\pm 0.03$&$  2113\pm 592$&$  5.96\pm0.32$&FeII  \cr
2013-10e&Bronze&$ 11.48\pm 1.82$&$    9.6\pm   5.5$&$ -9.96\pm 0.35$&$ 1.23\pm 0.05$&$  2483\pm 839$&$  5.08\pm0.50$&\dots \cr
2013-10h&Bronze&$ 22.70\pm 3.10$&$   14.5\pm   3.7$&$-10.97\pm 0.19$&$ 1.22\pm 0.03$&$  3020\pm 813$&$  6.32\pm0.27$&FeII  \cr
2013-12b&Gold  &$ 22.49\pm 3.23$&$    7.0\pm   2.8$&$-10.67\pm 0.22$&$ 1.29\pm 0.03$&$  4018\pm1175$&$  5.60\pm0.32$&FeIIb \cr
2014-01a&Gold  &$ 15.85\pm 2.28$&$   11.6\pm   2.2$&$-10.49\pm 0.21$&$ 1.22\pm 0.02$&$  2780\pm 714$&$  5.76\pm0.28$&FeII  \cr
2014-06b&Gold  &$  7.94\pm 1.26$&$   27.6\pm   5.2$&$ -9.94\pm 0.35$&$ 1.08\pm 0.03$&$  1500\pm 489$&$  5.67\pm0.45$&He/N: \cr
2014-12a&Gold  &$ 49.66\pm 6.79$&$   10.8\pm   3.9$&$-11.53\pm 0.23$&$ 1.32\pm 0.04$&$  5070\pm2206$&$  6.43\pm0.36$&\dots \cr
2015-09c&Gold  &$ 90.37\pm12.99$&$    4.7\pm   0.8$&$-11.68\pm 0.14$&$ 1.40\pm 0.01$&$  8017\pm4275$&$  6.13\pm0.23$&FeII  \cr
2015-09d&Silver&$ 16.60\pm 2.63$&$   13.6\pm  10.2$&$-10.61\pm 0.36$&$ 1.21\pm 0.07$&$  2679\pm1043$&$  5.97\pm0.55$&\dots \cr
2016-05a&Silver&$  5.50\pm 0.87$&$   24.3\pm   4.5$&$ -9.00\pm 0.92$&$ 1.13\pm 0.07$&$  1186\pm 668$&$  4.29\pm1.23$&FeII  \cr
2016-08b&Gold  &$  7.24\pm 1.15$&$   42.6\pm   4.3$&$-10.05\pm 0.44$&$ 1.00\pm 0.08$&$  1288\pm 467$&$  6.05\pm0.58$&\dots \cr
2016-08d&Gold  &$ 36.98\pm 5.06$&$   11.8\pm   1.1$&$-11.33\pm 0.25$&$ 1.29\pm 0.03$&$  4178\pm1457$&$  6.41\pm0.35$&FeII  \cr
2016-11a&Gold  &$ 21.68\pm 3.12$&$    8.6\pm   2.2$&$-10.71\pm 0.20$&$ 1.27\pm 0.02$&$  3648\pm 978$&$  5.76\pm0.28$&FeIIb \cr
2016-11b&Gold  &$ 17.70\pm 2.42$&$   12.6\pm   1.8$&$-10.65\pm 0.19$&$ 1.22\pm 0.02$&$  2838\pm 721$&$  5.96\pm0.26$&\dots \cr
2016-12a&Silver&$  6.03\pm 0.91$&$   41.4\pm   9.0$&$ -9.65\pm 0.48$&$ 1.02\pm 0.06$&$  1180\pm 437$&$  5.51\pm0.63$&\dots \cr
2016-12c&Gold  &$  9.29\pm 1.40$&$   22.3\pm   6.0$&$-10.08\pm 0.31$&$ 1.12\pm 0.03$&$  1718\pm 536$&$  5.72\pm0.41$&\dots \cr
2017-01d&Gold  &$ 13.80\pm 2.18$&$   19.1\pm   9.7$&$-10.55\pm 0.28$&$ 1.15\pm 0.05$&$  2163\pm 669$&$  6.14\pm0.39$&\dots \cr
2017-11e&Silver&$ 12.59\pm 1.99$&$    4.7\pm   1.2$&$ -9.75\pm 0.29$&$ 1.30\pm 0.02$&$  3273\pm 924$&$  4.42\pm0.39$&He/N  \cr
\enddata
\end{deluxetable*}

\begin{deluxetable*}{llrrccccl}
\tablenum{4}
\tablecolumns{9}
\tabletypesize{\scriptsize}
\tablecaption{Jitter Lightcurve Nova Parameters\label{tab4}}
\tablehead{\colhead{Nova} & \colhead{} & \colhead{$L_4$} & \colhead{$t_2$} & \colhead{log $\dot M$} & \colhead{$M_\mathrm{WD}$} & \colhead{$V_\mathrm{max}$} & \colhead{log $P_\mathrm{rec}$} & \colhead{} \\ \colhead{(M31N)} & \colhead{Quality} & \colhead{($10^4$L$_{\odot}$)} & \colhead{(days)} & \colhead{(M$_{\odot}$~yr$^{-1}$)}  & \colhead{(M$_{\odot}$)} & \colhead{(km~s$^{-1}$)} & \colhead{(yr)} & \colhead{Type}}
\startdata
2006-10a&Bronze&$  3.16\pm 0.50$&$   67.6\pm  34.4$&$ -7.48\pm 1.17$&$ 1.14\pm 0.16$&$   453\pm 477$&$  2.18\pm1.37$&FeII  \cr
2008-07a&Bronze&$  1.51\pm 0.40$&$  482.5\pm 103.5$&$ -7.01\pm 0.44$&$ 0.96\pm 0.14$&$   201\pm 106$&$  2.19\pm0.51$&FeII  \cr
2009-08a&Bronze&$  3.47\pm 0.55$&$   92.0\pm  36.8$&$ -8.27\pm 0.95$&$ 1.03\pm 0.17$&$   594\pm 457$&$  3.72\pm1.39$&FeII  \cr
2010-06b&Bronze&$  3.98\pm 0.63$&$   69.7\pm  62.6$&$ -8.77\pm 0.91$&$ 1.01\pm 0.18$&$   773\pm 528$&$  4.43\pm1.30$&FeII  \cr
2010-06d&Bronze&$  3.87\pm 0.61$&$   95.0\pm  91.5$&$ -9.00\pm 0.97$&$ 0.93\pm 0.19$&$   776\pm 459$&$  4.93\pm1.31$&FeII  \cr
2010-10b&Bronze&$  4.17\pm 0.66$&$  211.4\pm 151.1$&$ -9.44\pm 0.95$&$ 0.65\pm 0.27$&$  1079\pm 514$&$  5.67\pm1.14$&FeII  \cr
2010-11a&Bronze&$  2.63\pm 0.42$&$   44.9\pm  25.6$&$ -7.00:       $&$ 1.22\pm 0.10$&$   397\pm 272$&$  1.10\pm0.74$&He/Nn \cr
2011-05a&Bronze&$  7.24\pm 1.61$&$   74.3\pm  25.2$&$-10.61\pm 0.46$&$ 0.75\pm 0.13$&$  1282\pm 384$&$  7.06\pm0.53$&\dots \cr
2011-11a&Silver&$  3.63\pm 0.57$&$   33.2\pm   8.9$&$ -7.12\pm 1.21$&$ 1.24\pm 0.11$&$   481\pm 491$&$  1.16\pm1.13$&FeII  \cr
2012-05b&Bronze&$  2.40\pm 0.45$&$   60.2\pm  28.0$&$ -7.00:       $&$ 1.20\pm 0.09$&$   363\pm 213$&$  1.22\pm0.62$&\dots \cr
2013-10b&Bronze&$  5.01\pm 0.95$&$   27.1\pm   7.5$&$ -8.77\pm 0.84$&$ 1.13\pm 0.07$&$  1050\pm 593$&$  3.99\pm1.14$&\dots \cr
2013-12a&Bronze&$  2.63\pm 0.42$&$  166.8\pm  58.9$&$ -7.67\pm 0.97$&$ 1.03\pm 0.19$&$   390\pm 337$&$  2.90\pm1.29$&He/Nn \cr
2014-05a&Bronze&$  4.57\pm 0.87$&$   54.4\pm  23.1$&$ -9.09\pm 0.96$&$ 1.02\pm 0.14$&$   920\pm 563$&$  4.82\pm1.30$&\dots \cr
2014-10a&Bronze&$  3.84\pm 0.55$&$   91.1\pm  55.0$&$ -8.91\pm 0.85$&$ 0.95\pm 0.15$&$   757\pm 416$&$  4.77\pm1.15$&FeII  \cr
2014-11a&Bronze&$  3.80\pm 0.72$&$  110.2\pm  32.1$&$ -9.09\pm 0.85$&$ 0.89\pm 0.14$&$   782\pm 383$&$  5.13\pm1.10$&\dots \cr
2014-12b&Bronze&$  4.57\pm 0.72$&$   38.1\pm  32.1$&$ -8.74\pm 0.91$&$ 1.10\pm 0.13$&$   923\pm 642$&$  4.11\pm1.30$&\dots \cr
2015-03a&Bronze&$  2.88\pm 0.46$&$  127.1\pm  72.0$&$ -7.72\pm 1.01$&$ 1.05\pm 0.18$&$   426\pm 385$&$  2.89\pm1.36$&\dots \cr
2015-04a&Bronze&$  3.47\pm 0.55$&$  116.5\pm  49.0$&$ -8.67\pm 0.80$&$ 0.94\pm 0.14$&$   659\pm 361$&$  4.49\pm1.09$&\dots \cr
2015-05c&Bronze&$  2.63\pm 0.50$&$   82.2\pm  24.2$&$ -7.00:       $&$ 1.18\pm 0.14$&$   336\pm 345$&$  1.33\pm1.09$&\dots \cr
2015-06b&Bronze&$  5.01\pm 0.79$&$   87.2\pm  72.6$&$ -9.84\pm 1.11$&$ 0.83\pm 0.23$&$  1033\pm 508$&$  6.13\pm1.35$&\dots \cr
2015-10b&Bronze&$  5.45\pm 0.82$&$  100.3\pm  75.8$&$-10.12\pm 0.93$&$ 0.72\pm 0.25$&$  1194\pm 431$&$  6.50\pm1.09$&FeII  \cr
2015-11c&Bronze&$  5.01\pm 0.79$&$   51.8\pm  18.2$&$ -9.34\pm 0.94$&$ 1.01\pm 0.14$&$  1007\pm 566$&$  5.17\pm1.25$&FeII  \cr
2016-03d&Bronze&$  4.17\pm 0.79$&$   80.1\pm  24.5$&$ -9.13\pm 0.90$&$ 0.95\pm 0.15$&$   839\pm 450$&$  5.06\pm1.19$&\dots \cr
2016-08e&Silver&$  6.61\pm 1.05$&$   60.6\pm  16.7$&$-10.14\pm 0.45$&$ 0.92\pm 0.13$&$  1167\pm 403$&$  6.36\pm0.59$&FeII  \cr
2016-09a&Bronze&$  4.57\pm 0.72$&$   57.0\pm  35.1$&$ -9.13\pm 0.95$&$ 1.01\pm 0.16$&$   918\pm 563$&$  4.91\pm1.29$&FeII  \cr
2016-09b&Silver&$  5.50\pm 0.87$&$   43.9\pm  17.3$&$ -9.46\pm 0.83$&$ 1.03\pm 0.11$&$  1099\pm 554$&$  5.27\pm1.10$&FeII  \cr
2016-10a&Silver&$  3.80\pm 0.60$&$   35.5\pm  13.0$&$ -7.39\pm 1.12$&$ 1.21\pm 0.10$&$   538\pm 501$&$  1.71\pm1.20$&\dots \cr
\enddata
\end{deluxetable*}

\begin{deluxetable*}{llrrccccclc}
\tablenum{5}
\tablecolumns{11}
\tabletypesize{\scriptsize}
\tablecaption{Recurrent Nova Parameters\label{tab5}}
\tablehead{\colhead{RN} & \colhead{} & \colhead{$L_4$} & \colhead{$t_2$} & \colhead{log $\dot M$} & \colhead{$M_\mathrm{WD}$} & \colhead{$V_\mathrm{max}$} & \colhead{$P_\mathrm{rec}$} & \colhead{$P_\mathrm{rec}$(obs)} & \colhead{} & \colhead{} \\ \colhead {(M31N)} & \colhead{Quality} & \colhead{($10^4$L$_{\odot}$)} & \colhead{(days)} & \colhead{(M$_{\odot}$~yr$^{-1}$)}  & \colhead{(M$_{\odot}$)} & \colhead{(km~s$^{-1}$)} & \colhead{(yr)} & \colhead{(yr)} & \colhead{Type} & \colhead{Notes\tablenotemark{a}}}
\startdata
1923-12c&Silver&$  4.17\pm 0.79$&$   12.6\pm   0.9$&$ -7.01\pm 0.91$&$ 1.32\pm 0.07$&$   641\pm 490$&$      3.2$&$   9.5$&He/N  &  1 \cr
1926-07c&Gold  &$  4.37\pm 1.01$&$   11.1\pm   1.4$&$ -7.00:       $&$ 1.33\pm 0.07$&$   667\pm 556$&$      2.7$&$   2.7$&He/Nn &  2 \cr
1960-12a&Bronze&$  3.47\pm 0.55$&$    4.7\pm   0.2$&$ -7.00:       $&$ 1.39\pm 0.02$&$   920\pm 398$&$      1.1$&$   6.2$&He/N  &  3 \cr
1963-09c&Bronze&$  3.02\pm 0.48$&$    4.0\pm   0.5$&$ -7.00:       $&$ 1.37\pm 0.00$&$   826\pm 335$&$      1.5$&$   5.0$&He/N  &  4 \cr
1966-09e&Silver&$  1.82\pm 0.34$&$   13.0\pm   1.5$&$ -7.01\pm 0.41$&$ 1.31\pm 0.05$&$   586\pm 295$&$      4.2$&$  40.9$&FeII: &  5 \cr
1984-07a&Silver&$ 15.00\pm 3.33$&$    9.7\pm   1.5$&$-10.34\pm 0.26$&$ 1.24\pm 0.02$&$  2877\pm 749$&$ 3.16$E$5  $&$   9.4$&FeIIb &  6 \cr
1990-10a&Silver&$  4.33\pm 0.68$&$   11.1\pm   1.7$&$ -7.00:       $&$ 1.33\pm 0.06$&$   670\pm 504$&$      2.7$&$   5.3$&FeIIb &  7 \cr
1997-11k&Silver&$  1.66\pm 0.26$&$    110\pm    29$&$ -7.01\pm 0.84$&$ 1.12\pm 0.15$&$   283\pm 229$&$     39.9$&$   4.0$&FeII: &  8 \cr
2006-11c&Bronze&$ 11.07\pm 1.51$&$    1.8\pm   0.5$&$ -9.00\pm 0.39$&$ 1.38\pm 0.01$&$  3750\pm1357$&$    814.7$&$   8.2$&He/N  &  9 \cr
2007-10b&Bronze&$  2.09\pm 0.43$&$    2.7\pm   0.6$&$ -7.00:       $&$ 1.39\pm 0.03$&$   899\pm 397$&$      1.2$&$  10.2$&He/N  & 10 \cr
2007-11f&Bronze&$  5.30\pm 0.80$&$   10.2\pm   2.6$&$ -7.83\pm 0.68$&$ 1.28\pm 0.04$&$  1042\pm 628$&$     86.9$&$   9.1$&He/Nn & 11 \cr
2008-12a&Gold  &$  2.17\pm 0.21$&$    2.2\pm   0.1$&$ -7.08\pm 0.04$&$ 1.40\pm 0.01$&$  1047\pm 438$&$      1.2$&$   1.0$&He/N  & 12 \cr
2013-10c&Gold  &$ 19.95\pm 3.16$&$    5.5\pm   1.7$&$-10.44\pm 0.23$&$ 1.31\pm 0.02$&$  4130\pm1160$&$ 1.75$E$5  $&$  10.1$&\dots & 13 \cr
2017-01e&\dots &$  3.47\pm 0.71$&$    5.2\pm   1.6$&$ -7.00:       $&$ 1.38\pm 0.02$&$   885\pm 383$&$      1.2$&$   2.5$&He/N  & 14 \cr
\enddata
\tablenotetext{a}{With the exception of M31N 2017-01e,
outburst data are from \citet{Clark2024} and are based on the following outbursts:
(1) 2012-01b;
(2) 2020-01b, 2022-09a; 
(3) 2013-05b;
(4) 2010-10e, 2015-10c;
(5) 2007-08d;
(6) 2012-09a, 2022-11b;
(7) 2016-07e;
(8) 2009-11b;
(9) 2015-02b;
(10) 2007-10b, 2017-12a;
(11) 2007-11f, 2016-12e;
(12) 2008-12a, and subsequent outbursts from \citet{Burris2023};
(13) 2013-10c;
(14) 2024-08c \citep{Shafter2024b}.
}
\end{deluxetable*}

\begin{deluxetable}{lrrrrrrr}
\tablenum{6}
\tablecolumns{8}
\tabletypesize{\scriptsize}
\tablecaption{M31 Nova Parameters\label{tab6}}
\tablehead{\colhead{Parameter} & \colhead{Linear} & \colhead{Break} & \colhead{Jitter} & \colhead{Fe II} & \colhead{He/N\tablenotemark{a}} & \colhead{RNe} & \colhead{All M31}
}
\startdata
$\langle M_\mathrm{WD} \rangle$ (M$_{\odot}$) & $1.15\pm0.11$ & $1.18\pm0.12$ & $1.00\pm0.15$ & $1.12\pm0.14$ & $1.26\pm0.12$ &$1.33\pm0.08$ & $1.15\pm0.14$  \cr
$\langle M_\mathrm{WD} \rangle \mathrm{(Gold)}$ (M$_{\odot}$) & $1.16\pm0.09$ & $1.20\pm0.10$ & $\dots$ & $\dots$ & $\dots$ & $1.35\pm0.05$ & $1.19\pm0.10$  \cr
$\langle M_\mathrm{WD} \rangle \mathrm{(Silver)}$ (M$_{\odot}$) & $1.13\pm0.08$ & $1.14\pm0.13$ & $1.10\pm0.15$ & $\dots$ & $\dots$ & $1.26\pm0.09$ & $1.14\pm0.11$  \cr
$\langle M_\mathrm{WD} \rangle \mathrm{(Bronze)}$ (M$_{\odot}$) & $1.16\pm0.15$ & $1.12\pm0.16$ & $0.99\pm0.15$ &$\dots$ & $\dots$ & $1.36\pm0.05$ & $1.11\pm0.18$ \cr
$\langle \mathrm{log}~{\dot M} \rangle$ (M$_{\odot}~\mathrm{yr}^{-1}$) & $-9.27\pm1.39$ & $-10.04\pm1.09$ & $-8.59\pm1.08$ & $-9.34\pm1.40$ & $-8.77\pm1.65$ & $-7.69\pm1.27$ & $-9.27\pm1.41$ \cr
$\langle V_\mathrm{max} \rangle$ (km~s$^{-1}$) & $1568\pm1215$ & $2457\pm1560$ & $757\pm306$ & $1610\pm1390$ & $1903\pm1501$& $1373\pm1241$ & $1690\pm1361$ \cr
$\langle \mathrm{log}~P_\mathrm{rec} \rangle$ (yr) & $4.44\pm2.01$ & $5.28\pm1.49$ & $4.04\pm1.79$ & $4.61\pm1.94$& $3.17\pm2.54$ & $1.40\pm1.88$ & $4.39\pm2.06$ \cr
\enddata
\tablenotetext{a}{Including Fe IIb (Hybrid) novae}
\end{deluxetable}

\begin{deluxetable*}{lrcrcr}
\tablenum{7}
\tablecolumns{6}
\tabletypesize{\scriptsize}
\tablecaption{Variation of Nova Parameters with Isophotal Radius\tablenotemark{a}\label{tab7}}
\tablehead{\colhead{Parameter} & \colhead{Mean (bulge)} & \colhead{$\sigma_\mathrm{mean}$ (bulge)} & \colhead{Mean (disk)} & \colhead{$\sigma_\mathrm{mean}$ (disk)} & \colhead{Significance}
}
\startdata
$\log L_4$ & $0.897\pm0.346$ & 0.0300 & $0.739\pm0.343$ & 0.0523 & $2.6\sigma$ \cr
$\log t_2$ & $1.375\pm0.388$  & 0.0338 & $1.355\pm0.435$ & 0.0663 & $0.27\sigma$ \cr
$M_\mathrm{WD}$ &  $1.145\pm0.134$ & 0.0117 & $1.167\pm0.159$ & 0.0243 & $0.82\sigma$ \cr
$\log \dot M$  &  $-9.451\pm1.358$ & 0.118 & $-8.719\pm1.444$ & 0.220 & $2.9\sigma$ \cr
$\log V_\mathrm{max}$  &  $3.141\pm0.314$ & 0.0273  & $3.025\pm0.320$ &  0.0487 & $2.1\sigma$ \cr
$\log P_\mathrm{rec}$  &  $4.659\pm1.945$ & 0.169 & $3.557\pm2.210$ &  0.337 & $2.9\sigma$ \cr
\enddata
\tablenotetext{a}{We define the bulge to include isophotal radii
$a\leq15'$ and the disk for $a>15'$.}
\end{deluxetable*}

%\facilities{}

\newpage

\bibliography{novarefs}{}

\begin{thebibliography}{}
\expandafter\ifx\csname natexlab\endcsname\relax\def\natexlab#1{#1}\fi
\providecommand{\url}[1]{\href{#1}{#1}}
\providecommand{\dodoi}[1]{doi:~\href{http://doi.org/#1}{\nolinkurl{#1}}}
\providecommand{\doeprint}[1]{\href{http://ascl.net/#1}{\nolinkurl{http://ascl.net/#1}}}
\providecommand{\doarXiv}[1]{\href{https://arxiv.org/abs/#1}{\nolinkurl{https://arxiv.org/abs/#1}}}

\bibitem[{{Arp}(1956)}]{Arp1956}
{Arp}, H.~C. 1956, \aj, 61, 15, \dodoi{10.1086/107284}

\bibitem[{{Aydi} {et~al.}(2024){Aydi}, {Chomiuk}, {Strader}, {Sokolovsky},
  {Williams}, {Buckley}, {Ederoclite}, {Izzo}, {Kyer}, {Linford}, {Kniazev},
  {Metzger}, {Miko{\l}ajewska}, {Molaro}, {Molina}, {Mukai}, {Munari}, {Orio},
  {Panurach}, {Shappee}, {Shen}, {Sokoloski}, {Urquhart}, \&
  {Walter}}]{Aydi2024}
{Aydi}, E., {Chomiuk}, L., {Strader}, J., {et~al.} 2024, \mnras, 527, 9303,
  \dodoi{10.1093/mnras/stad3342}

\bibitem[{{Bland-Hawthorn} \& {Gerhard}(2016)}]{Bland-Hawthorn2016}
{Bland-Hawthorn}, J., \& {Gerhard}, O. 2016, \araa, 54, 529,
  \dodoi{10.1146/annurev-astro-081915-023441}

\bibitem[{{Burris} {et~al.}(2023){Burris}, {Shafter}, \&
  {Hornoch}}]{Burris2023}
{Burris}, W.~A., {Shafter}, A.~W., \& {Hornoch}, K. 2023, Research Notes of the
  American Astronomical Society, 7, 179, \dodoi{10.3847/2515-5172/acf220}

\bibitem[{{Ciardullo} {et~al.}(1987){Ciardullo}, {Ford}, {Neill}, {Jacoby}, \&
  {Shafter}}]{Ciardullo1987}
{Ciardullo}, R., {Ford}, H.~C., {Neill}, J.~D., {Jacoby}, G.~H., \& {Shafter},
  A.~W. 1987, \apj, 318, 520, \dodoi{10.1086/165388}

\bibitem[{{Clark} {et~al.}(2024){Clark}, {Hornoch}, {Shafter},
  {Ku{\v{c}}{\'a}kov{\'a}}, {Vra{\v{s}}til}, {Ku{\v{s}}nir{\'a}k}, \&
  {Wolf}}]{Clark2024}
{Clark}, J.~G., {Hornoch}, K., {Shafter}, A.~W., {et~al.} 2024, \apjs, 272, 28,
  \dodoi{10.3847/1538-4365/ad3c39}

\bibitem[{{Cohen} {et~al.}(2025){Cohen}, {Guetta}, {Hillman}, {Della Valle},
  {Izzo}, {Perdelwitz}, \& {Livio}}]{Cohen2025}
{Cohen}, A., {Guetta}, D., {Hillman}, Y., {et~al.} 2025, \apj, 981, 198,
  \dodoi{10.3847/1538-4357/adb628}

\bibitem[{{Craig} {et~al.}(2025){Craig}, {Aydi}, {Chomiuk}, {Strader}, {Stone},
  {Sokolovsky}, {Mukai}, {Kawash}, {Fl{\'o}}, {Boussin}, {Charbonnel}, \&
  {Garde}}]{Craig2025}
{Craig}, P., {Aydi}, E., {Chomiuk}, L., {et~al.} 2025, \mnras, 538, 2339,
  \dodoi{10.1093/mnras/staf385}

\bibitem[{{Curtin} {et~al.}(2015){Curtin}, {Shafter}, {Pritchet}, {Neill},
  {Kundu}, \& {Maccarone}}]{Curtin2015}
{Curtin}, C., {Shafter}, A.~W., {Pritchet}, C.~J., {et~al.} 2015, \apj, 811,
  34, \dodoi{10.1088/0004-637X/811/1/34}

\bibitem[{{Darnley} {et~al.}(2014){Darnley}, {Williams}, {Bode}, {Henze},
  {Ness}, {Shafter}, {Hornoch}, \& {Votruba}}]{Darnley2014}
{Darnley}, M.~J., {Williams}, S.~C., {Bode}, M.~F., {et~al.} 2014, \aap, 563,
  L9, \dodoi{10.1051/0004-6361/201423411}

\bibitem[{{Darnley} {et~al.}(2004){Darnley}, {Bode}, {Kerins}, {Newsam}, {An},
  {Baillon}, {Novati}, {Carr}, {Cr{\'e}z{\'e}}, {Evans}, {Giraud-H{\'e}raud},
  {Gould}, {Hewett}, {Jetzer}, {Kaplan}, {Paulin-Henriksson}, {Smartt},
  {Stalin}, \& {Tsapras}}]{Darnley2004}
{Darnley}, M.~J., {Bode}, M.~F., {Kerins}, E., {et~al.} 2004, \mnras, 353, 571,
  \dodoi{10.1111/j.1365-2966.2004.08087.x}

\bibitem[{{Darnley} {et~al.}(2006){Darnley}, {Bode}, {Kerins}, {Newsam}, {An},
  {Baillon}, {Belokurov}, {Calchi Novati}, {Carr}, {Cr{\'e}z{\'e}}, {Evans},
  {Giraud-H{\'e}raud}, {Gould}, {Hewett}, {Jetzer}, {Kaplan},
  {Paulin-Henriksson}, {Smartt}, {Tsapras}, \& {Weston}}]{Darnley2006}
---. 2006, \mnras, 369, 257, \dodoi{10.1111/j.1365-2966.2006.10297.x}

\bibitem[{{Darnley} {et~al.}(2016){Darnley}, {Henze}, {Bode}, {Hachisu},
  {Hernanz}, {Hornoch}, {Hounsell}, {Kato}, {Ness}, {Osborne}, {Page},
  {Ribeiro}, {Rodr{\'\i}guez-Gil}, {Shafter}, {Shara}, {Steele}, {Williams},
  {Arai}, {Arcavi}, {Barsukova}, {Boumis}, {Chen}, {Fabrika}, {Figueira},
  {Gao}, {Gehrels}, {Godon}, {Goranskij}, {Harman}, {Hartmann}, {Hosseinzadeh},
  {Horst}, {Itagaki}, {Jos{\'e}}, {Kabashima}, {Kaur}, {Kawai}, {Kennea},
  {Kiyota}, {Ku{\v{c}}{\'a}kov{\'a}}, {Lau}, {Maehara}, {Naito}, {Nakajima},
  {Nishiyama}, {O'Brien}, {Quimby}, {Sala}, {Sano}, {Sion}, {Valeev},
  {Watanabe}, {Watanabe}, {Williams}, \& {Xu}}]{Darnley2016}
{Darnley}, M.~J., {Henze}, M., {Bode}, M.~F., {et~al.} 2016, \apj, 833, 149,
  \dodoi{10.3847/1538-4357/833/2/149}

\bibitem[{{De} {et~al.}(2021){De}, {Kasliwal}, {Hankins}, {Sokoloski}, {Adams},
  {Ashley}, {Babul}, {Bagdasaryan}, {Delacroix}, {Dekany}, {Greffe}, {Hale},
  {Jencson}, {Karambelkar}, {Lau}, {Mahabal}, {McKenna}, {Moore}, {Ofek},
  {Sharma}, {Smith}, {Soon}, {Soria}, {Srinivasaragavan}, {Tinyanont},
  {Travouillon}, {Tzanidakis}, \& {Yao}}]{De2021}
{De}, K., {Kasliwal}, M.~M., {Hankins}, M.~J., {et~al.} 2021, \apj, 912, 19,
  \dodoi{10.3847/1538-4357/abeb75}

\bibitem[{{de Vaucouleurs} {et~al.}(1991){de Vaucouleurs}, {de Vaucouleurs},
  {Corwin}, {Buta}, {Paturel}, \& {Fouque}}]{Devaucouleurs1991}
{de Vaucouleurs}, G., {de Vaucouleurs}, A., {Corwin}, Herold~G., J., {et~al.}
  1991, {Third Reference Catalogue of Bright Galaxies}

\bibitem[{{Della Valle} \& {Izzo}(2020{\natexlab{a}})}]{2020A&ARv..28....3D}
{Della Valle}, M., \& {Izzo}, L. 2020{\natexlab{a}}, \aapr, 28, 3,
  \dodoi{10.1007/s00159-020-0124-6}

\bibitem[{{Della Valle} \& {Izzo}(2020{\natexlab{b}})}]{DellaValle2020}
---. 2020{\natexlab{b}}, \aapr, 28, 3, \dodoi{10.1007/s00159-020-0124-6}

\bibitem[{{Della Valle} \& {Livio}(1998)}]{DellaValle1998}
{Della Valle}, M., \& {Livio}, M. 1998, \apj, 506, 818, \dodoi{10.1086/306275}

\bibitem[{{Della Valle} {et~al.}(1994){Della Valle}, {Rosino}, {Bianchini}, \&
  {Livio}}]{DellaValle1994}
{Della Valle}, M., {Rosino}, L., {Bianchini}, A., \& {Livio}, M. 1994, \aap,
  287, 403

\bibitem[{{Ford} {et~al.}(2013){Ford}, {Gear}, {Smith}, {Eales}, {Baes},
  {Bendo}, {Boquien}, {Boselli}, {Cooray}, {De Looze}, {Fritz}, {Gentile},
  {Gomez}, {Gordon}, {Kirk}, {Lebouteiller}, {O'Halloran}, {Spinoglio},
  {Verstappen}, \& {Wilson}}]{Ford2013}
{Ford}, G.~P., {Gear}, W.~K., {Smith}, M. W.~L., {et~al.} 2013, \apj, 769, 55,
  \dodoi{10.1088/0004-637X/769/1/55}

\bibitem[{{Ford}(1978)}]{Ford1978}
{Ford}, H.~C. 1978, \apj, 219, 595, \dodoi{10.1086/155819}

\bibitem[{{Hillman} {et~al.}(2020){Hillman}, {Shara}, {Prialnik}, \&
  {Kovetz}}]{Hillman2020}
{Hillman}, Y., {Shara}, M.~M., {Prialnik}, D., \& {Kovetz}, A. 2020, Nature
  Astronomy, 4, 886, \dodoi{10.1038/s41550-020-1062-y}

\bibitem[{{Hornoch} {et~al.}(2022){Hornoch}, {Kucakova}, \&
  {Shafter}}]{Hornoch2022}
{Hornoch}, K., {Kucakova}, H., \& {Shafter}, A.~W. 2022, The Astronomer's
  Telegram, 15545, 1

\bibitem[{{Hornoch} \& {Shafter}(2015)}]{Hornoch2015}
{Hornoch}, K., \& {Shafter}, A.~W. 2015, The Astronomer's Telegram, 7116, 1

\bibitem[{{Hubble}(1929)}]{Hubble1929}
{Hubble}, E.~P. 1929, \apj, 69, 103, \dodoi{10.1086/143167}

\bibitem[{{Kato} {et~al.}(2014){Kato}, {Saio}, {Hachisu}, \&
  {Nomoto}}]{Kato2014}
{Kato}, M., {Saio}, H., {Hachisu}, I., \& {Nomoto}, K. 2014, \apj, 793, 136,
  \dodoi{10.1088/0004-637X/793/2/136}

\bibitem[{{Kawash} {et~al.}(2022){Kawash}, {Chomiuk}, {Strader}, {Sokolovsky},
  {Aydi}, {Kochanek}, {Stanek}, {Kostrzewa-Rutkowska}, {Hodgkin}, {Mukai},
  {Shappee}, {Jayasinghe}, {Rizzo Smith}, {Holoien}, {Prieto}, \&
  {Thompson}}]{Kawash2022}
{Kawash}, A., {Chomiuk}, L., {Strader}, J., {et~al.} 2022, \apj, 937, 64,
  \dodoi{10.3847/1538-4357/ac8d5e}

\bibitem[{{Knigge}(2011)}]{Knigge2011}
{Knigge}, C. 2011, in Astronomical Society of the Pacific Conference Series,
  Vol. 447, Evolution of Compact Binaries, ed. L.~{Schmidtobreick}, M.~R.
  {Schreiber}, \& C.~{Tappert}, 3, \dodoi{10.48550/arXiv.1108.4716}

\bibitem[{{Kurtenkov} {et~al.}(2015){Kurtenkov}, {Pessev}, {Tomov},
  {Barsukova}, {Fabrika}, {Vida}, {Hornoch}, {Ovcharov}, {Goranskij}, {Valeev},
  {Moln{\'a}r}, {S{\'a}rneczky}, {Kostov}, {Nedialkov}, {Valenti}, {Geier},
  {Wiersema}, {Henze}, {Shafter}, {Mu{\~n}oz Dimitrova}, {Popov}, \&
  {Stritzinger}}]{Kurtenkov2015}
{Kurtenkov}, A.~A., {Pessev}, P., {Tomov}, T., {et~al.} 2015, \aap, 578, L10,
  \dodoi{10.1051/0004-6361/201526564}

\bibitem[{{Licquia} \& {Newman}(2015)}]{Licquia2015}
{Licquia}, T.~C., \& {Newman}, J.~A. 2015, \apj, 806, 96,
  \dodoi{10.1088/0004-637X/806/1/96}

\bibitem[{{Mandel} {et~al.}(2023){Mandel}, {Shara}, {Zurek}, {Conroy}, \& {van
  Dokkum}}]{Mandel2023}
{Mandel}, S., {Shara}, M.~M., {Zurek}, D., {Conroy}, C., \& {van Dokkum}, P.
  2023, \mnras, 518, 5279, \dodoi{10.1093/mnras/stac2960}

\bibitem[{{McConnachie} {et~al.}(2005){McConnachie}, {Irwin}, {Ferguson},
  {Ibata}, {Lewis}, \& {Tanvir}}]{McConnachie2005}
{McConnachie}, A.~W., {Irwin}, M.~J., {Ferguson}, A.~M.~N., {et~al.} 2005,
  \mnras, 356, 979, \dodoi{10.1111/j.1365-2966.2004.08514.x}

\bibitem[{{Nomoto} {et~al.}(2007){Nomoto}, {Saio}, {Kato}, \&
  {Hachisu}}]{Nomoto2007}
{Nomoto}, K., {Saio}, H., {Kato}, M., \& {Hachisu}, I. 2007, ApJ, 663, 1269

\bibitem[{{Pecaut} \& {Mamajek}(2013)}]{Pecaut2013}
{Pecaut}, M.~J., \& {Mamajek}, E.~E. 2013, \apjs, 208, 9,
  \dodoi{10.1088/0067-0049/208/1/9}

\bibitem[{{Pritchet} \& {van den Bergh}(1987)}]{Pritchet1987}
{Pritchet}, C.~J., \& {van den Bergh}, S. 1987, \apj, 318, 507,
  \dodoi{10.1086/165387}

\bibitem[{{Rappaport} {et~al.}(1983){Rappaport}, {Verbunt}, \&
  {Joss}}]{Rappaport1983}
{Rappaport}, S., {Verbunt}, F., \& {Joss}, P.~C. 1983, \apj, 275, 713,
  \dodoi{10.1086/161569}

\bibitem[{{Rector} {et~al.}(2022){Rector}, {Shafter}, {Burris}, {Walentosky},
  {Viafore}, {Strom}, {Cool}, {Sola}, {Crayton}, {Pilachowski}, {Jacoby},
  {Corbett}, {Rene}, \& {Hernandez}}]{Rector2022}
{Rector}, T.~A., {Shafter}, A.~W., {Burris}, W.~A., {et~al.} 2022, \apj, 936,
  117, \dodoi{10.3847/1538-4357/ac87ad}

\bibitem[{{Ritchey}(1917)}]{Ritchey1917}
{Ritchey}, G.~W. 1917, \pasp, 29, 210, \dodoi{10.1086/122638}

\bibitem[{{Rosino}(1964)}]{Rosino1964}
{Rosino}, L. 1964, Annales d'Astrophysique, 27, 498

\bibitem[{{Rosino}(1973)}]{Rosino1973}
---. 1973, \aaps, 9, 347

\bibitem[{{Schaefer}(2018)}]{Schaefer2018}
{Schaefer}, B.~E. 2018, \mnras, 481, 3033, \dodoi{10.1093/mnras/sty2388}

\bibitem[{{Schaefer}(2022)}]{Schaefer2022}
---. 2022, \mnras, 517, 6150, \dodoi{10.1093/mnras/stac2900}

\bibitem[{{Schaefer}(2025)}]{Schaefer2025}
---. 2025, \apj, 993, 232, \dodoi{10.3847/1538-4357/ae0616}

\bibitem[{{Selvelli} \& {Gilmozzi}(2019)}]{Selvelli2019}
{Selvelli}, P., \& {Gilmozzi}, R. 2019, \aap, 622, A186,
  \dodoi{10.1051/0004-6361/201834238}

\bibitem[{{Shafter}(2017)}]{Shafter2017}
{Shafter}, A.~W. 2017, \apj, 834, 196, \dodoi{10.3847/1538-4357/834/2/196}

\bibitem[{{Shafter}(2019)}]{2019enhp.book.....S}
---. 2019, {Extragalactic Novae; A historical perspective},
  \dodoi{10.1088/2514-3433/ab2c63}

\bibitem[{{Shafter} \& {Hornoch}(2025)}]{Shafter2025}
{Shafter}, A.~W., \& {Hornoch}, K. 2025, The Astronomer's Telegram, 17182, 1

\bibitem[{{Shafter} {et~al.}(2012){Shafter}, {Hornoch}, {Ciardullo}, {Darnley},
  \& {Bode}}]{Shafter2012}
{Shafter}, A.~W., {Hornoch}, K., {Ciardullo}, J. V.~R., {Darnley}, M.~J., \&
  {Bode}, M.~F. 2012, The Astronomer's Telegram, 4503, 1

\bibitem[{{Shafter} \& {Irby}(2001)}]{Shafter2001}
{Shafter}, A.~W., \& {Irby}, B.~K. 2001, \apj, 563, 749, \dodoi{10.1086/324044}

\bibitem[{{Shafter} {et~al.}(2017){Shafter}, {Kundu}, \&
  {Henze}}]{Shafter2017b}
{Shafter}, A.~W., {Kundu}, A., \& {Henze}, M. 2017, Research Notes of the
  American Astronomical Society, 1, 11, \dodoi{10.3847/2515-5172/aa9847}

\bibitem[{{Shafter} {et~al.}(2022{\natexlab{a}}){Shafter}, {Taguchi}, {Zhao},
  \& {Hornoch}}]{Shafter2022c}
{Shafter}, A.~W., {Taguchi}, K., {Zhao}, J., \& {Hornoch}, K.
  2022{\natexlab{a}}, Research Notes of the American Astronomical Society, 6,
  241, \dodoi{10.3847/2515-5172/aca2a6}

\bibitem[{{Shafter} {et~al.}(2011){Shafter}, {Darnley}, {Hornoch},
  {Filippenko}, {Bode}, {Ciardullo}, {Misselt}, {Hounsell}, {Chornock}, \&
  {Matheson}}]{Shafter2011}
{Shafter}, A.~W., {Darnley}, M.~J., {Hornoch}, K., {et~al.} 2011, \apj, 734,
  12, \dodoi{10.1088/0004-637X/734/1/12}

\bibitem[{{Shafter} {et~al.}(2015){Shafter}, {Henze}, {Rector}, {Schweizer},
  {Hornoch}, {Orio}, {Pietsch}, {Darnley}, {Williams}, {Bode}, \&
  {Bryan}}]{Shafter2015}
{Shafter}, A.~W., {Henze}, M., {Rector}, T.~A., {et~al.} 2015, \apjs, 216, 34,
  \dodoi{10.1088/0067-0049/216/2/34}

\bibitem[{{Shafter} {et~al.}(2022{\natexlab{b}}){Shafter}, {Hornoch},
  {Ku{\v{c}}{\'a}kov{\'a}}, {Zhao}, {Zhang}, {Gao}, {Costa}, {Burris}, {Clark},
  {Wolf}, \& {Zasche}}]{Shafter2022a}
{Shafter}, A.~W., {Hornoch}, K., {Ku{\v{c}}{\'a}kov{\'a}}, H., {et~al.}
  2022{\natexlab{b}}, Research Notes of the American Astronomical Society, 6,
  214, \dodoi{10.3847/2515-5172/ac9ab9}

\bibitem[{{Shafter} {et~al.}(2022{\natexlab{c}}){Shafter}, {Hornoch}, {Zhao},
  {Tu}, {Xu}, {Zhang}, {Ruan}, {Sun}, {Ding}, \& {Gao}}]{Shafter2022b}
{Shafter}, A.~W., {Hornoch}, K., {Zhao}, J., {et~al.} 2022{\natexlab{c}}, The
  Astronomer's Telegram, 15729, 1

\bibitem[{{Shafter} {et~al.}(2024{\natexlab{a}}){Shafter}, {Zhao}, {Hornoch},
  {Ku{\v{c}}{\'a}kov{\'a}}, {Taguchi}, {Zhang}, {You}, {Wang}, {Xu}, {Wang},
  {Ren}, {Ding}, {Yan}, {Zhang}, {Wang}, {Bond}, {Williams}, \&
  {Zeimann}}]{Shafter2024b}
{Shafter}, A.~W., {Zhao}, J., {Hornoch}, K., {et~al.} 2024{\natexlab{a}},
  Research Notes of the American Astronomical Society, 8, 256,
  \dodoi{10.3847/2515-5172/ad84d5}

\bibitem[{{Shafter} {et~al.}(2024{\natexlab{b}}){Shafter}, {Hornoch},
  {Ku{\v{c}}{\'a}kov{\'a}}, {Fatka}, {Zhao}, {Gao}, {Yaqup}, {Zhong},
  {Esamdin}, {Bai}, {Wang}, {Benni}, {Luo}, \& {Yousuf}}]{Shafter2024a}
{Shafter}, A.~W., {Hornoch}, K., {Ku{\v{c}}{\'a}kov{\'a}}, H., {et~al.}
  2024{\natexlab{b}}, Research Notes of the American Astronomical Society, 8,
  5, \dodoi{10.3847/2515-5172/ad19de}

\bibitem[{{Shara}(1981)}]{Shara1981}
{Shara}, M.~M. 1981, \apj, 243, 926, \dodoi{10.1086/158657}

\bibitem[{{Shara}(1989)}]{Shara1989}
---. 1989, \pasp, 101, 5, \dodoi{10.1086/132400}

\bibitem[{{Shara} {et~al.}(1986){Shara}, {Livio}, {Moffat}, \&
  {Orio}}]{Shara1986}
{Shara}, M.~M., {Livio}, M., {Moffat}, A. F.~J., \& {Orio}, M. 1986, \apj, 311,
  163, \dodoi{10.1086/164762}

\bibitem[{{Shara} {et~al.}(2018){Shara}, {Prialnik}, {Hillman}, \&
  {Kovetz}}]{Shara2018}
{Shara}, M.~M., {Prialnik}, D., {Hillman}, Y., \& {Kovetz}, A. 2018, \apj, 860,
  110, \dodoi{10.3847/1538-4357/aabfbd}

\bibitem[{{Shara} {et~al.}(2016){Shara}, {Doyle}, {Lauer}, {Zurek}, {Neill},
  {Madrid}, {Miko{\l}ajewska}, {Welch}, \& {Baltz}}]{Shara2016}
{Shara}, M.~M., {Doyle}, T.~F., {Lauer}, T.~R., {et~al.} 2016, \apjs, 227, 1,
  \dodoi{10.3847/0067-0049/227/1/1}

\bibitem[{{Shara} {et~al.}(2023){Shara}, {Lessing}, {Hounsell}, {Mandel},
  {Zurek}, {Darnley}, {Graur}, {Hillman}, {Meyer}, {Mikolajewska}, {Neill},
  {Prialnik}, \& {Sparks}}]{Shara2023}
{Shara}, M.~M., {Lessing}, A.~M., {Hounsell}, R., {et~al.} 2023, \apjs, 269,
  42, \dodoi{10.3847/1538-4365/ad02fd}

\bibitem[{{Sick} {et~al.}(2015){Sick}, {Courteau}, {Cuillandre}, {Dalcanton},
  {de Jong}, {McDonald}, {Simard}, \& {Tully}}]{Sick2015}
{Sick}, J., {Courteau}, S., {Cuillandre}, J.-C., {et~al.} 2015, in IAU
  Symposium, Vol. 311, Galaxy Masses as Constraints of Formation Models, ed.
  M.~{Cappellari} \& S.~{Courteau}, 82--85, \dodoi{10.1017/S1743921315003440}

\bibitem[{{Sin} {et~al.}(2017){Sin}, {Henze}, {Sala}, {Ederoclite}, {Hernanz},
  {Jose}, {Hornoch}, {Conseil}, \& {Kucakova}}]{Sin2017}
{Sin}, P., {Henze}, M., {Sala}, G., {et~al.} 2017, The Astronomer's Telegram,
  10001, 1

\bibitem[{{Socia} {et~al.}(2018){Socia}, {Henze}, {Shafter}, \&
  {Horst}}]{Socia2018}
{Socia}, Q., {Henze}, M., {Shafter}, A.~W., \& {Horst}, J.~C. 2018, Research
  Notes of the American Astronomical Society, 2, 190,
  \dodoi{10.3847/2515-5172/aae7ce}

\bibitem[{{Starrfield} {et~al.}(2016){Starrfield}, {Iliadis}, \&
  {Hix}}]{Starrfield2016}
{Starrfield}, S., {Iliadis}, C., \& {Hix}, W.~R. 2016, \pasp, 128, 051001,
  \dodoi{10.1088/1538-3873/128/963/051001}

\bibitem[{{Tang} {et~al.}(2014){Tang}, {Bildsten}, {Wolf}, {Li}, {Kong}, {Cao},
  {Cenko}, {De Cia}, {Kasliwal}, {Kulkarni}, {Laher}, {Masci}, {Nugent},
  {Perley}, {Prince}, \& {Surace}}]{Tang2014}
{Tang}, S., {Bildsten}, L., {Wolf}, W.~M., {et~al.} 2014, \apj, 786, 61,
  \dodoi{10.1088/0004-637X/786/1/61}

\bibitem[{{Torres}(2010)}]{Torres2010}
{Torres}, G. 2010, \aj, 140, 1158, \dodoi{10.1088/0004-6256/140/5/1158}

\bibitem[{{Townsley} \& {Bildsten}(2004)}]{Townsley2004}
{Townsley}, D.~M., \& {Bildsten}, L. 2004, \apj, 600, 390,
  \dodoi{10.1086/379701}

\bibitem[{{Townsley} \& {Bildsten}(2005)}]{Townsley2005}
---. 2005, \apj, 628, 395, \dodoi{10.1086/430594}

\bibitem[{{Warner}(2003)}]{Warner2003}
{Warner}, B. 2003, {Cataclysmic Variable Stars},
  \dodoi{10.1017/CBO9780511586491}

\bibitem[{{Williams} {et~al.}(2015){Williams}, {Dalcanton}, {Dolphin}, {Weisz},
  {Lewis}, {Lang}, {Bell}, {Boyer}, {Fouesneau}, {Gilbert}, {Monachesi}, \&
  {Skillman}}]{Williams2015}
{Williams}, B.~F., {Dalcanton}, J.~J., {Dolphin}, A.~E., {et~al.} 2015, \apj,
  806, 48, \dodoi{10.1088/0004-637X/806/1/48}

\bibitem[{{Williams}(1992)}]{Williams1992}
{Williams}, R.~E. 1992, \aj, 104, 725, \dodoi{10.1086/116268}

\bibitem[{{Williams} \& {Shafter}(2004)}]{Williams2004}
{Williams}, S.~J., \& {Shafter}, A.~W. 2004, \apj, 612, 867,
  \dodoi{10.1086/422833}

\bibitem[{{Wolf} {et~al.}(2013){Wolf}, {Bildsten}, {Brooks}, \&
  {Paxton}}]{Wolf2013}
{Wolf}, W.~M., {Bildsten}, L., {Brooks}, J., \& {Paxton}, B. 2013, \apj, 777,
  136, \dodoi{10.1088/0004-637X/777/2/136}

\bibitem[{Yaron {et~al.}(2005)Yaron, Prialnik, Shara, \& Kovetz}]{Yaron2005}
Yaron, O., Prialnik, D., Shara, M.~M., \& Kovetz, A. 2005, The Astrophysical
  Journal, 623, 398, \dodoi{10.1086/428436}

\bibitem[{{Yungelson} {et~al.}(1997){Yungelson}, {Livio}, \&
  {Tutukov}}]{Yungelson1997}
{Yungelson}, L., {Livio}, M., \& {Tutukov}, A. 1997, \apj, 481, 127,
  \dodoi{10.1086/304020}

\bibitem[{{Zorotovic} \& {Schreiber}(2020)}]{Zorotovic2020}
{Zorotovic}, M., \& {Schreiber}, M.~R. 2020, Advances in Space Research, 66,
  1080, \dodoi{10.1016/j.asr.2019.08.044}

\end{thebibliography}
\bibliographystyle{aasjournal}

\appendix

For completeness, in this appendix we summarize in Table~\ref{tabA1} the observed properties,
including equatorial coordinates (J2000), of the 22 known or
suspected M31 RNe as of 2026 January 1. Two of these systems, M31N 1953-09b and 1961-11a, will require confirmation before their
recurrent nature can be firmly established. Photometric data sufficient to establish the peak absolute magnitude, $M_R (\mathrm{max})$,
and $t_2$ time are
available for just 14 of the 22 systems. These are the 14 RNe that have been analyzed in this paper (see Table~\ref{tab5}).

\begin{deluxetable*}{lccrrccrcr}
\tabletypesize{\scriptsize}
\tablenum{A1}
%\tablewidth{0pt}
\tablecolumns{10}
\tablecaption{M31 Recurrent Novae\label{tabA1}}
\tablehead{\colhead{Recurrent Nova} & \colhead{R.A. (h,m,s)} & \colhead{Decl. ($^{\circ},',''$)} & \colhead{$N_\mathrm{rec}$} & \colhead{$\langle\Delta t\rangle$ (yr)} & \colhead{$\langle M_\mathrm{R}\rangle$ (max)} & \colhead{$\langle t_2\rangle$ (d)} & \colhead{$a$ ($'$)} & \colhead{Type} & \colhead{References\tablenotemark{a}}
}
\startdata
M31N1919-09a & 0 43 28.65 & 41 21 42.1 & 1 & 78.7 &\dots         &\dots         &  10.50 &  \dots & 1  \cr
M31N1923-12c & 0 42 38.07 & 41 08 41.4 & 2 &  9.5 &$-6.96\pm0.18$&$12.6\pm0.9$  &  10.99 &  He/N  & 1,2  \cr
M31N1926-06a & 0 41 40.66 & 41 03 33.7 & 1 & 36.5 &\dots         &$\grtsim$5    &  17.43 &  \dots & 1  \cr
M31N1926-07c & 0 42 52.37 & 41 16 12.8 & 4 &  2.7 &$-7.01\pm0.20$&$11.1\pm1.4$  &   1.68 &  He/Nn & 1,2,3  \cr
M31N1945-09c & 0 41 28.58 & 40 53 13.7 & 1 & 27.0 &\dots         &\dots         &  29.19 &  \dots & 1  \cr
M31N1953-09b:& 0 42 20.69 & 41 16 07.9 & 1 & 50.8 &\dots         &\dots         &   5.77 &  \dots & 1  \cr
M31N1960-12a & 0 42 55.71 & 41 14 12.5 & 2 &  6.2 &$-6.76\pm0.14$&$4.7\pm0.2$   &   4.40 &  He/N  & 1,2  \cr
M31N1961-11a:& 0 42 31.36 & 41 16 21.0 & 1 & 43.6 &\dots         &\dots         &   3.03 &  \dots & 1  \cr
M31N1963-09c & 0 42 57.75 & 41 08 12.3 & 6 &  5.0 &$-6.61\pm0.14$&$4.0\pm0.5$   &  17.60 &  He/N  & 1,2  \cr
M31N1966-09e & 0 39 30.33 & 40 29 14.0 & 1 & 40.9 &$-6.06\pm0.18$&$13.0\pm1.5$  &  59.56 &  Fe II & 1,2  \cr
M31N1982-08b & 0 46 06.68 & 42 03 49.3 & 1 & 14.9 &\dots         &\dots         &  60.83 &  \dots & 1  \cr
M31N1984-07a & 0 42 47.15 & 41 16 19.6 & 3 &  9.4 &$-8.35\pm0.11$&$9.7\pm1.5$   &   0.57 &  Fe IIb& 1,2  \cr
M31N1990-10a & 0 43 04.00 & 41 17 08.1 & 3 &  5.3 &$-7.00\pm0.14$&$11.1\pm1.7$  &   4.28 &  Fe IIb& 1,2  \cr
M31N1997-11k & 0 42 39.60 & 41 09 02.9 & 3 &  4.0 &$-5.96\pm0.14$&$110\pm29$    &  10.57 &  Fe II & 1,2  \cr
M31N2001-11a & 0 44 14.55 & 41 22 04.2 & 1 & 23.2 &$\sim -7.5$   &$\sim 8$      &  27.90 &  \dots & 4   \cr
M31N2005-10a & 0 44 20.70 & 41 23 11.0 & 1 & 16.8 &\dots         &\dots         &  29.43 &  \dots & 5  \cr
M31N2006-11c & 0 41 33.17 & 41 10 12.4 & 1 &  8.2 &$-8.02\pm0.11$&$1.8\pm0.5$   &  19.82 &  He/N  & 2,6  \cr
M31N2007-10b & 0 43 29.47 & 41 17 13.9 & 1 & 10.2 &$-6.21\pm0.20$&$2.7\pm0.6$   &  12.90 &  He/N  & 2,7  \cr
M31N2007-11f & 0 41 31.54 & 41 07 13.6 & 1 &  9.1 &$-7.22\pm0.13$&$10.2\pm2.6$  &  18.90 &  He/Nn & 2,8  \cr
M31N2008-12a & 0 45 28.80 & 41 54 10.1 & 21&  1.0 &$-6.25\pm0.04$&$2.2\pm0.1$   &  48.92 &  He/N  & 1,2,9,10,11 \cr
M31N2013-10c & 0 43 09.54 & 41 15 39.9 & 1 & 10.1 &$-8.66\pm0.14$&$5.5\pm1.7$   &   6.38 &  \dots & 2,12  \cr
M31N2017-01e & 0 44 10.72 & 41 54 22.1 & 5 &  2.5 &$-6.76\pm0.20$&$5.2\pm1.6$   &  54.19 &  He/N  & 13,14,15 \cr
\enddata
\tablenotetext{a}{(1) \citet{Shafter2015}; (2) \citet{Clark2024}; (3) \citet{Shafter2022a};
(4) \citet{Shafter2025}; (5) \citet{Hornoch2022}; (6) \citet{Hornoch2015}; (7) \citet{Socia2018};
(8) \citet{Sin2017}; (9) \citet{Shafter2012}; (10) \citet{Darnley2016};
(11) \citet{Burris2023}; (12) \citet{Shafter2024a};
(13) \citet{Shafter2022b}; (14) \citet{Shafter2022c}; (15) \citet{Shafter2024b}
 }
\end{deluxetable*}

\end{document}